\documentclass[prb,aps,eqsecnum]{revtex4}
\usepackage{graphicx,psfrag,amsmath,amssymb,float}
\usepackage{color}
\newcommand{\be}{\begin{equation}}
\newcommand{\ee}{\end{equation}}
\newcommand{\bea}{\begin{eqnarray}}
\newcommand{\eea}{\end{eqnarray}}
\begin{document}

\title{Bipairing and The Stripe Phase in 4-Leg Hubbard Ladders}
\author{Ming-Shyang Chang}
\affiliation{Department of Physics \& Astronomy, University of
British Columbia, Vancouver, B.C., Canada, V6T 1Z1}
\author{Ian Affleck}
\affiliation{Department of Physics \& Astronomy, University of
British Columbia, Vancouver, B.C., Canada, V6T 1Z1}
\date{\today}

\begin{abstract}
Density Matrix Renormalization Group (DMRG) calculations on 4-leg $t-J$ and
Hubbard ladders have found a phase exhibiting ``stripes'' at intermediate
doping. Such behavior can be viewed as generalized Friedel oscillations,
with wavelength equal to the inverse hole density, induced by the open
boundary conditions. So far, this phase has not been understood using the
conventional weak coupling bosonization approach. Based on studies from
a general bosonization proof, finite size spectrum, an improved analysis of
weak coupling renormalization group equations and the decoupled 2-leg ladders
limit, we here find new types of phases of 4-leg ladders which exhibit
``stripes''. They also inevitably exhibit ``bipairing'', meaning that there
is a gap to add 1 or 2 electrons (but not 4) and that both single electron
and electron pair correlation functions decay exponentially while
correlation functions of charge 4 operators exhibit power-law decay. Whether
or not bipairing occurs in the stripe phase found in DMRG is an important
open question.
\end{abstract}
\maketitle

\section{INTRODUCTION}

There is clear experimental evidence for ``static stripes'' i.e. charge
density waves (CDW's), in some cuprate superconductors at commensurate
doping (where the superconductivity is suppressed) [\onlinecite{Tranquada,
Tranquada04}]. It has been suggested that ``fluctuating stripes'' may occur
at incommensurate doping and may be responsible for superconductivity. It
has also been suggested that stripes only occur in models where the long
range Coulomb interactions are kept [\onlinecite{Kivelson}]. On the other
hand, numerical evidence for stripes has been found in Hubbard and $t$-$J$
ladders, which contain only short range interactions [\onlinecite{White2D,
White2leg, White3leg, White4leg, White4leg2, White6leg, Jeckelmann6leg}]. An understanding of the occurrence of stripes in these models,
whether or not long-range Coulomb interactions are required for their
existence and their connection with superconductivity are important open
questions.

There are many experimental realizations of ladder systems [%
\onlinecite{Ladder1, Ladder2, Ladder3, Ladder4}]. Recently, stripe has been
observed by a resonant X-ray scattering technique in a 2-leg ladder compound [%
\onlinecite{Sawatzky}]. Numerical evidence for stripes comes from density
matrix renormalization group (DMRG) work [\onlinecite{White2D, White2leg,
White3leg, White4leg, White4leg2, White6leg, Jeckelmann6leg}].
Since this method, as originally formulated, is intrinsically
one-dimensional (1D), the results have been presented for ``ladders'' i.e.
finite systems in which the number of rungs is considerably greater than the
number of legs. (For instance, results exhibiting stripes have been
presented for systems of size 6 $\times $ 21 [\onlinecite{Jeckelmann6leg}%
].) In the limit where the length of the ladders (number of rungs) is much
larger than their width (number of legs) they become 1D systems and a
corresponding arsenal of field theory methods, such as bosonization, can be
applied. Combining DMRG results with field theory methods to extrapolate to
the limit of infinitely long ladders is an important step towards reaching
the two dimensional (2D) limit. Of course, an extrapolation in the number of
legs must finally be taken.

DMRG works much more efficiently with open boundary conditions (OBC) and
most work on ladders has used OBC in the leg direction. Such boundary
conditions can induce ``generalized Friedel oscillations'', meaning
oscillations in the electron density which decay away from the boundary with
a non-trivial power law and oscillate with an incommensurate wave-vector,
often related to the hole density [\onlinecite{Ian2leg}]. While sometimes
regarded as an unphysical nuisance, we regard OBC as a useful diagnostic
tool. According to bosonization results, the density-density correlation
function for an infinite length ladder decays with twice the exponent, the
same wave-vector and the square of the amplitude, governing the Friedel
oscillations. Thus the Friedel oscillations are giving information about
correlations in the infinite system. Furthermore, in the 2D limit of an
infinite number of legs, these Friedel oscillations could turn into a static
incommensurate CDW or else fluctuating stripes.

While a true long-range CDW is possible at commensurate filling even in 1D,
bosonization/field theory methods suggest that it is not possible at
incommensurate filling in 1D, giving way instead to boundary induced Friedel
oscillations [\onlinecite{Ian2leg, Friedel}]. (A long-range CDW at $p/q$
filling, where $p$ and $q$ are integers, becomes increasingly suppressed as $%
q$ increases.) This assertion is related to Coleman's or Mermin-Wager's
theorem about the impossibility of spontaneous breaking of continuous
symmetries (and the impossibility of the existence of the corresponding
Goldstone modes) in Lorentz invariant 1D systems even at zero temperature
(or correspondingly 2D classical systems at any finite temperature). A true
long-range incommensurate CDW leads to spontaneous breaking of translational
symmetry by any integer number of lattice spacings, no matter how large,
whereas a commensurate CDW only breaks a finite dimensional symmetry. Taking
the low energy continuum limit, translational symmetry at incommensurate
filling is promoted to a true continuous $U(1)$ symmetry and Coleman's
theorem apparently applies.

``Stripes'' or boundary-induced Friedel oscillations, have been observed in
the $t$-$J$ model, related to the $U\to \infty $ limit of the Hubbard model,
on 2-leg ladders [\onlinecite{Ian2leg}]. These Friedel oscillations, at $%
x\gg 1$, are of the form:
\begin{equation}
\sum_{a=1}^{2}<n_{a}(x)>\to {\frac{A\cos (2\pi nx+\alpha )}{|x|^{2K_{+\rho }}%
}}.  \label{2leg}
\end{equation}
Here the average electron density is,
\begin{equation}
n\equiv N_{e}/(2L),
\end{equation}
where $N_{e}$ is the total number of electrons. $K_{+\rho }$ is the
Luttinger parameter for the charge boson ($\rho $) which is the sum ($+$) of
the 2 charge bosons corresponding to the two bands in a weak coupling
analysis. $A$ and $\alpha $ are constants. Note that we may replace the
oscillation wave-vector by:
\begin{equation}
2\pi n\to -2\pi \delta ,
\end{equation}
where $\delta \equiv 1-n$, is the hole density, measured from half-filling,
since $x$ is always integer. ``Snapshots'' of typical configurations within
the DMRG calculations show pairs of nearby holes, one from each leg,
well-separated from other pairs. An appealing picture is that the holes are
pairing into bosons, 1 hole from each leg. (Here we refer to bosons with a
conserved particle number, such as atoms, not the bosons arising from
bosonization.) This phase is of C1S0 type, indicating that only 1 gapless
charge boson survives and zero gapless spin bosons, out of the 2 charge and
2 spin bosons introduced in bosonizing the 2 leg ladder. This phase exhibits
exponential decay for the single electron Green's function but power law
decay for the electron pair Green's function. One may approximately map the
2-leg fermionic ladder into a (single leg) bosonic chain. The standard
superfluid phase of this boson model exhibits Friedel oscillations at
wave-vector $2\pi \delta $ where $\delta =N_{b}/L$, is the number of bosons
(i.e. hole pairs) per unit length and the number of bosons is: $%
N_{b}=N_{h}/2 $, were $N_{h}$ is the number of holes. These Friedel
oscillations in the bosonic model just correspond to a sort of quasi-solid
behavior. We can think of the bosons as almost forming a solid near the
boundary with a uniform spacing between all nearest neighbor bosons. This,
of course, coexists with quasi-superfluid behavior since the phase
correlations also decay with a power law. Thus it could be called
quasi-supersolid behavior and is typical of many 1D systems. In general the
density profile will contain other Fourier modes besides the $2\pi n$ mode
kept in Eq. (\ref{2leg}). This $2\pi n$ mode dominates in the sense that it
has both the smallest wave-vector and also the smallest power law decay
exponent. Note that this behavior is quite different than what we might
expect in a C2S0 phase, for example, or which occurs at zero interaction
strength in the C2S2 phase. Then we expect density oscillations at
wave-vector $2k_{F,e}$ and also $2k_{F,o}$ where $k_{,Fe/o}$ are the Fermi
wave-vectors for the even and odd bands. (They are even or odd under the
parity transformation that interchanges the two legs.) In the C1S0 phase
that is observed in 2-leg ladders, the oscillation wave-vector can be
written as $2\pi n=2(k_{F,e}+k_{F,o})=4\overline{k}_{F}$ where $\overline{k}%
_{F}$ is the average Fermi wave-vector.

DMRG works on the doped 4-leg $t$-$J$ model exhibited two phases, both of
which appear to have a spin gap [\onlinecite{White4leg, White4leg2,
Troyer, Capponi}]. At low doping, the dominant Friedel oscillation
wave-vector appears to be $4\pi n$, where $n$, the average electron density
is now:
\begin{equation}
n\equiv N_{e}/(4L).
\end{equation}
Above a critical doping $\delta _{c}$, corresponding to $\delta _{c}\approx
1/8$, for $J=0.35t$ and $0.5t$ , the oscillation wave-vector changes to $%
2\pi n$. DMRG ``snapshots'' of typical configurations suggest well separated
pair holes in the lower density phase but 4-hole clusters (1 hole on each
leg) in the higher density phase. This is consistent with the $2\pi n$
Friedel oscillation wave-vector since the average separation along the
ladder of the equally spaced 4-hole clusters would be $1/n$. This higher
density phase with $2\pi n$ oscillation wave-vector has been referred to as a
stripe phase.

In the standard weak coupling approach we assume that all the interactions
are small compared to the hopping. Thus we first solve for the band
structure of the non-interacting model and then take the continuum limit of
the interacting model, yielding right and left moving fermions from each
band. These continuum limit fermions are then bosonized. Letting $k_{Fi}$ be
the Fermi wave-vector of the 4 bands ($i=1,2,3$ or $4$) the number of
electrons in each band (summing over both spins) is:
\begin{equation}
N_{e}^{i}=L(2k_{Fi}/\pi ),\ \ (i=1,2,3,4).
\end{equation}
Thus we see that the electron density is:
\begin{equation}
n=N/4L=\sum_{i=1}^{4}k_{Fi}/(2\pi ),
\end{equation}
so
\begin{equation}
2\pi n=\sum_{i=1}^{4}k_{Fi}=4\bar{k}_{F}.  \label{n4kF}
\end{equation}
Thus the stripe phase again corresponds to Friedel oscillations at a
wave-vector of $4\bar{k}_{F}$. Using the equivalence of $2\pi n$ with $-2\pi
\delta $, the physical picture of the stripe phase is well-separated
clusters of 4 holes (one on each leg). If such 4-hole clusters are equally
spaced we obtain a Friedel oscillation wave-vector of $2\pi n$, since the
average separation along the legs of the clusters is $1/n$. On the other
hand, the lower density phase shows no evidence for Friedel oscillations at
wave-vector $2\pi n$ and instead $4\pi n$ ($8\bar{k}_{F}$) oscillations
appear to dominate. (We note that, when interactions are included, we expect
the Fermi wave-vectors to be renormalized or lose their significance
entirely. However, the sum of all Fermi wave-vectors, $4\bar{k}_{F}=2\pi n$,
is known to still be a meaningful and ``unrenormalized'' wave-vector even in
the presence of interactions. This follows from the 1D version of
Luttinger's theorem, proven in Ref. [\onlinecite{Yamanaka}]). This stripe
phase appears to have a spin gap. Limited DMRG results have been presented
on the decay of the pair correlation function. Pairing correlations appear
to go through a maximum, as a function of doping, at a somewhat higher
doping than $\delta _{c}$ where their behavior appears consistent with
power-law decay [\onlinecite{White4leg, White4leg2}].

A simpler picture of the stripe phase of 4 leg ladders was obtained by
mapping each hole pair on the upper 2 legs into a boson moving on a single
chain, and likewise mapping each hole pair on the lower 2 legs onto a boson
moving on a different single chain. Thus the 4-leg fermionic ladder is
mapped into a 2-leg bosonic ladder [\onlinecite{Troyer}].

Bosonization analysis of the 4-leg ladder it is not an easy task. A total of
8 left and right-moving bosonic fields must be introduced, one for each spin
component in each band (or rung). In a general phase any of these boson fields or
their duals might be pinned leading to $3^{8}$ possible phases! Taking into
account that the points where the fields are pinned are of physical
significance, leads to even higher estimates of the number of possible
phases. Additional phases occur if it is assumed that some of the bands are
completely filled (or empty). Keeping only Lorentz invariant, non-Umklapp
interactions in the continuum limit there are 32 different interactions
terms and hundreds of non-zero terms in the RG $\beta $-functions at
quadratic order. Analysis of these RG equations generally indicates that
some couplings flow to the strong coupling region where the equations break
down. The bosonization analysis of [\onlinecite{NlegRG}] predicted more
than 10 different phases for the 4-leg ladder with open boundary conditions
in the leg direction, as hoppings and hole doping are varied. However, only
two of these phases have a spin gap and neither of them has the right
Friedel oscillation wave-vector to describe the stripe phase seen in DMRG
work.

An alternative bosonization approach was introduced in [%
\onlinecite{Ledermann}]. This was based on the limit of low doping with
small but finite on-site interaction $U$. Although details were sketchy, it
seemed to correctly reproduce the two phases found by DMRG for the 4-leg
ladder.

The purpose of this note is to reexamine bosonization approaches to 4-leg
ladders and the nature of the stripe phase found in DMRG. We first give a
general discussion of the conventional weak-coupling bosonization approach
and of Friedel oscillations in 4-leg ladders. We restrict our attention to
C1S0 phases where there are various candidates for the stripe phase. We
sketch a proof that \textit{all} possible phases with stripes inevitably have
exponentially decaying pair correlations within this approach. However, they
can have bipairing, corresponding to quartets of fermions. The finite size spectrum
(FSS) of the stripe phase, which indicates the total charge excitations
relative to the ground state can only be multiples of four, is also
consistent with bipairing. On the other hand, a pairing phase should
correspond to $8\bar{k}_{F}$ density oscillations instead of stripes. Then
we show that the stripe phase could arise from an improved analysis of the so-called ``fixed ray solutions'' of the renormalization group (RG) equations. This new
analysis [\onlinecite{LinPrivate}] is based upon the potential structure of
the RG equations [\onlinecite{Chen, MSChang}] and gives the C1S0 phases with
correlations consistent with our general discussions based on bosonization
and FSS.

We then discuss a different bosonization approach. We start with two identical
decoupled 2-leg ladders and then turn on the interleg hopping $t_{2,\perp }$ and
interleg interaction $V_{2,\perp }$, which are much smaller than the other
energy scales in the 2-leg Hamiltonian. Now the system becomes a pair of
weakly coupled identical 2-leg ladders. Each 2-leg ladder is in the C1S0
phase [\onlinecite{2leg}] and we also assume that $t_{2,\perp }$ and $%
V_{2,\perp }$ are small compared to the gaps to the other 3 modes in each
2-leg ladder. The model then becomes equivalent to the continuum limit of a
2-leg ladder of bosons, rather than fermions, of the type studied in [%
\onlinecite{Orignac}]. This model has two phases: 4$\pi \rho _{0}$ density
oscillation (with superfluidity) and 2$\pi \rho _{0}$ density oscillation (with
boson-pair superfluidity) that correspond to the 2 phases seen in DMRG, where $%
\rho _{0}$ is the average boson density. The Friedel oscillations correspond
in the two models and the stripe phase has boson-pairs, i.e. bipairing.
Interestingly, these two phases also have the same features as those found
in the improved weak coupling RG, where 2$\pi \rho _{0}$, 4$\pi \rho _{0}$,
superfluid and boson-pair superfluid in two-leg bosonic ladders correspond
to $4\bar{k}_{F}$, $8\bar{k}_{F}$, pairing and bipairing, respectively, in
4-leg fermoinic ladders.

This approach is very close to that of [\onlinecite{Troyer}] which maps the
4-leg fermionic model onto a 2-leg bosonic model in a more phenomenological
way, based on exact diagonalization and DMRG. Our results are consistent
with those of that paper which did not, however, point out the occurance of
bipairing in the striped phase. This approach, and our conclusions are also
closely related to those of [\onlinecite{Ledermann}], as we will discuss. We
point out that the numerical results [\onlinecite{Troyer}] on the bosonic model which is argued to
correspond to the 4-leg fermionic model in its stripe phase show some
evidence for boson pairing. However, published results on the 4-leg
fermionic ladder so far do not show a gap for pairs as should occur in the
bipairing phase. Since the ladder length in Ref. [\onlinecite{Troyer}] was only $L=24$, the problem may be
that the pair-gap is too small compared to the finite size gap at this
ladder length. More DMRG results on longer ladders may be required to
confirm that the stripe phase indeed has bipairing. It is similarly unclear
whether pair correlations decay exponentially, as would occur in a bipairing
phase, or with a power law as in a pairing phase. More DMRG work should shed
light on this question.

If this bipairing assumption is correct it finally yields a remarkably
simple picture of the stripe phase in 4-leg ladders. It is a phase in which
electrons do not form pairs, but rather bipairs. While the usual pairing
does not occur in this phase, it tends towards a more exotic form of
supercondutivity based on condensed charge-4 objects. There are proposals in
the literature that some systems like a frustrated Josephson junction chain [%
\onlinecite{4eJosephsonPRL, 4eJosephsonPRB}], strongly coupled fermions [%
\onlinecite{4eNozieres}], cuprates with strong phase fluctuations [\onlinecite{4eUnpublished}] and a spin 3/2 fermionic chain [\onlinecite{Wu}] can have unusual pairs composed of 2 Cooper pairs or four-particle condensations. Experimental evidences that Cooper pairs tunnel in pairs across a
$(100)/(110)$ interface of two $d-$wave superconductors are also reported [\onlinecite{4eExp1, 4eExp2}]. They are very similar to but not exactly the same as the bipairing here. Starting with
a four band system, the phase we propose is more like a generalized
Luther-Emery phase with four-holes.

We note that the stripes observed in 6-leg ladders [\onlinecite{White6leg,
Jeckelmann6leg}] appear to have a more dynamical character, containing
less than 6 holes per stripe. Thus they may exhibit behavior closer to that
proposed in the 2D case.

In the next section we review some features of bosonization of 4-leg ladders
including Friedel oscillations and sketch a proof of one of our main results: within the
usual bosonization framework any phase with stripes does not have pairing
(but may have bipairing). This result applies very generally and does not
depend on the detailed form of the Hamiltonian. In Sec. III we study the
finite size spectrum corresponding to a stripe phase. In Sec. IV we study
weak coupling RG equations, showing that a stripe phase can arise from
studying fixed ray solutions. In Sec. V we discuss the case of small $%
t_{\perp ,2}$, $V_{\perp ,2}$ and the analogy with a 2 leg bosonic ladder.
Sec. VI contains conclusions. In Appendix A, we give some more details about the proof that stripes and pairing can not coexist. In Appendix B we derive in detail, in the weakly coupled Hubbard
model, how the density operator obtains terms leading to $4k_{F}$ oscillations. The initial conditions of RG equations in terms of the bare interactions and Fermi velocities are given in Appendix C.

\section{continuum limit and bosonization}

\subsection{Bosonization}

In this section and the next we focus on a weak coupling treatment of the
Hamiltonian on a ladder with $N$ legs and $L$ rungs with open boundary
conditions in both rung and leg directions. The Hamilonian is $%
H=H_{0}+H_{int}$, with
\begin{equation}
H_{0}=-\sum_{\alpha =\pm }\left[ \sum_{x=1}^{L-1}\sum_{a=1}^{N}tc_{a,\alpha
}^{\dagger }(x)c_{a,\alpha }(x+1)+\sum_{x=1}^{L}\sum_{a=1}^{N-1}t_{\perp
}c_{a,\alpha }^{\dagger }(x)c_{a+1,\alpha }(x)\right] +h.c.  \label{H0}
\end{equation}
and
\begin{equation}
H_{int}=U\sum_{x=1}^{N}\sum_{a=1}^{L}n_{a,\uparrow }(x)n_{a,\downarrow }(x)
\label{Hint}
\end{equation}
Here $n_{a,\alpha }\equiv c_{a,\alpha }^{\dagger }c_{a,\alpha }$. We first
diagonalize the rung hopping terms in the Hamiltonian by transforming to the
band basis, $\psi _{i,\alpha }$, for $i=1$, $2$, $3$, $4$:
\begin{equation}
\psi _{i,\alpha }\equiv \sum_{a=1}^{4}S_{ia}c_{a,\alpha },
\label{bandFermion}
\end{equation}
where $S$ is a unitary matrix:
\begin{equation}
S_{ia}=\sqrt{\frac{2}{5}}\sin \left( \frac{\pi }{5}\ ia\right) .
\label{S_matrix}
\end{equation}
The dispersion relation for the $j^{\hbox{th}}$ band is $\epsilon
_{j}(k)=-2t\cos k-2t_{\perp }\cos (k_{yj}),$ where $k_{yj}=j\pi /5$. We may
take the continuum limit, to study the low energy physics, by introducing
right and left moving fermions fields (sometimes we call them chiral
fermions) which contain wave-vectors of the original band fermion fields
near the Fermi points,
\begin{equation}
\psi _{j\alpha }(x)\;\sim \;\psi _{Rj\alpha }(x)\;e^{ik_{Fj}x}\;+\;\psi
_{Lj\alpha }(x)\;e^{-ik_{Fj}x},  \label{chiral}
\end{equation}
where $k_{Fj}$ is the Fermi wave vector for band $j$. Then we can bosonize
these right and left fermions by the dictionary
\begin{equation}
\psi _{R/Li\alpha }\sim \eta _{i}e^{i\sqrt{4\pi }\varphi _{R/Li\alpha }}
\label{bosonize}
\end{equation}
where $\eta _{i}$ are Klein factors with $\left\{ \eta _{i},\eta
_{j}\right\} =2\delta _{ij}$. The commutation relations of the boson fields
are
\begin{eqnarray}
\left[ \varphi _{Ri\alpha }(x),\varphi _{Lj\beta }(y)\right]  &=&\frac{i}{4}%
\delta _{ij}\delta _{\alpha \beta }, \\
\left[ \varphi _{Ri\alpha }(x),\varphi _{Rj\beta }(y)\right]  &=&\frac{i}{4}%
sgn(x-y)\delta _{ij}\delta _{\alpha \beta }, \\
\left[ \varphi _{Li\alpha }(x),\varphi _{Lj\beta }(y)\right]  &=&\frac{-i}{4}%
sgn(x-y)\delta _{ij}\delta _{\alpha \beta }.
\end{eqnarray}
Now we have two chiral bosons and it's more convenient to describe physics
by the conventional bosonic field $\phi $ and its dual field $\theta $,
\begin{equation}
\phi _{i\alpha }=\varphi _{Ri\alpha }+\varphi _{Li\alpha },\theta _{i\alpha
}=\varphi _{Ri\alpha }-\varphi _{Li\alpha },  \label{RL}
\end{equation}
since the partial derivatives of $\theta $ and $\phi $ are density and current,
respectively. We can also separate the charge $\rho $ and spin $\sigma $
degrees of freedom by introducing the following fields,
\begin{eqnarray}
\phi _{i\rho } &=&\frac{1}{\sqrt{2}}(\phi _{i\uparrow }+\phi _{i\downarrow
}),  \label{band_boson1} \\
\phi _{i\sigma } &=&\frac{1}{\sqrt{2}}(\phi _{i\uparrow }-\phi _{i\downarrow
})  \label{band_boson2}
\end{eqnarray}
and similarly for $\theta $ fields. Then Eq. (\ref{H0}) becomes
\begin{equation}
H_{0}=\sum_{i,\nu }\frac{v_{i}}{2}\int dx[(\partial _{x}\phi _{i\nu
})^{2}+(\partial _{x}\theta _{i\nu })^{2}],  \label{H_quadratic}
\end{equation}
where $\nu =\rho $ or $\sigma $ and $v_{i}$ is the Fermi velocity of band $i$%
. We define new fields from the linear combinations of the bosons from
different bands:
\begin{equation}
\phi _{ij}^{\rho ,\sigma \pm }=\frac{1}{\sqrt{2}}(\phi _{i\rho ,\sigma }\pm
\phi _{j\rho ,\sigma }),  \label{2-leg BF}
\end{equation}
and the same for $\theta _{ij}^{\rho ,\sigma \pm }$ and $\theta _{ij}^{\sigma \pm }$%
. Eq. (\ref{2-leg BF}) is convenient when we consider the bosonized
interactions but is not necessary the best basis to describe the system. The basis
to describe the system should be the one in which boson fields get pinned, in
principle determined by the interactions. We know that there should be four
mutually orthogonal charge and spin bosons for 4-leg ladders. That is to
say, starting with bosons in the band basis, Eq. (\ref{band_boson1}) and (%
\ref{band_boson2}), guided by the interactions, we can find the proper new
basis in which some linear combinations of band bosons are pinned. The new and
band basis only differ by an orthogonal transformation. In general, the
transformations of charge and spin fields don't have to be same.

Due to the different Fermi velocities in Eq. (\ref{H_quadratic}), after
changing to a new basis, there will be mixing terms between the derivative
of boson fields. These mixing derivative terms are only up to quadratic order in the boson fields.
Spinless fermions on 2-leg ladders were studied in Ref. [\onlinecite{LeHur}]
and they found that the mixing terms only modify the exponents of
correlation functions within the conventional bosonization analysis. We are
mainly concerned about whether an operator is power law or exponentially
decaying but not its exponent. In the later sections, we only discuss the
mixing terms if it's necessary.

\subsection{Stripes and Bipairing}

In this section and Appendix A, we wish to sketch a proof of a general result that would apply to any of
these phases that are candidates for a stripe phase, regardless of what
basis for the boson fields we use. Stripes are incommensurate density
oscillations with the lowest wave-vector $2\pi n$ (e.g. no $\pi n$ density
oscillations) [\onlinecite{White4leg, White4leg2}]. The wave-vector $2\pi n$ is
equivalent to $4\bar{k}_{F}$ via 1D Luttinger theorem [\onlinecite{Yamanaka}].
Stripes thus can be described as the generalized Friedel oscillations
induced by the boundary [\onlinecite{Ian2leg}]. A derivation of the $4\bar{k}_{F}$
components of density operators is given in Appendix B. At incommensurate
filling, the 1D Luttinger theorem suggests that at least one gapless charge mode
always exists. Stripe phases appear to have a spin gap in DMRG works and we
expect that phases with more than one gapless charge mode are generically
unstable. So we restrict out attention to C1S0 phases.

We then argue that \textit{any} phase exhibiting stripes (i.e. $2\pi n$
oscillations) cannot exhibit pairing (i.e. power law pair correlations and
gapless pair excitations) but can exhibit bipairing (charge four operators).

In a C1S0 phase, 7 out of 8 boson fields are pinned due to interactions. We
will consider all possible four fermions interactions written in terms of
right and left moving fields $\psi _{R/L}$, including those that are ignored
in the conventional weak coupling analysis. Similar to what we discussed in
the previous section, after bosonizing, if the interaction is relevant, the
boson fields in the interaction will tend to be pinned to constants. Before
we proceed, we should explain explicitly what we mean by a boson field $%
\theta $ (or $\phi $) being pinned. Consider a vertex operator of a boson
field, $e^{iq\theta }$, where $a$ is a constant. If $\left\langle
e^{iq\theta }\right\rangle $ $=$ const $\neq 0$, then we say $\theta $ is
``pinned'' to some constant modulo $2\pi /q$. On the other hand, we say $%
\theta $ is ``unpinned'' if $\left\langle e^{iq\theta }\right\rangle $ $=0$.
There could be three situations for $\theta $ to be unpinned. First, $\theta
$ is gapless. Second, a part of it is gapless and the rest is pinned. For
example, if $(\theta _{1\rho }+\theta _{2\rho })/\sqrt{2}$ is gapless and $%
(\theta _{3\rho }+\theta _{4\rho })/\sqrt{2}$ is pinned, then $\left\langle
e^{iq\Theta _{1\rho }}\right\rangle $ is still zero since we can replace the
pinned fields by their pinned values. Thirdly, its dual field $\phi $ or
part of its dual field is pinned and $\theta $ will fluctuate violently.

There are many such phases characterized by which fields are pinned. In
general, it is not appropriate to simply label the pinned fields as $\phi
_{i\nu }$ or $\theta _{i\nu }$ in terms of the band basis. The interaction
terms in the Hamiltonian involve various linear combinations of these fields
and it is generally necessary to characterize phases by first going to a
different basis of boson fields:
\begin{equation}
\vec{\theta}_{\rho }^{\prime }=R_{\rho }\vec{\theta}_{\rho },\ \vec{\phi}%
_{\rho }^{\prime }=R_{\rho }\vec{\phi}_{\rho },\ \vec{\theta}_{\sigma
}^{\prime }=R_{\sigma }\vec{\theta}_{\sigma },\ \vec{\phi}_{\sigma }^{\prime
}=R_{\sigma }\vec{\phi}_{\sigma },
\end{equation}
where $\vec{\theta}_{\rho }=(\theta _{1\rho },\theta _{2\rho },\theta
_{3\rho },\theta _{4\rho })$ etc and $R_{\rho }$ and $R_{\sigma }$ are
orthogonal matrices. We will refer to the basis $(\theta _{i\rho }^{\prime
},\phi _{i\rho }^{\prime })$ and $(\theta _{i\sigma }^{\prime },\phi
_{i\sigma }^{\prime })$ as the ``pinning basis''. One of the 4 charge bosons
in the new basis, $(\theta _{i\rho }^{\prime },\phi _{i\rho }^{\prime })$
remains gapless in a C1S0 phase. All of the other 7 bosons are gapped.
However, for each of these 7 bosons we must specify whether it is $\theta
_{i\nu }^{\prime }$ or $\phi _{i\nu }^{\prime }$ which is pinned. We refer
to this choice of which bosons are pinned as a ``pinning pattern''. Each
pinning pattern corresponds to a distinct phase. (Actually the number of
distinct phases is even larger than this since the points, $c_{i\nu }$ at
which a boson is pinned, eg. $\left\langle \theta _{i\nu }^{\prime
}\right\rangle =c_{i\nu }$, can also characterize the phase.)

Following conventional bosonization analysis, we will assume that all C1S0 phases are
characterized by such a pinning pattern. Our basic strategy is to check through all
possible pinning patterns for a C1S0 phase and see if any pinning pattern can make the correlation functions
of pair and $4\bar{k}_{F}$ density operators decay with a power-law at the
same time. Given a pinning pattern, it's easy to know whether an operator has
power-law decaying correlations or not. We can replace the pinned bosons by
constants inside the correlation function. The correlation function will
decay exponentially if the vertex operator contains the dual of a pinned
boson. It will decay with a power-law if the vertex operator only contains
the gapless mode and pinned bosons. However, there are three main
complications here. The first one is: what is the basis for the boson fields
(including the gapless mode)? Even though we know the basis, there are still
too many pinning patterns (more than $2^{7}$). The last problem is that the
wave-vectors of $4k_{F}$ density operators can be changed and actually
correspond to that of stripes if some Fermi momentum renormalization occurs.
For example, the wave-vector $k_{F1}+k_{F2}+2k_{F3}=4\bar{k}_{F}$ if the
condition $k_{F3}=k_{F4}$ is satisfied. In this subsection, we try to solve
these problems. As a matter of fact, $R_{\rho }$ and $R_{\sigma }$ are not
arbitrarily orthogonal matrices and they have to satisfy some constraints
due to the symmetries. Then we can classify the pinning patterns according
to the possible gapless charge modes. For each case with the different
gapless charge mode, we can systematically examine whether pairing and
stripes can coexist or not by the inner product argument which will be
introduced later. When the pinning patterns with all the possible gapless
charge modes are checked, we complete the proof. In order to know what field
can be a gapless charge mode, we should study symmetries first.

We now want to discuss the consequences of symmetries. We would like
to introduce a specific combination of band bosons Eq. (\ref{band_boson1})
and (\ref{band_boson2}), which is relevant to symmetry. Define the total
charge (or spin) field as
\begin{equation}
\Theta _{1\nu }=\frac{1}{2}(\theta _{1\nu }+\theta _{2\nu }+\theta _{3\nu
}+\theta _{4\nu }),  \label{4leg-BF}
\end{equation}
where $\nu =\rho $ or $\sigma $ and similarly for $\Phi _{1\nu }$ if we
replace $\theta $ by $\phi $.

We can somewhat restrict the possible pinning patterns by symmetry
considerations. Charge conservation symmetry, $\psi \to e^{i\gamma }\psi $
for all fermion fields, $\psi $, corresponds to the translation:
\begin{equation}
\phi _{j\rho }\to \phi _{j\rho }+\sqrt{2/\pi }\gamma .  \label{phi_j_rho}
\end{equation}
Thus, the allowed pinned $\phi _{i\rho }^{\prime }$ must be the linear
combinations of $\phi _{ij}^{\rho -},$ in other words, any pinned $\phi
_{i\rho }^{\prime }$ fields must be orthogonal to $\Phi _{1\rho }$.
Similarly, the subgroup of $SU(2)$ symmetry, $\psi _{\alpha }\to e^{i\alpha
\epsilon }\psi _{\alpha }$ where $\alpha =+$ or $-$ for spin up or down,
respectively, corresponds to:
\begin{equation}
\phi _{j\sigma }\to \phi _{j\sigma }+\sqrt{2/\pi }\epsilon .
\end{equation}
Therefore, any pinned $\phi _{i\sigma }^{\prime }$ can only be the linear
combinations of $\phi _{ij}^{\sigma -}$ in order not to violate the $SU(2)$
symmetry. We also expect translation symmetry to be unbroken at
incommensurate filling, as discussed in Sec. I. We see from Eq. (\ref{chiral}%
), that translation by one site: $x\to x+1$, corresponds to the symmetry:
\begin{equation}
\psi _{R/Lj\nu }\to e^{\pm ik_{Fj}}\psi _{R/Lj\nu },\ \ (j=1,2,3,4)
\end{equation}
corresponding to:
\begin{equation}
\theta _{j\rho }\to \theta _{j\rho }+\sqrt{2/\pi }k_{Fj},\ \ (j=1,2,3,4).
\label{theta_j_rho}
\end{equation}
Consider a vertex operator of a boson field, $e^{iq\theta }$, where $q$ is a
constant. Since in general the factors $\sqrt{2/\pi }k_{Fj}$ are not zero
modulo $2\pi /q$, this may forbid pinning of \textit{any} of the $\theta
_{j\rho }^{\prime }$ fields. However, it may happen, due perhaps to some
renormalization phenomenon, that 2 or more of the $k_{Fj}$ are equal. In
that case it might be possible for some of the $\theta _{j\rho }^{\prime }$
bosons to be pinned. For example if $k_{Fi}=k_{Fj}$, then $\theta
_{ij}^{\rho -}$ could be pinned. We will allow for that possibility since
there is no reason to exclude it in the strong coupling region. However, $%
\Theta _{1\rho }$ can \textit{never} be pinned since it transforms under
translation by one site as:
\begin{equation}
\Theta _{1\rho }\to \Theta _{1\rho }+4\bar{k}_{F}/\sqrt{2\pi },
\end{equation}
and $4\bar{k}_{F}/2\pi =n$ is always non-zero modulo $2\pi /q$ at
incommensurate filling. The fact that $\Theta _{1\rho }$ can never be pinned
is not exactly equivalent to the statement that the gapless boson in a
general C1S0 phase at incommensurate filling has to be $(\Theta _{1\rho
},\Phi _{1\rho })$. There could be exceptions if some pinned $\theta _{i\rho
}^{\prime }$ fields are \textit{not} orthogonal to $\Theta _{1\rho }$. If
that happens, then $\Theta _{1\rho }$ can't even be chosen as a basis field.
As we know, all the pinned $\phi _{i\rho }^{\prime }$ must be orthogonal to $%
\Phi _{1\rho }$ due to the charge conservation. However, translational
symmetry doesn't demand all the pinned $\theta _{i\rho }^{\prime }$ must be
orthogonal to $\Theta _{1\rho }$ and in general whether a $\theta _{i\rho
}^{\prime }$ is pinned or not depends on the Fermi momentum. Thus to know
whether a $\theta _{i\rho }^{\prime }$ can be pinned without violating
translational symmetry, we have to discuss the possible Fermi momentum
renormalization.

However, even though some $\theta _{i\rho }^{\prime }$ field is allowed to
be pinned by symmetry, in practice, whether it really gets pinned or not
depends on the interactions in the Hamiltonian. In the weak coupling
treatment, the renormalizations of $k_{Fj}$ are assumed not to happen and $%
k_{Fj}$ are in general all different. The interactions involving $\theta
_{i\rho }^{\prime }$ fields won't be present in the Hamiltonian due to the
fast oscillating factors in front of them. However, in the strong coupling
region, if some renormalization of Fermi momentum occurs such that the
oscillating factor becomes a constant, then new interaction will appear in
the Hamiltonian. For example, if $2(k_{Fi}+k_{Fj})=2\pi $, the interaction
involving a $\theta _{i\rho }^{\prime }$ field, such as $%
e^{-i2(k_{Fi}+k_{Fj})x}\psi _{Ri\alpha }^{\dagger }\psi _{Rj\overline{\alpha
}}^{\dagger }\psi _{Li\overline{\alpha }}\psi _{Lj\alpha }\propto e^{-i\sqrt{%
4\pi }(\theta _{ij}^{\rho +}\pm \phi _{ij}^{\sigma -})}$, can be present in
the Hamiltonian since the oscillating factor $e^{-i2(k_{Fi}+k_{Fj})x}$
becomes a constant at each lattice site $x$. In this case the shift of $%
\theta _{ij}^{\rho +}$ under translation is $2\pi /\sqrt{4\pi }$, which
becomes $2\pi $ for $\sqrt{4\pi }\theta _{ij}^{\rho +}$. Thus if this
interaction is relevant, then then the field $\theta _{ij}^{\rho +}$ will be
pinned.

This is an important point for our proof. We will discuss whether a $\theta
_{i\rho }^{\prime }$ can be pinned based on the Fermi momentum
renormalization that results in the new interaction involving $\theta
_{i\rho }^{\prime }$ in the Hamiltonian. The reason is that the number of
different oscillating factors is finite and we can certainly check all of them.
In the proof, we first want to know what could be the gapless charge mode. $%
\Theta _{1\rho }$ and $\Phi _{1\rho }$ are always unpinned but this doesn't
mean that the gapless mode has to be $(\Theta _{1\rho },\Phi _{1\rho })$.
We have to know when $\theta _{i\rho }^{\prime }$ fields \textit{not} orthogonal to $\Theta _{1\rho }$ can be pinned. We can find all such situations by checking through the interactions containing $\theta _{i\rho
}^{\prime }$ fields not orthogonal to $\Theta _{1\rho }$ and then conclude what are the
gapless charge modes. For example if $\theta _{ij}^{\rho +}$, not orthogonal
to $\Theta _{1\rho }$, is pinned, then the gapless mode won't be $\Theta
_{1\rho }$ though $\Theta _{1\rho }$ is still unpinned. This can
happen if $2(k_{Fi}+k_{Fj})=2\pi $, as we discussed above. So we just have
to check those Fermi momentum renormalizations which will make the
interactions, containing $\theta _{i\rho }^{\prime }$ fields not orthogonal
to $\Theta _{1\rho }$, appear in the Hamiltonian.

Besides these issues related to symmetry and interactions, the
renormalizations of Fermi momenta have another effect on our discussions.
Our goal is to answer whether stripes and pairing can coexist or not.
Stripes correspond to the $4\overline{k}_{F}$ density operators. However, a
general $4k_{F}$ density operator may correspond to the $4\overline{k}_{F}$
one if some suitable Fermi momentum renormalization happens. For example,
the wave-vector $2k_{F1}+k_{F2}+k_{F3}$ is the same as $4\bar{k}_{F}$ if $%
k_{F1}=k_{F4}$. This means that for each pinning pattern we have to
carefully discuss all the $4k_{F}$ density operators with some Fermi
momentum renormalizations where their wave-vectors become that of stripes.

We have checked through all the possible pinning patterns to see if stripes
and pairing can coexist. This seems difficult to do since there is in
principle an infinite number of pinning patterns. The fact that the pinning
basis is determined by the orthogonal matrices $R_{\rho }$ and $R_{\sigma }$
greatly reduces the difficulty. We will show that whether two operators can
both decay with a power-law for a given pinning pattern only depends on the operator
forms written in terms of band basis and what is the gapless mode.

First of all, we have to know the conditions so that two (or more) operators
can (or can't) be power law decaying at the same time. To find these
conditions, consider two arbitrary operators expressed in terms of the band
basis and let's rewrite them in the following way:
\begin{eqnarray}
&&O_{A}\sim e^{i\left( \vec{u}_{\rho A}\cdot \vec{\theta}_{\rho }+\vec{v}%
_{\rho A}\cdot \vec{\phi}_{\rho }+\vec{u}_{\sigma A}\cdot \vec{\theta}%
_{\sigma }+\vec{v}_{\sigma A}\cdot \vec{\phi}_{\sigma }\right) },
\label{O_vector1} \\
&&O_{B}\sim e^{i\left( \vec{u}_{\rho B}\cdot \vec{\theta}_{\rho }+\vec{v}%
_{\rho B}\cdot \vec{\phi}_{\rho }+\vec{u}_{\sigma B}\cdot \vec{\theta}%
_{\sigma }+\vec{v}_{\sigma B}\cdot \vec{\phi}_{\sigma }\right) }.
\label{O_vector2}
\end{eqnarray}
We can represent any vertex operator $O_{A}$ by four coefficient vectors $%
\vec{u}_{\rho A},$ $\vec{v}_{\rho A},$ $\vec{u}_{\sigma A}$ and $\vec{v}%
_{\sigma A}$. For example, if $O_{A}\sim e^{i\sqrt{2\pi }(\phi _{1}^{\rho
}+\theta _{1}^{\sigma })}$, then $\vec{v}_{\rho A}=(\sqrt{2\pi },0,0,0)$, $%
\vec{u}_{\sigma A}=(\sqrt{2\pi },0,0,0)$ and both $\vec{u}_{\rho A}$ and $%
\vec{v}_{\sigma A}$ are $(0,0,0,0)$.

Eq. (\ref{O_vector1}) and (\ref{O_vector2}) are written in terms of the
operators in the band basis, which is not necessarily the basis in which
boson fields are pinned. At this stage, we may not know what the new basis
should be but we know at least that the fields in the transformed basis have
to be orthogonal to each other. Then Eq. (\ref{O_vector1}) and (\ref
{O_vector2}) can be rewritten as
\begin{eqnarray}
&&O_{A}\sim e^{i\left( \vec{u}_{\rho A}^{\prime }\cdot \vec{\theta}_{\rho
}^{\prime }+\vec{v}_{\rho A}^{\prime }\cdot \vec{\phi}_{\rho }^{\prime }+%
\vec{u}_{\sigma A}^{\prime }\cdot \vec{\theta}_{\sigma }^{\prime }+\vec{v}%
_{\sigma A}^{\prime }\cdot \vec{\phi}_{\sigma }^{\prime }\right) },
\label{O'_vector1} \\
&&O_{B}\sim e^{i\left( \vec{u}_{\rho B}^{\prime }\cdot \vec{\theta}_{\rho
}^{\prime }+\vec{v}_{\rho B}^{\prime }\cdot \vec{\phi}_{\rho }^{\prime }+%
\vec{u}_{\sigma B}^{\prime }\cdot \vec{\theta}_{\sigma }^{\prime }+\vec{v}%
_{\sigma B}^{\prime }\cdot \vec{\phi}_{\sigma }^{\prime }\right) },
\label{O'_vector2}
\end{eqnarray}
where $\vec{\theta}_{\rho }^{\prime }=R_{\rho }\vec{\theta}_{\rho }$, $\vec{%
\theta}_{\sigma }^{\prime }=R_{\sigma }\vec{\theta}_{\sigma }$ and similarly
for $\vec{\phi}_{\rho }^{\prime }=R_{\rho }\vec{\phi}_{\rho }$ and $\vec{\phi%
}_{\sigma }^{\prime }=R_{\sigma }\vec{\phi}_{\sigma }$. Here $R_{\rho }$ and
$R_{\sigma }$ are two orthogonal 4 by 4 matrices that transform the boson
fields from band basis to a new one. The new corresponding coefficient
vectors are $\vec{u}_{\rho A}^{\prime }=\vec{u}_{\rho A}R_{\rho }^{T}$, $%
\vec{u}_{\sigma A}^{\prime }=\vec{u}_{\sigma A}R_{\sigma }^{T}...$ here $%
R_{\rho }^{T}$ and $R_{\sigma }^{T}$ are the transpose of $R_{\rho }$ and $%
R_{\sigma }$.

Now assume the bosons get pinned in this new basis. We know only one of $%
\theta _{i}^{\prime }$ and $\phi _{i}^{\prime }$ can be pinned and the
presence of the dual of a pinned boson will result in exponential decay.
Consider a simple situation in which all bosons are pinned, i.e. a C0S0
phase and the pinned bosons are $(\phi _{1\rho }^{\prime },\theta _{2\rho
}^{\prime },\theta _{3\rho }^{\prime },\phi _{4\rho }^{\prime })$ in the
charge channel and $(\theta _{1\sigma }^{\prime },\phi _{2\sigma }^{\prime
},\phi _{3\sigma }^{\prime },\theta _{4\sigma }^{\prime })$ in the spin
channel, even though we don't know explicitly the transformations to the new
basis.

This set of pinned bosons enforces some constraints on the coefficient
vectors $\vec{u}_{\rho A/B},$ $\vec{v}_{\rho A/B},$ $\vec{u}_{\sigma A/B}$
and $\vec{v}_{\sigma A/B}$ so that $O_{A}$ and $O_{B}$ don't decay
exponentially. The constraints are simply that the coefficients for the dual
of each pinned boson must be zero. In this case, for $X=A$ or $B$:
\begin{eqnarray*}
\vec{u}_{\rho X}^{\prime } &=&(0,-,-,0), \\
\vec{v}_{\rho X}^{\prime } &=&(-,0,0,-), \\
\vec{u}_{\sigma X}^{\prime } &=&(-,0,0,-), \\
\vec{v}_{\sigma X}^{\prime } &=&(0,-,-,0),
\end{eqnarray*}
where ``$-$'' means no constraint. These constraints actually imply the
following equations:
\begin{eqnarray}
\vec{u}_{\rho A}^{\prime }\cdot \vec{v}_{\rho B}^{\prime } &=&\vec{u}%
_{\sigma A}^{\prime }\cdot \vec{v}_{\sigma B}^{\prime }=0,  \label{IP'1} \\
\vec{v}_{\rho A}^{\prime }\cdot \vec{u}_{\rho B}^{\prime } &=&\vec{v}%
_{\sigma A}^{\prime }\cdot \vec{u}_{\sigma B}^{\prime }=0,  \label{IP'2} \\
\vec{u}_{\rho A}^{\prime }\cdot \vec{v}_{\rho A}^{\prime } &=&\vec{u}%
_{\sigma A}^{\prime }\cdot \vec{v}_{\sigma A}^{\prime }=0,  \label{IP'3} \\
\vec{u}_{\rho B}^{\prime }\cdot \vec{v}_{\rho B}^{\prime } &=&\vec{u}%
_{\sigma B}^{\prime }\cdot \vec{v}_{\sigma B}^{\prime }=0.  \label{IP'4}
\end{eqnarray}
One can easily see that different sets of pinned bosons will imply the same
equations for inner products. So the actually pinning patterns of boson
fields are irrelevant for the constraints here and the crucial point is that
all the fields are pinned.

Eq. (\ref{IP'1})-(\ref{IP'4}) are for the coefficient vectors in the new
primed basis. Therefore, we don't really know their components. However, we know
inner products are invariant under any orthogonal transformation. Thus, the
coefficient vectors in the band basis should satisfy the same Eq. (\ref{IP'1}%
)-(\ref{IP'4}) but without the primes. So for any two operators $O_{A}$ and $%
O_{B}$ written in the band basis, we know the necessary condition on their
coefficient vectors for $O_{A}$ and $O_{B}$ not to be exponentially decaying.

However, things are slightly different for a C1S0 phase, which is the case
we are really interested in. Now the inner products $\vec{u}_{\rho
A}^{\prime }\cdot \vec{v}_{\rho B}^{\prime }$ and $\vec{v}_{\rho A}^{\prime
}\cdot \vec{u}_{\rho B}^{\prime }$ in the charge channel are not necessary
zero since the overlap is allowed in the subspace of gapless boson fields.
For example none of $O_{A}\sim e^{i\phi _{1\rho }^{\prime }}$ and $O_{B}\sim
e^{i\theta _{1\rho }^{\prime }}$ are exponentially decaying if ($\theta
_{1\rho }^{\prime },\phi _{1\rho }^{\prime }$) is gapless even though $\vec{v%
}_{\rho A}^{\prime }\cdot \vec{u}_{\rho B}^{\prime }=1$ in this example. In
principle, we even don't know what's the value for the nonzero inner
products if the gapless field is arbitrary. However, here we first discuss
the case in which gapless mode is the total charge bosons ($\Theta _{1\rho
},\Phi _{1\rho }$). This means, choosing $\phi _{1\rho }^{\prime }=\Phi
_{1\rho }$, that the first row of the orthogonal matrix $R_{\rho }$ can be
chosen to be $(1/2,1/2,1/2,1/2)$. Roughly speaking, for each charge boson
field in the band basis, we have
\begin{eqnarray*}
\theta _{i\rho } &=&\frac{1}{2}\Theta _{1\rho }+\cdots , \\
\phi _{i\rho } &=&\frac{1}{2}\Phi _{1\rho }+\cdots .
\end{eqnarray*}
Therefore, we find that for all the pairing $\Delta $ and $4\bar{k}_{F}$
density operators $n_{4\bar{k}_{F}}$, the coefficients of $\Theta _{1\rho }$
and $\Phi _{1\rho }$ are fixed:
\begin{eqnarray}
\Delta &\sim &e^{i\left( \vec{u}_{\rho \Delta }^{\prime }\cdot \vec{\theta}%
_{\rho }^{\prime }+\vec{v}_{\rho \Delta }^{\prime }\cdot \vec{\phi}_{\rho
}^{\prime }+\vec{u}_{\sigma \Delta }^{\prime }\cdot \vec{\theta}_{\sigma
}^{\prime }+\vec{v}_{\sigma \Delta }^{\prime }\cdot \vec{\phi}_{\sigma
}^{\prime }\right) }=e^{i\sqrt{\frac{\pi }{2}}\Phi _{1\rho }+\cdots },
\label{pairing'} \\
n_{4\bar{k}_{F}} &\sim &e^{i\left( \vec{u}_{\rho n}^{\prime }\cdot \vec{%
\theta}_{\rho }^{\prime }+\vec{v}_{\rho n}^{\prime }\cdot \vec{\phi}_{\rho
}^{\prime }+\vec{u}_{\sigma n}^{\prime }\cdot \vec{\theta}_{\sigma }^{\prime
}+\vec{v}_{\sigma n}^{\prime }\cdot \vec{\phi}_{\sigma }^{\prime }\right)
}=e^{-i\sqrt{2\pi }\Theta _{1\rho }+\cdots }.  \label{4KF'}
\end{eqnarray}
As we mentioned before, the only nonzero part of inner products must come
from the gapless boson ($\Theta _{1\rho },\Phi _{1\rho }$). Although we
don't know the full representation of pairing and $4\bar{k}_{F}$ density
operators in the new basis, nonetheless, from Eq. (\ref{pairing'}) and (\ref
{4KF'}), it's sufficient to conclude that the only nonzero inner product
between $(\vec{v}^{\prime })^{\prime }s$ and $(\vec{u}^{\prime })^{\prime }s$
has to be exactly
\begin{equation}
\vec{v}_{\rho \Delta }^{\prime }\cdot \vec{u}_{\rho n}^{\prime }=\pm \pi ,
\label{PNpi}
\end{equation}
where the positive sign occurs when taking the Hermitian conjugate of one of
the operators. In the following convention, we choose $4\bar{k}_{F}$ density
operators so that $-\pi $ is taken in Eq. (\ref{PNpi}). Finally, we get the
necessary conditions on the coefficient vectors in the band basis so that a
C1S0 phase can have both pairing and stripe correlations:
\begin{eqnarray}
\vec{v}_{\rho \Delta }\cdot \vec{u}_{\rho n} &=&-\pi ,  \label{IP1} \\
\vec{u}_{\rho \Delta }\cdot \vec{v}_{\rho n} &=&0,  \label{IP2} \\
\vec{u}_{\sigma \Delta }\cdot \vec{v}_{\sigma n} &=&\vec{v}_{\sigma \Delta
}\cdot \vec{u}_{\sigma n}=0,  \label{IP3} \\
\vec{u}_{\rho \Delta }\cdot \vec{v}_{\rho \Delta } &=&\vec{u}_{\sigma \Delta
}\cdot \vec{v}_{\sigma \Delta }=0,  \label{IP4} \\
\vec{u}_{\rho n}\cdot \vec{v}_{\rho n} &=&\vec{u}_{\sigma n}\cdot \vec{v}%
_{\sigma n}=0.  \label{IP5}
\end{eqnarray}

What about the cases when the gapless field is not the total charge mode?
We don't know the value for this inner product if we don't know what
the gapless field is. However, we do know one thing; the inner products, Eq.
(\ref{IP1})-(\ref{IP5}), between the coefficient vectors can only be nonzero
due to the contribution from the gapless field. In other words, the inner
products must be zero after the gapless mode is projected out.
Mathematically, we mean $\vec{v}_{\rho \Delta }^{\perp }\cdot \vec{u}_{\rho
n}^{\perp }=0$ where $\vec{v}_{\rho A}^{\perp }=\vec{v}_{\rho A}-(\vec{v}%
_{\rho A}\cdot \vec{g})\vec{g}$ and $\vec{u}_{\rho A}^{\perp }=\vec{u}_{\rho
A}-(\vec{u}_{\rho A}\cdot \vec{g})\vec{g}$ where the gapless charge mode is $%
\vec{g}\cdot \vec{\phi}_{\rho }$ ($\vec{g}$ is a unit vector). Similarly, we
also have $\vec{u}_{\rho \Delta }^{\perp }\cdot \vec{v}_{\rho n}^{\perp }=0$%
, $\vec{u}_{\rho \Delta }^{\perp }\cdot \vec{v}_{\rho \Delta }^{\perp }=0$
and $\vec{u}_{\rho n}^{\perp }\cdot \vec{v}_{\rho n}^{\perp }=0$. The
conditions on the spin fields are still the same since they have nothing to
do with the gapless charge mode. The new conditions on the coefficient
vectors after projecting out the gapless charge modes are
\begin{eqnarray}
\vec{v}_{\rho \Delta }^{\perp }\cdot \vec{u}_{\rho n}^{\perp } &=&\vec{u}%
_{\rho \Delta }^{\perp }\cdot \vec{v}_{\rho n}^{\perp }=0,  \label{IPP1} \\
\vec{u}_{\rho \Delta }^{\perp }\cdot \vec{v}_{\rho \Delta }^{\perp } &=&\vec{%
u}_{\rho n}^{\perp }\cdot \vec{v}_{\rho n}^{\perp }=0.  \label{IPP2}
\end{eqnarray}
Once the gapless charge mode is known, it's not difficult to check Eq. (\ref
{IPP1}) and (\ref{IPP2}) for the pair and $4\bar{k}_{F}$ density operators.

To illustrate what we mean by Eq. (\ref{IPP1}) and (\ref{IPP2}), we take a
two band system for example. The operator $O_{A}\sim e^{i\sqrt{2\pi }(\theta
_{1}^{\rho }+\theta _{2}^{\rho }+\phi _{1}^{\rho }-\phi _{2}^{\rho })}$ has
the coefficient vectors $\vec{v}_{\rho A}=\sqrt{2\pi }(1,1)$ and $\vec{u}%
_{\rho A}=\sqrt{2\pi }(1,-1)$. It's easy to see that this operator satisfies
the condition $\vec{v}_{\rho A}\cdot \vec{u}_{\rho A}=0$. If $(\theta
_{12}^{\rho +},\phi _{12}^{\rho +})$ is the gapless charge mode, then $\vec{g%
}=(1,1)/\sqrt{2}$. Only $\theta $ fields contain the gapless mode. The
gapless mode won't appear in the $\phi $ fields since in this case it's
orthogonal to the $\phi $ field, $\phi _{1}^{\rho }-\phi _{2}^{\rho }$. Then
there is no contribution to the inner product from the gapless mode and we
have $\vec{v}_{\rho A}^{\perp }\cdot \vec{u}_{\rho A}^{\perp }=0$. This
operator could have power-law decaying correlations. However, if $(\theta
_{1}^{\rho },\phi _{1}^{\rho })$ is the gapless field, we have to exclude
the part of inner product due to the gapless field since the overlap between
$\theta $ and $\phi $ fields is allowed for the gapless mode. Now $\vec{g}%
=(1,0)$, we find the inner product $\vec{v}_{\rho A}^{\perp }\cdot \vec{u}%
_{\rho A}^{\perp }=-2\pi \neq 0$. Then the correlation function of this
operator will decay exponentially since $\theta _{2}^{\rho }$ and $\phi
_{2}^{\rho }$ can't be pinned at the same time.

It should be clear now that once we write down the bosonized expressions for
pair and $4\bar{k}_{F}$ density operators in the band basis, we can tell
whether stripes and pairing can coexist or not by checking the inner
products of coefficient vectors. First consider pairing. By ``pairing'' we mean the existence of
any operator of charge $2$ whose correlation functions exhibit power-law
decay. Any pair operator that only contains right or left fermions, such as $%
\psi _{Ra\uparrow }\psi _{Rb\downarrow },$ can never exhibit power law decay
since from Eq. (\ref{RL}) we know $\varphi _{Ra\alpha }=(\phi _{a\alpha
}+\theta _{a\alpha })/2$ and this will result in $\vec{u}_{\rho \Delta
}\cdot \vec{v}_{\rho \Delta }$ and $\vec{u}_{\sigma \Delta }\cdot \vec{v}%
_{\sigma \Delta }$ $\neq 0$. (Eq. (\ref{IP4}) is violated.) Furthermore, any
pair operator with non-zero total $z$-component of spin will contain a
factor with the exponential of $\Phi _{1\sigma }$ and hence exhibit
exponential decay.

So, there are only two types of charge $2$ operators, containing only two
fermion fields, which are candidates for power-law decay. Ignoring Klein
factors, these are:
\begin{eqnarray}
\psi _{Ra\alpha }\psi _{La\bar{\alpha}} &\sim &e^{i\sqrt{2\pi }(\phi _{a\rho
}\pm \theta _{a\sigma })}  \label{P1} \\
\psi _{Ra\alpha }\psi _{Lb\bar{\alpha}} &\sim &e^{i\sqrt{\pi }(\theta
_{ab}^{\rho -}+\phi _{ab}^{\rho +}\pm \theta _{ab}^{\sigma +}\pm \phi
_{ab}^{\sigma -})}\ \ \ (a\neq b)  \label{P2}
\end{eqnarray}
Here $a$ and $b$ are band indices and $\alpha =\uparrow $ or $\downarrow ,$ $%
\bar{\alpha}\equiv -\alpha $. The $+$ or $-$ sign occurs for $\alpha
=\uparrow $ or $\downarrow $ respectively. Even though Eq. (\ref{P2})
carries non-zero momentum for $k_{F_{a}}\neq k_{F_{b}}$, there is no reason
to exclude it.

We now consider the $4\overline{k}_{F}$ density operators, corresponding to stripes.
The following two operators are the most general $4k_{F}$ density operators:
\begin{eqnarray}
\psi _{Ri\alpha }^{\dagger }\psi _{Lj\alpha }\psi _{Rk\alpha }^{\dagger
}\psi _{Ll\alpha } &\sim &e^{-i\sqrt{\pi }[(\theta _{ij}^{\rho +}+\theta
_{kl}^{\rho +})+(\phi _{ij}^{\rho -}+\phi _{kl}^{\rho -})\pm (\theta
_{ij}^{\sigma +}+\theta _{kl}^{\sigma +})\pm (\phi _{ij}^{\sigma -}+\phi
_{kl}^{\sigma -})]},  \label{4KF1} \\
\psi _{Ri\alpha }^{\dagger }\psi _{Lj\alpha }\psi _{Rk\overline{\alpha }%
}^{\dagger }\psi _{Ll\overline{\alpha }} &\sim &e^{-i\sqrt{\pi }[(\theta
_{ij}^{\rho +}+\theta _{kl}^{\rho +})+(\phi _{ij}^{\rho -}+\phi _{kl}^{\rho
-})\pm (\theta _{ij}^{\sigma +}-\theta _{kl}^{\sigma +})\pm (\phi
_{ij}^{\sigma -}-\phi _{kl}^{\sigma -})]},  \label{4KF2}
\end{eqnarray}
Here $i$, $j$, $k$ and $l$ are arbitrary band indices and the $+$ or $-$
sign is for $\alpha =\uparrow $ or $\downarrow $ respectively. Whether $i$, $%
j$, $k$ and $l$ are all different or not actually doesn't change the
conclusion. When all the band indices are different, the operators
correspond to the oscillation wave-vector of $4\bar{k}_{F}\equiv
k_{F1}+k_{F2}+k_{F3}+k_{F4}$ which is $2\pi n$, the wave-vector of stripes.
However, if some special renormalizations of Fermi momentum, such as $%
k_{F1}=k_{F3}$ and $k_{F2}=k_{F4}$ occur, then the wave vector $%
2(k_{F1}+k_{F2})$ also corresponds to $2\pi n$. Therefore, we have to
consider all $4k_{F}$ density operators with the proper Fermi momentum
renormalizations such that the operators correspond to stripes. That's the
reason why we consider arbitrary rather than only all different band indices
in Eq. (\ref{4KF1}) and (\ref{4KF2}).

In practice, we start with a pinning pattern where only the gapless
charge mode is determined. Then we check if the pair operators Eq. (\ref{P1}), (%
\ref{P2}) and $4k_{F}$ density operators Eq. (\ref{4KF1}), (\ref{4KF2}) can
satisfy the inner product conditions (Eq. (\ref{IP1})-(\ref{IP5}) if ($%
\Theta _{1\rho },\Phi _{1\rho }$) is gapless otherwise the modified Eq. (\ref
{IPP1})-(\ref{IPP2})). In this procedure, many conditions on Fermi momenta
will be involved. They are the conditions to make an interaction containing a
$\theta _{i\rho }^{\prime }$ appear in the Hamiltonian (if $\theta _{i\rho
}^{\prime }$ is not orthogonal to $\Theta _{1\rho }$, then the gapless mode
is not ($\Theta _{1\rho },\Phi _{1\rho }$)), and the conditions to make a $%
4k_{F}$ density operator corresponding to stripes. Recall that the condition
$4\bar{k}_{F}=2\pi n$ is always satisfied. It's also important to keep track
on the consistency of all the Fermi momentum conditions. We have carefully
checked all possible cases and concluded that stripes and pairing can
not coexist in any C1S0 phases. The whole proof is somewhat lengthy but
rather straightforward [\onlinecite{thesis}]. Some more details are given in Appendix A.

Now we want to ask further if \textit{any} operator of non-zero charge can
have a power-law decaying correlation function in a stripe phase. Consider
the case when ($\Theta _{1\rho },\Phi _{1\rho }$) is gapless. Inspired by
the real space picture of stripes and the finite size spectrum analysis in
the next section, we find that there is always some charge-four bipairing
operator, which does so. The most general $S_{z}=0$ non-chiral bipairing operators
are:
\begin{equation}
\psi _{Rs\alpha }\psi _{Lt\overline{\alpha }}\psi _{Ru\beta }\psi _{Lv%
\overline{\beta }}\sim e^{i\sqrt{\pi }[(\phi _{st}^{\rho +}+\phi _{uv}^{\rho
+})+(\theta _{st}^{\rho -}+\theta _{uv}^{\rho -})\pm (\phi _{st}^{\sigma
+}\pm \phi _{uv}^{\sigma +})\pm (\theta _{st}^{\sigma -}\pm \theta
_{uv}^{\sigma -})]}.  \label{bipair_1}
\end{equation}
Here $s$, $t$, $u$ and $v$ are arbitrary band indices. Although Eq. (\ref
{bipair_1}) are the possible bipairing operators in the most general sense,
only when $s=t$ and $u=v$ or $s=v$ and $t=u$ does Eq. (\ref{bipair_1})
carry zero momentum and have no real space modulation in the correlation
functions. On the other hand, we have no reason to exclude the possibility
that Eq. (\ref{bipair_1}) does decay with a power law with an oscillating
factor at this stage.

Following the previous discussion, now we will prove that any C1S0 phase
with $4\bar{k}_{F}$ density oscillations, also has bipairing correlation.
The conditions for bipairing and $4\bar{k}_{F}$ density operators to coexist
are almost the same as Eq. (\ref{IP1})-(\ref{IP5}) but now with $\vec{v}%
_{\rho ,bi}\cdot \vec{u}_{\rho ,n}=-2\pi $, where we use a subscript ``$bi$%
'' for bipairing operators. Now for simplicity, let's focus on the following
two types of bipairing operators carrying zero momentum:

\begin{eqnarray}
\psi _{Rs\alpha }\psi _{Ls\alpha }\psi _{Rt\overline{\alpha }}\psi _{Lt%
\overline{\alpha }} &\sim &e^{i\sqrt{4\pi }(\phi _{st}^{\rho +}\pm \phi
_{st}^{\sigma -})},  \label{bipair_2} \\
\psi _{Rs\alpha }\psi _{Ls\overline{\alpha }}\psi _{Rt\beta }\psi _{Lt%
\overline{\beta }} &\sim &e^{i\sqrt{4\pi }(\phi _{st}^{\rho +}\pm \theta
_{st}^{\sigma \pm })}.  \label{bipair_3}
\end{eqnarray}

We find that Eq. (\ref{4KF1}), (\ref{4KF2}) and Eq. (\ref{bipair_2}) can
coexist if we choose $\{s,t\}=$ $\{i,j\}$ or $\{k,l\}$. If we have $\alpha
=\beta $ in Eq. (\ref{bipair_3}), i.e. $\theta _{st}^{\sigma +}$ is present,
then Eq. (\ref{4KF1}), (\ref{4KF2}) and Eq. (\ref{bipair_3}) can coexist
with the choice that $\{s,t\}=$ $\{i,j\}$ or $\{k,l\}$. If we have $\alpha =%
\overline{\beta }$ in Eq. (\ref{bipair_3}), then $\theta _{st}^{\sigma -}$
is present. Eq. (\ref{4KF1}) and (\ref{bipair_3}) can coexist with the
choice that $\{s,t\}=$ $\{i,k\}$ or $\{j,l\}$. Eq. (\ref{4KF2}) and (\ref
{bipair_3}) can coexist if $\{s,t\}=$ $\{i,l\}$ or $\{j,k\}$.

\subsection{Completely Empty or Filled Bands}

Another possible type of stripe phase has one or more of the bands
completely empty (or completely filled). We now show that a standard
treatment of these phases does not lead to any with coexisting pairing and
stripes. Suppose that two bands are completely empty. Without loss of
generality, we may choose then to be bands $3$ and $4$ so that all the
electrons go into bands $1$ and $2$. It then follows from Eq. (\ref{n4kF})
that
\begin{equation}
k_{F1}+k_{F2}=2\pi n=2\pi (1-\delta ).
\end{equation}
So stripes, i.e. Friedel oscillations at wave-vector $2\pi n$, corresponds
to $2k_{F}$ oscillations in the effective 2-band model. There are two cases
to consider:
\begin{eqnarray}
\psi _{Li\alpha }^{\dagger }\psi _{Ri\alpha } &\sim &e^{{i}\sqrt{2\pi }%
(\theta _{i}^{\rho }\pm \theta _{i}^{\sigma })}  \label{2kF1} \\
\psi _{Li\alpha }^{\dagger }\psi _{Rj\alpha } &\sim &e^{{i}\sqrt{\pi }\left(
\theta _{ij}^{\rho +}-\phi _{ij}^{\rho -}\pm \theta _{ij}^{\sigma +}\mp \phi
_{ij}^{\sigma -}\right) }.  \label{2kF2}
\end{eqnarray}
For the above situation, we have $\{i,j\}=\{1,2\}$ in Eq. (\ref{2kF1}) and (%
\ref{2kF2}). Following the similar method, we can prove the incompatibility
of pairing and $2k_{F}$ oscillations. If instead bands $3$ and $4$ are
completely filled, Eq. (\ref{n4kF}) now implies:
\begin{equation}
k_{F1}+k_{F2}+2\pi =2\pi n,
\end{equation}
but this is equivalent to the case of the two bands being empty since $\exp
(2\pi inx)=\exp [2\pi i(n-1)x]$ for any lattice site, $x$. If only one band (%
$4$) is empty (or filled) then stripes would correspond to oscillations at
wave-vector $k_{F1}+k_{F2}+k_{F3}$ which can never occur since any operator
which occurs in the continuum representation of the density operator must
contain an even number of fermion fields. Similarly stripes could not occur
in a phase with 3 empty (or filled) bands since this would require
oscillations at $k_{F1}$.

\subsection{Summary and Generalization}

Therefore, for the C1S0 phase with stripes, there is no pairing but bipairing
occurs. In Sec. V we will discuss the case of small $t_{\perp ,2}$, $%
V_{\perp ,2}$ and the analogy with a 2-leg bosonic ladder. The bipairing
correlation in the four-band fermion system may correspond to the boson pair
superfluid (BPSF) phase in Ref. [\onlinecite{Orignac}].

Now we would like to generalize this argument a little bit further. What
about any other generalized charge two operators such as
\begin{equation}
\psi _{Ra\alpha }\psi _{Lb\bar{\alpha}}\psi _{Li\alpha }^{\dagger }\psi
_{Rj\alpha }\sim e^{i\sqrt{\pi }[(\theta _{ab}^{\rho -}+\theta _{ij}^{\rho
+})+(\phi _{ab}^{\rho +}-\phi _{ij}^{\rho -})\pm (\theta _{ab}^{\sigma
+}+\theta _{ij}^{\sigma +})\pm (\phi _{ab}^{\sigma -}-\phi _{ij}^{\sigma -})%
]}.  \label{GP}
\end{equation}
Although the correlation amplitude should be smaller compared to the usual
pairing operators, still, is it possible that such generalized pairing
operators after renormalized by some density-like (charge neutral) operators
can coexist with stripes? There will be a lot more such charge two operators
since Eq. (\ref{P1}), (\ref{P2}) and other unconventional pair operators can
be combined with any charge neutral operator, as long as in the end we have
a non-chiral, spin zero and charge two operator. So far we have checked up
to charge two operators composed of six fermions and concluded none of them
can coexist with stripes as expected. But what about charge two operators
composed of more fermions? This endless question may require another
approach for its resolution. Instead, we will study the finite size spectrum
of a C1S0 phase, from which a connection between charge operator and the
lowest density oscillation is established.

\section{Finite Size Spectrum}

In this section we will establish a general connection between charge and density
operators through the consistency for the finite size spectrum of a C1SO phase. Assume that
the total charge field is the gapless charge mode. The low energy effective Hamiltonian in a C1S0 phase
is simply that of a free boson.
\begin{equation}
H-\mu N_{e}=\frac{v_{1\rho }}{2}\int dx\left[ K_{1\rho }(\partial _{x}\Phi
_{1\rho })^{2}+{\frac{1}{K_{1\rho }}}(\partial _{x}\Theta _{1\rho
})^{2}\right] .  \label{Hamtheta1}
\end{equation}
In the rest of this subsection we only discuss the boson $\Phi _{1\rho }$
(and its conjugate boson $\Theta _{1\rho }$) so, for convenience, in this
section we drop the superscript $\rho $ and the subscript $1$. Here $%
v\partial _{x}\Theta =K\partial _{t}\Phi =\Pi $, where $\Pi $ is the
canonical momentum variable conjugate to $\Phi $. It is natural to regard $%
\Phi $ as a periodic variable. Consider first a pairing phase where a charge
two operator:
\begin{equation}
\Delta (x)\sim e^{i\sqrt{\frac{\pi }{2}}\Phi },  \label{Delta}
\end{equation}
has power law decay. Only keeping operators in the low energy Hilbert space
which exhibit power-law decay , we expect that all such operators will have
even charge, involving only the exponential of integer multiples of $i\sqrt{%
\frac{\pi }{2}}\Phi $. It is then natural to assume that we should make the
periodic identification:
\begin{equation}
\Phi \leftrightarrow \Phi +2\sqrt{2\pi }.  \label{Phiid}
\end{equation}
That is to say, we regard $\sqrt{\pi /2}\Phi $ as an angular variable. We
now wish to argue that consistent quantization of the free boson requires
that $\Theta $ also be regarded as a periodic variable with:
\begin{equation}
\Theta \leftrightarrow \Theta +\sqrt{\frac{\pi }{2}}.  \label{Thetaid}
\end{equation}
As we shall see, this, in turn implies that the minimum Friedel oscillation
wavevector is $8\bar{k}_{F}$. Alternatively, in a bipairing phase, the
lowest dimension charge operators is $\exp [i\sqrt{2\pi }\Phi _{1}^{\rho }]$
and it is now natural to identify
\begin{equation}
\Phi \leftrightarrow \Phi +\sqrt{2\pi }.  \label{Phiid2}
\end{equation}
We then will argue that consistent quantization requires:
\begin{equation}
\Theta \leftrightarrow \Theta +\sqrt{2\pi },  \label{Thetaid2}
\end{equation}
which we will show implies that the minimum Friedel oscillation wavevector
is $4\bar{k}_{F}$. This approach confirms the conclusions arrive at by more
pedestrian means in the previous sub-section. As a biproduct of this
discussion, we will derive the finite size spectrum, with both periodic and
open BCs, in both pairing and bipairing (stripes) phases.

First consider a C1S0 pairing phase. We place the system in a box of size $L$
with periodic boundary conditions. The mode expansion for $\Phi (t,x)$ takes
the form:
\begin{equation}
\Phi (t,x)=\Phi _{0}+2\sqrt{2\pi }m{\frac{x}{L}}+\sqrt{\frac{\pi }{2}}{\frac{%
p}{K}}{\frac{vt}{L}}+\sum_{k=1}^{\infty }\sqrt{\frac{1}{4\pi kK}}\left(
a_{Rk}e^{-i(2\pi k/L)(vt-x)}+a_{Lk}e^{-i(2\pi k/L)(vt+x)}+h.c.\right) .
\label{mode}
\end{equation}
Here $m$ and $p$ are arbitrary integers; $a_{Rk}$ and $a_{Lk}$ are bosonic
annihilation operators for right and left movers. The normalization of the $%
mx/L$ term in the expansion is determined by the periodic BCs and the
identification Eq. (\ref{Phiid}). i.e. $\Phi (L)=\Phi (0)+2\sqrt{2\pi }m$ is
equivalent to PBC using Eq. (\ref{Phiid}). The (very important)
normalization of the $vt/L$ term requires more explanation. We may think of
this term as being proportional to a zero mode conjugate momentum operator $%
\hat{\Pi}_{0}$, which is canonically conjugate to $\hat{\Phi}_{0}$:
\begin{equation}
\lbrack \Phi _{0},\Pi _{0}]=i,  \label{com0}
\end{equation}
\begin{equation}
\Phi (t,x)=\Phi _{0}+{\frac{\hat{\Pi}_{0}vt}{KL}}+\ldots  \label{modePi0}
\end{equation}
$\hat{\Pi}_{0}$ is the zero momentum Fourier mode of the conjugate momentum
field $\Pi (x)$:
\begin{equation}
\hat{\Pi}_{0}\equiv \int_{0}^{L}\Pi (x).
\end{equation}
The $\hat{\Pi}_{0}vt/(KL)$ term in the mode expansion for $\Phi $ is
necessary in order that the canonical commutation relations are obeyed:
\begin{equation}
\lbrack \Phi (x),\partial _{t}\Phi (y)]={\frac{i}{KL}}\sum_{k=-\infty
}^{\infty }e^{i(2\pi k/L)(x-y)}={\frac{i}{K}}\delta _{P}(x-y),
\end{equation}
where $\delta _{P}(x-y)$ is the periodic Dirac $\delta $-function. The $k=0$
term in this Fourier expansion of the $\delta $-function comes from the
commutator in Eq. (\ref{com0}). Comparing the mode expansion in Eq. (\ref
{mode}) to (\ref{modePi0}), we see that the eigenvalues of the canonical
momentum operator, $\hat{\Pi}_{0}$ are:
\begin{equation}
\Pi _{0}=\sqrt{\frac{\pi }{2}}p,
\end{equation}
for integer $p$. That these are the correctly normalized eigenvalues follows
from the fact that the wave-functionals contain factors of the form:
\begin{equation}
\Psi (\Phi _{0})=e^{i\Pi _{0}\Phi _{0}}=e^{i\sqrt{\frac{\pi }{2}}p\Phi _{0}}.
\end{equation}
These wave-functions are single-valued under the identification of Eq. (\ref
{Phiid}). The eigenvalues of $\hat{\Pi}_{0}$, i.e. the normalization of the $%
vt/L$ term in the mode expansion of Eq. (\ref{mode}) determines the charge
quantum numbers of all low-lying states in the spectrum with periodic
boundary conditions. This follows from observing that the total electron
number operator is:
\begin{equation}
\hat{N}_{e}=\frac{2K}{v}\sqrt{\frac{2}{\pi }}\int_{0}^{L}dx\partial _{t}\Phi
.
\end{equation}
This in turn can be checked by confirming that:
\begin{equation}
\lbrack \hat{N}_{e},\Delta ]=2\Delta ,
\end{equation}
where $\Delta $ is the charge-2 operator in Eq. (\ref{Delta}). The mode
expansion of Eq. (\ref{mode}) then implies that the charges of all states in
the low energy spectrum are:
\begin{equation}
N_{e}=2p,
\end{equation}
even integers. This is an example of the general one-to-one correspondence
between operators and states in the finite size spectrum with PBC in a
conformal field theory. In a phase in which all operators have even charge,
all states in the spectrum also have even charge.

We can go further and deduce the Friedel oscillation wavevector from the
normalization of the $vt/L$ term in Eq. (\ref{mode}). This can be done by
using $K\partial _{t}\Phi =v\partial _{x}\Theta $, $\partial _{t}\Theta
=vK\partial _{x}\Phi $, to deduce the mode expansion for the field $\Theta
(t,x)$:
\begin{equation}
\Theta (t,x)=\Theta _{0}+\sqrt{\frac{\pi }{2}}p{\frac{x}{L}}+2\sqrt{2\pi }Km{%
\frac{vt}{L}}+\sum_{k=1}^{\infty }\sqrt{\frac{K}{4\pi k}}\left(
-a_{Rk}e^{-i(2\pi k/L)(vt-x)}+a_{Lk}e^{-i(2\pi k/L)(vt+x)}+h.c.\right) .
\label{modeT}
\end{equation}
From this mode expansion we see that $\Theta $ is periodically identified as
in Eq. (\ref{Thetaid}). Insert Eq. (\ref{mode}) and (\ref{modeT}) into the
Hamiltonian (\ref{Hamtheta1}), we obtain the finite size spectrum of a
pairing phase with PBC:
\begin{equation}
E-2\mu p={\frac{2\pi v}{L}}\left[ 2Km^{2}+{\frac{p^{2}}{8K}}%
+\sum_{k=1}^{\infty }k(n_{Lk}+n_{Rk})\right] ,
\end{equation}
where $n_{Lk}$ and $n_{Rk}$ are the occupation numbers for the left and
right moving states of momentum $\pm 2\pi k/L$.

Now consider the Friedel oscillations. Oscillations at wave-vector of the
form $2n\bar{k}_{F}$ (actually a sum of any $2n$ of the $k_{Fi}$'s) can only
occur if some operator of the form $(\psi _{R}^{\dagger }\psi _{L})^{n}$ has
power law decay. Upon bosonizing, all such operators are expressed as $\exp
\left( in\sqrt{\pi /2}\Theta _{1}^{\rho }\right) $ multiplied by an
exponential involving only pinned boson fields. However, not all such
operators can occur in the low energy spectrum since they must respect the
periodic identification in Eq. (\ref{Thetaid}). The lowest dimension
operator allowed by this identification has $n=4$ corresponding to $8\bar{k}%
_{F}$ oscillations. Again we are effectively using the relationship between
the finite size spectrum and the operator content. The $n=4$ operator
corresponds to the $p=1$ state in the mode expansion of Eq. (\ref{modeT}).

Now consider a bipairing phase where there are no charge $2$ operators in
the low energy spectrum, the lowest charge being $4$, corresponding to the
operator $\exp [i\sqrt{2\pi }\Phi ]$, leading to the periodicity condition
on $\Phi $ in Eq. (\ref{Phiid2}). The mode expansion for $\Phi $ is
therefore altered to:
\begin{equation}
\Phi (t,x)=\Phi _{0}+\sqrt{2\pi }m{\frac{x}{L}}+\sqrt{2\pi }{\frac{p}{K}}{%
\frac{vt}{L}}+\sum_{k=1}^{\infty }\sqrt{\frac{1}{4\pi kK}}\left(
a_{Rk}e^{-i(2\pi k/L)(vt-x)}+a_{Lk}e^{-i(2\pi k/L)(vt+x)}+h.c.\right) .
\label{mode2}
\end{equation}
The coefficient of $vt/L$ gets multiplied by a factor of $2$ since the
wave-functional $\exp [i\Pi _{0}\Phi _{0}]$ must now be invariant under the
shift of Eq. (\ref{Phiid2}), requiring the conjugate momentum, $\hat{\Pi}%
_{0} $ to have eigenvalues $\sqrt{2\pi }p$. Correspondingly the mode
expansion for $\Theta $ becomes:
\begin{equation}
\Theta (t,x)=\Theta _{0}+\sqrt{2\pi }p{\frac{x}{L}}+\sqrt{2\pi }Km{\frac{vt}{%
L}}+\sum_{k=1}^{\infty }\sqrt{\frac{K}{4\pi k}}\left( -a_{Rk}e^{-i(2\pi
k/L)(vt-x)}+a_{Lk}e^{-i(2\pi k/L)(vt+x)}+h.c.\right) ,  \label{modeT2}
\end{equation}
implying the periodic identification of Eq. (\ref{Thetaid2}). Now the lowest
dimension Friedel oscillation operator with power law decay is $\exp (i\sqrt{%
2\pi }\Theta _{1}^{\rho })$, which is a $4\bar{k}_{F}$ operator. The finite
size spectrum of a bipairing phase with PBC is:
\begin{equation}
E-4\mu p={\frac{2\pi v}{L}}\left[ \frac{K}{2}m^{2}+{\frac{p^{2}}{2K}}%
+\sum_{k=1}^{\infty }k(n_{Lk}+n_{Rk})\right] .
\end{equation}

These arguments show, based only on plausible assumptions about regarding
the fields $\Phi _{1}^{\rho }$ and $\Theta _{1}^{\rho }$ as periodic
variables, that C1S0 phases with pairing have $8\bar{k}_{F}$ oscillations
(and hence no stripes) but phases with bipairing have $4\bar{k}_{F}$
oscillations, corresponding to stripes.

With OBC, the boundary conditions:
\begin{equation}
\Theta (0)=\hbox{constant},\ \ \Theta (L)=\hbox{constant},
\end{equation}
are applied. This sets the quantum number $m=0$ and $a_{Rk}=a_{Lk}$ in the
mode expansion of Eq. (\ref{modeT}) and (\ref{modeT2}). Setting $Q=2p$ for
the pairing phase and $Q=4p$ for the biparing phase, the finite size
spectrum implied by these mode expansions and the Hamiltonian of Eq. (\ref
{Hamtheta1}) is:
\begin{equation}
E-\mu Q={\frac{\pi v}{L}}\left[ {\frac{Q^{2}}{16K}}+\sum_{k=1}^{\infty
}kn_{Lk}\right] ,
\end{equation}
where the charge, $Q$ (measured from a reference point like half-filling) is
restricted to all even integers in a pairing phase but is restricted to
integer multiples of $4$ in a bipairing phase. For even $Q$, in a bipairing
phase, there is a gap $\Delta _{E}$, to states with $Q/2$ odd, so we may
write, for any even $Q$ in a bipairing phase:
\begin{equation}
E-\mu Q=\Delta _{E}{\frac{[1-(-1)^{Q/2}]}{2}}+O({\frac{1}{L})}.
\end{equation}
The parameters $v$, $K$, $\mu $ and $\Delta _{E}$ all vary with density.
Nonetheless, this zigzag pattern of energies for even $Q$ should allow
unambiguous detection of a biparing phase for large enough $L$.

\section{RG FOR DOPED 4-LEG LADDER}

The combination of weak-coupling RG and bosonization is one
standard tool to study the phase diagram of $N$-leg systems [{\onlinecite{2leg, 3legRG, NlegRG, Ledermann}]. The results for doped $4$-leg ladder are mostly within the
context of the Hubbard model [\onlinecite{NlegRG, Ledermann}]. Here we
will show that the stripe phase can be found in the special
solution of RG equations. This phase doesn't have pairing but
bipairing, which is consistent with our bosonization argument.

The first step is to determine the relevant couplings according to
the RG flow since they will control which boson fields will get
``pinned'' and therefore will allow us to map out the phase
diagram in terms of bare interactions and doping. However, to
analyze the RG flow is a tricky task for there are 32 coupled
nonlinear differential equations. It seems that the RG ultimately
flows onto a special set of solutions, corresponding to some
direction in the multi-dimensional coupling constant space. These
special solutions are called ``fixed rays'' and different rays
usually indicate different phases [\onlinecite{NlegRG}]. Some fixed-ray
solutions may correspond to phases with higher symmetry than the
original Hamiltonian. Two-leg ladders at half-filling provide one
example of such symmetry enhancement in the low energy limit
[\onlinecite{Lin98}]. Later on, some subleading corrections were found
which make the RG flow deviate from the fixed ray. However, these
subleading terms don't grow fast enough to really spoil the fixed
ray in the undoped case but give some anomalous corrections to the
gap functions, vanishing in the weak coupling limit [\onlinecite{Lin98,
MSChang}]. Things become dramatically different in the doped
systems. Now these subleading terms are relevant perturbations for
the fragile gapless modes. They will generate gaps although these
may be much smaller compared to those driven by the ``fixed ray''.
For example, the weak coupling RG phase diagram for doped two-leg
ladders is modified after taking these terms into consideration
[\onlinecite{LinPrivate}]. Recently, a hidden potential structure of RG
equations in ladder systems was discovered. Everything can be
understood better within the framework of this ``RG potential''
which allows the RG flow to be viewed as the trajectory of a
particle finding a minimum in the coupling constant space
[\onlinecite{Chen, MSChang}]. Then the fixed ray is just like a
``valley/ridge'' in the ``mountains'' of the RG potential. The
topography near the vicinity of such a ``valley/ridge'' will
determine the stability of the fixed ray and give the exponents
governing the subleading terms.

\subsection{RG Potential and Its Implications}

The method we discuss is very general but we mainly focus on
$N=4$. As the
conventional starting point, we first diagonalize the hopping terms in Eq. (%
\ref{H0}) and obtain $4$ bands. Next we linearize each band around
different Fermi points in the low energy limit, and introduce
$SU(2)$ scalar and vector current operators
\begin{equation}
J_{ij}^{L/R}=\frac{1}{2}\;\psi _{L/Ri\alpha }^{\dag }\;\psi
_{L/Rj\alpha
}\;,\;\;\;\;\mbox{\boldmath$J$}_{ij}^{L/R}=\frac{1}{2}\;\psi
_{L/Ri}^{\alpha \dag }\;\vec{\sigma}_{\alpha }^{\beta }\;\psi
_{L/Rj\beta }\;.  \label{SU2}
\end{equation}
(Note the unconventional factor of $1/2$ in the scalar operators,
introduced for later convenience.) We can rewrite the interactions
in Eq. (\ref{Hint}) in terms of the current operators in Eq.
(\ref{SU2}):
\begin{eqnarray}
H_{int} &=&\tilde{c}_{ij}^{\rho
}J_{ij}^{R}J_{ij}^{L}-\tilde{c}_{ij}^{\sigma
}\mbox{\boldmath$J$}_{ij}^{R}\cdot \mbox{\boldmath$J$}_{ij}^{L}  \nonumber \\
&&+\tilde{f}_{ij}^{\rho }J_{ii}^{R}J_{jj}^{L}-\tilde{f}_{ij}^{\sigma }%
\mbox{\boldmath$J$}_{ii}^{R}\cdot \mbox{\boldmath$J$}_{jj}^{L},
\label{Hintcf}
\end{eqnarray}
where $\tilde{f}_{ij}$ and $\tilde{c}_{ij}$ denote the forward and
Cooper scattering amplitudes, respectively, between bands $i$ and
$j$. Many repeated indices appear in this section, such as $i$ and
$j$ in Eq. (\ref {Hintcf}) and they are always implicitly summed
over. In Eq. (\ref{Hintcf}), we only keep the Lorentz invariant
interactions involving the product of a left current and a right
current. The LL and RR terms don't contribute to the RG equations
at second order and are expected to only shift the
velocities of the various modes. Note that $\tilde{c}_{ii}$ and $\tilde{f}%
_{ii}$ describe the same vertex so we set $\tilde{f}_{ii}=0$.
Also,
symmetries imply $\tilde{c}_{ij}=\tilde{c}_{ji}$ and $\tilde{f}_{ij}=\tilde{f%
}_{ji}$ [\onlinecite{NlegRG}]. That's how we get 32 different couplings in
doped 4-leg ladders. Then one can derive RG equations by the
operator product expansions of these $SU(2)$ scalar and vector
current operators.

Provided that the RG equations are known [\onlinecite{NlegRG}], how to
analyze the
RG flow is still non-trivial. Since all the interactions in Eq. (\ref{Hintcf}%
) are marginal at first glance, if one numbers all the couplings $\tilde{f}%
_{ij}$ and $\tilde{c}_{ij}$ and rename them as $\tilde{g}_{i}$
($i$ from 1 to 32)$,$ within the one-loop calculations, the
coupled non-linear RG equations can be written in the concise
form:
\begin{equation}
\frac{d\tilde{g}_{i}}{dl}=\tilde{M}_{i}^{jk}\
\tilde{g}_{j}\tilde{g}_{k}, \label{RGE}
\end{equation}
where $\tilde{g}_{i}$ is some coupling and the coefficient matrices $\tilde{M}%
_{i}^{jk}=$ $\tilde{M}_{i}^{kj}$ are symmetric in indices $j$ and
$k$ by construction. [$l$ is the logarithm of the ratio of a
characteristic length scale to the lattice scale, $l\equiv \ln
(L/a)$.] Recently, an unexpected potential structure of Eq.
(\ref{RGE}) was proven [\onlinecite{Chen, MSChang}].
After a proper rescaling to new couplings $g_{i}=\alpha _{i}%
\tilde{g}_{i},$ where $\alpha _{i}$ are constants, Eq. (\ref{RGE})
can be reduced to
\begin{equation}
\frac{dg_{i}}{dl}=M_{i}^{jk}\ g_{j}g_{k}=-\frac{\partial V(\vec{g})}{%
\partial g_{i}}.  \label{RGP}
\end{equation}
where $M_{i}^{jk}$ is totally symmetric in indices $i,j$ and $k$ and $V(\vec{%
g})$ is the so called RG potential [\onlinecite{MSChang}]. The scaling
constants $\alpha _{i}$ and the explicit RG potential form can be
found in Appendix C. It now provides a geometric picture for the
RG flows of Eq. (\ref{RGP}), which can be regarded as the
trajectory of an overdamped particle searching for a potential
minimum in the multi-dimensional coupling space. Thus, the
ultimate fate of the flow would either rest on the fixed points or
flow along some directions as the ``valleys/ridges'' of the
potential profile but there is only a trivial fixed point (all
$\tilde{g}_{i}=0$) within one-loop order.

Precisely, these directions are special sets of analytic solutions of Eq. (%
\ref{RGP}):
\begin{equation}
g_{i}(l)=\frac{G_{i}}{l_{d}-l},  \label{FR}
\end{equation}
if the constants $G_{i}$ satisfy the algebraic constraint,
\begin{equation}
G_{i}=M_{i}^{jk}G_{j}G_{k}.  \label{FR_solution}
\end{equation}
Eq. (\ref{FR}) is only valid for $l<l_{d}=\ln \xi /a$ where $\xi $
is a characteristic length scale where the coupling constants
become large. These special analytic solutions are referred as
``fixed rays'' because they grow under RG with the fixed ratios.
Sometimes, the specific ratios of the fixed ray reflect extra
symmetry in the Hamiltonian and the fixed ray is called a
``symmetric ray'' [\onlinecite{Lin98}].

In general, it's very unlikely that the bare values of $g_{i}(0)$
are proportional to the constants $G_{i}$. Then we have to check
whether these fixed rays are stable against deviations
[\onlinecite{MSChang}]. As long as the deviations grow slower than Eq.
(\ref{FR}), then the fixed ray is stable. Within the RG potential
picture, the stability of each fixed ray is determined by the
local topography along the direction.

In the vicinity of the fixed ray, if there are some small
deviations away from it
\begin{equation}
g_{i}(l)={\frac{G_{i}}{l_{d}-l}}+\Delta g_{i}(l),
\label{deviation}
\end{equation}
where $\Delta g_{i}(l)\ll g_{i}(l)$. The equations which describe
the deviations $\Delta g_{i}$ are
\begin{equation}
\frac{d}{dl}(\Delta g_{i})\ =\frac{B_{ij}}{l_{d}-l}\ \Delta g_{j}\
, \label{RG_deviation}
\end{equation}
where $B_{ij}=2M_{jk}^{i}G_{k}$. Since $M_{jk}^{i}$ is totally symmetric in $%
i,j$ and $k$, the matrix $B_{ij}$ is symmetric in $i$ and $j$.
$B_{ij}$ can be diagonalized by an orthogonal matrix $O_{nm}$ so
that Eq. (\ref {RG_deviation}) will decouple into independent
equations,
\begin{equation}
\frac{d}{dl}(\delta g_{n})\ =\frac{\lambda _{n}}{l_{d}-l}\ \delta
g_{n}\ , \label{D_deviation}
\end{equation}
where $\delta g_{n}=O_{ni}$ $\Delta g_{i}$, are the couplings
after the
linear transformation and $\lambda _{n}$ are the eigenvalues of the matrix $%
B_{ij}$. If the initial bare couplings $\delta g_{n}(0)\ll
G_{n}/l_{d}$, then the solutions of Eq. (\ref{D_deviation}) are
\begin{equation}
\delta g_{n}(l)=\delta g_{n}(0)\left( \frac{l_{d}}{l_{d}-l}\right)
^{\lambda _{n}}\ \sim \frac{G_{n}^{\prime }}{(l_{d}-l)^{\lambda
_{n}}},  \label{deltag}
\end{equation}
where $G_{n}^{\prime }\sim O((U/t)^{1-\lambda _{n}})$ is generally
non-universal depending on the initial couplings. Therefore, the
appropriate ansatz for the RG flows should be
\begin{eqnarray}
g_{i}(l) &\simeq &\frac{G_{i}}{l_{d}-l}+\frac{O_{in}G_{n}^{\prime }}{%
(l_{d}-l)^{\lambda _{n}}},  \label{NewAnsatz1} \\
&\simeq &\frac{G_{i}}{l_{d}-l}+\frac{G_{i}^{\prime \prime }}{%
(l_{d}-l)^{\lambda _{i}^{\max }}}+\cdots ,  \label{NewAnsatz2}
\end{eqnarray}
where $\lambda _{i}^{\max }$ in Eq. (\ref{NewAnsatz2}) is the
largest one among $\lambda _{n}$'s with nonzero coefficients
$O_{in}G_{n}^{\prime
}\equiv G_{i}^{\prime \prime }$ and the divergent behavior is dominated by $%
\lambda _{i}^{\max }$. Although Eq. (\ref{NewAnsatz2}) is derived
from the stability analysis near the fixed ray, it seems to be
the general behavior of the RG flow but the values of some $\lambda
_{i}^{\max }$ may vary from the eigenvalues of $B_{ij}$ when away
from the fixed ray. In fact, such power-law divergent solutions
were suggested before [\onlinecite{NlegRG, 3legRG}] but the analysis only
focused on the most relevant terms with exponent one, i.e.
$G_{i}\neq 0$ terms. Note that Eq. (\ref{NewAnsatz2}) is still
not the exact solutions of RG equations but it captures the
divergent part correctly and is enough to determine the phase
diagram.

For $\lambda _{i}^{\max }<0$, these deviations are irrelevant whereas if $%
\lambda _{i}^{\max }>1$, the deviations grow faster than the fixed
ray, so that the phase associated with the fixed ray is fragile.
For $0<\lambda _{i}^{\max }\leq 1$, the deviations actually grow
although not strongly enough to spoil the asymptotic fixed ray, as
illustrated in Fig.~(\ref{fig:stability}). Nevertheless, it
doesn't mean they won't affect anything since they are also
relevant couplings in the conventional classification but just
less relevant than the fixed ray ones.

\begin{figure}[t]
\begin{center}
\includegraphics[width=0.7\textwidth,clip]{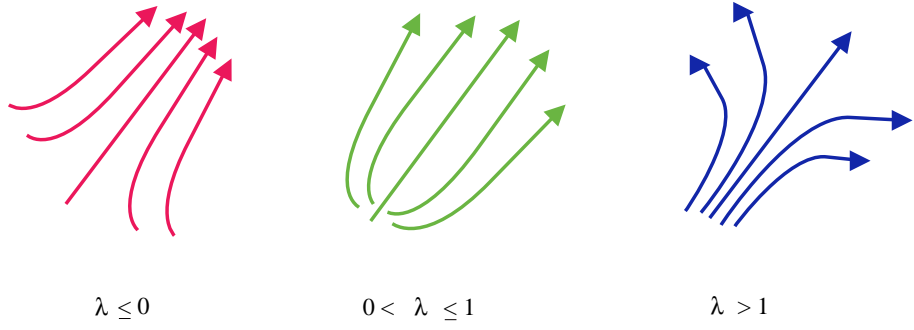}
\end{center}
\caption{The topography of RG flows near the fixed ray with $\lambda \leq 0$, $%
0<\lambda \leq 1$ and $\lambda >1$. It's clear that the deviation
is irrelevant for $\lambda \leq 0$ and relevant for $\lambda
>1$. The analysis of RG flow is more subtle for  $0<\lambda \leq
1$. In this case, although the deviation from fixed ray is
growing, the fixed ratios still remains. Therefore, RG still flows
onto the fixed ray but the phase is not only determined by the
fixed ray couplings.} \label{fig:stability}
\end{figure}

The effects of these subleading divergent terms on the RG flow of
a
particular coupling $g_{i}$ are dramatically different depending on whether $%
G_{i}=0$ or not. We can separate the effective Hamiltonian at the
cutoff length scale into the most relevant fixed ray part and the
subleading deviations as a perturbation. For $G_{i}\neq 0$, $g_{i}$
is relevant and it will lead to a gap. The subleading perturbations
will not destroy the original ground state but only modify the gap
function by giving rise to anomalous scaling [\onlinecite{MSChang}]. On the
other hand, if $G_{i}=0$, the subleading perturbations become
important if $0<\lambda _{i}^{\max }$ and will generate a small but
non-zero gap out of the initial gapless modes. Note that only when
initial deviations away from the fixed ray are small, are $\lambda
_{i}^{\max }$ universal and can be obtained from the eigenvalues of
$B_{ij}=2M_{jk}^{i}G_{k}$. However, from numerical solutions of the
RG equations we find that the couplings always diverge with power
law behavior like Eq. (\ref{NewAnsatz2}) though $\lambda _{i}^{\max
}$ is not the same as the eigenvalues of $B_{ij}$. We can extract
$\lambda _{i}^{\max } $ directly from the numerical solution of the
RG equations for those terms with $G_{i}=0$ and thus determine the
phase diagram from these relevant interactions. Surprisingly, as we
will see in the following sub-sections, the fixed ray and subleading
terms with universal $\lambda _{i}^{\max }$ from $B_{ij}$ already
are sufficient to determine the pinned bosons uniquely within
bosonization.

To recap, even in the weak-coupling RG analysis, there are
different energy scales. The fixed ray only represents the most
relevant couplings. The subleading couplings should be treated as
perturbations to the effective Hamlitonian corresponding to the
fixed ray and they are relevant enough to drive the effective
Hamiltonian into a phase in which the fragile gapless modes become
gapped but these gaps are small compared to those driven by the
fixed ray couplings.

\subsection{The Stripe Phase}

Now we know the ultimate fate of weak coupling RG must be a fixed
ray due to the existence of the RG potential [\onlinecite{Chen, MSChang}]. The
fixed ray indicates a direction in which the interactions will be
renormalized in the strong coupling region. If one can survey all
the fixed ray solutions in Eq. (\ref {FR_solution}) and the
corresponding subleading terms determined by the topography, then in
principle, all the phases in the weak coupling RG are obtained.
Following this idea, here we try to find the fixed ray whose
corresponding phase gives the stripe density oscillations. It turns
out that following fixed ray will do so:
\begin{eqnarray}
\sqrt{2}c_{11}^{\rho } &=&\sqrt{2}c_{44}^{\rho }=-\frac{c_{14}^{\rho }}{2}=%
\frac{c_{14}^{\sigma }}{2\sqrt{3}}=-f_{14}^{\rho }=\frac{f_{14}^{\sigma }}{2%
\sqrt{3}}=  \nonumber \\
\sqrt{2}c_{22}^{\rho } &=&\sqrt{2}c_{33}^{\rho }=-\frac{c_{23}^{\rho }}{2}=%
\frac{c_{23}^{\sigma }}{2\sqrt{3}}=-f_{23}^{\rho }=\frac{f_{23}^{\sigma }}{2%
\sqrt{3}}=-\frac{1}{16(l_{d}-l)}.  \label{FRstripe}
\end{eqnarray}
Eq. (\ref{FRstripe}) will be the solution of Eq. (\ref{FR_solution}) if $%
v_{1}=v_{4}$ and $v_{2}=v_{3}$. So if RG really flows onto this
fixed ray from some initial set of bare couplings, the
interpretation is that the fermi velocities get renormalized in the
corresponding phase [\onlinecite{Lin98}]. The upper and lower line in Eq.
(\ref{FRstripe}) correspond to the CDW fixed ray on effective 2-leg
systems composed of band pairs (1,4) and (2,3), respectively. In
principle, the fixed ray as the permutation of band indices in Eq.
(\ref{FRstripe}) also exists. The reason why we favor Eq. (\ref
{FRstripe}) is motivated by the fixed ray Eq. (\ref{FRU}), found in
the Hubbard model as we will see in the later sections.

Now we know that the fixed ray solution isn't the whole story for
the RG flow. The phase should be determined by all the relevant
interactions, including the subleading divergent ones. As long as
$G_{i}$ is given by the Eq. (\ref{FRstripe}) and with the known RG
matrix $M_{i}^{jk}$, then the largest divergent exponent $\lambda
_{i}^{\max }$ can be deduced
analytically from the eigenvalues and eigenvectors of the matrix $%
B_{ij}=2M_{jk}^{i}G_{k}$ in the vicinity of the fixed ray. There are two $%
5/8,$ four $1/2$, four $1/8$, four $-1/2,$ two $-3/8$, and 0s for
the
eigenvalues $\lambda _{n}$ of $B_{ij}$. The couplings are divergent for $%
\lambda _{n}>0$ even though they are small compared to the fixed ray
in the critical region. We should take more care about the terms
with eigenvalues $5/8,$ $1/2$ and $1/8$. To see what's their
influence, we have to know the direction corresponding to $\delta
g_{n}$, Eq. (\ref{deltag}), in the multi-dimensional space expanded
in the coupling basis $\Delta g_{i}$. The eigenvectors of $B_{ij}$
give this information. The subleading terms $\delta g_{n}$
corresponding to two $\lambda _{n}=5/8$ eigenvectors are in the
directions having non-zero projection on $c_{12}^{\rho
},c_{13}^{\rho },c_{24}^{\rho },c_{34}^{\rho },c_{12}^{\sigma
},c_{13}^{\sigma },c_{24}^{\sigma },$ and $c_{34}^{\sigma }$. The
eigenvectors of those corresponding to four $\lambda _{n}=1/2$ have
components on $c_{11}^{\rho },c_{22}^{\rho },c_{33}^{\rho
},c_{44}^{\rho },c_{14}^{\rho },c_{23}^{\rho },c_{11}^{\sigma
},c_{22}^{\sigma },c_{33}^{\sigma },c_{44}^{\sigma },c_{14}^{\sigma
},c_{23}^{\sigma },f_{14}^{\rho },f_{23}^{\rho },f_{14}^{\sigma }$
and $f_{23}^{\sigma }$. The terms with $\lambda _{n}=1/8$ have
components on $c_{12}^{\rho },c_{13}^{\rho },c_{24}^{\rho
},c_{34}^{\rho },c_{12}^{\sigma },c_{13}^{\sigma },c_{24}^{\sigma },$ and $%
c_{34}^{\sigma }$. Provided with this information, we know the
largest divergent exponent for each coupling, that is, $\lambda
_{i}^{\max }$ for the non-fixed ray couplings. Table
(\ref{C1S0_stripe}) summarizes $\lambda _{n}$ and the projections
of corresponding $\delta g_{n}$ in terms of coupling basis
$g_{i}$.

\begin{table}[ht]
\begin{tabular}{|c|c|}
\hline & nonzero component \\ \hline $5/8$ & $c_{12}^{\rho
},c_{13}^{\rho },c_{24}^{\rho },c_{34}^{\rho },c_{12}^{\sigma
},c_{34}^{\sigma }$ \\ \hline $5/8$ & $c_{12}^{\rho },c_{13}^{\rho
},c_{24}^{\rho },c_{34}^{\rho },c_{13}^{\sigma },c_{24}^{\sigma }$
\\ \hline $1/2$ & $c_{11}^{\rho },c_{44}^{\rho },c_{14}^{\rho
},c_{11}^{\sigma },c_{44}^{\sigma },f_{14}^{\rho },f_{14}^{\sigma
}$ \\ \hline $1/2$ & $c_{22}^{\rho },c_{33}^{\rho },c_{23}^{\rho
},c_{22}^{\sigma },c_{33}^{\sigma },f_{23}^{\rho },f_{23}^{\sigma
}$ \\ \hline $1/2$ & $c_{14}^{\rho },c_{11}^{\sigma
},c_{44}^{\sigma },c_{14}^{\sigma }$
\\ \hline
$1/2$ & $c_{23}^{\rho },c_{22}^{\sigma },c_{33}^{\sigma
},c_{23}^{\sigma }$
\\ \hline
$1/8$ & $c_{12}^{\rho },c_{34}^{\rho }$ \\ \hline $1/8$ &
$c_{13}^{\rho },c_{24}^{\rho }$ \\ \hline $1/8$ & $c_{12}^{\sigma
},c_{34}^{\sigma }$ \\ \hline $1/8$ & $c_{13}^{\sigma
},c_{24}^{\sigma }$ \\ \hline
\end{tabular}
\caption{This table summarizes the topography in the vicinity of
the fixed ray Eq.(\ref{FRstripe}). It shows the eigenvalues
$\lambda _{n}>0$ of the matrix $B_{ij}$ and their corresponding
eigen-direction in terms of the RG couplings.} \label{C1S0_stripe}
\end{table}

Now we would like to check numerically whether RG will really flow
onto this fixed ray Eq. (\ref{FRstripe}). It's very illuminating
to plot log[$\left| g_{i}(l)\right| $] v.s. log[($l_{d}-l$)] from
the numerical solution of the RG equations, where the absolute
value makes sure there won't be problems for those with $G_{i}<0$.
In the scaling region, if Eq. (\ref{NewAnsatz2}) is correct, then
we should see a straight line for each coupling $g_{i}(l)$, whose
slope indicates the exponent controlling the divergence. The
slopes will be negative one for the fixed ray, $-\lambda
_{i}^{\max }$ for the subleading terms and zero for irrelevant
terms.

If we choose the initial bare couplings with the ratios in Eq.
(\ref {FRstripe}) and with Fermi velocities $v_{1}=v_{4}$ and
$v_{2}=v_{3}$, we do find all the couplings grow with the fixed
ratios under RG flow toward the fixed ray Eq. (\ref{FRstripe}). In
order to see the subleading terms, we add some small deviations to
the initial bare couplings and Fermi velocity. The log-log plot of
each coupling agrees very well with the predicted slopes in Table
(\ref{C1S0_stripe}). A few selected examples are shown in
Fig.(\ref {fig:C1S0_stripe}).

\begin{figure}[th]
\begin{center}
\includegraphics[width=\columnwidth,clip]{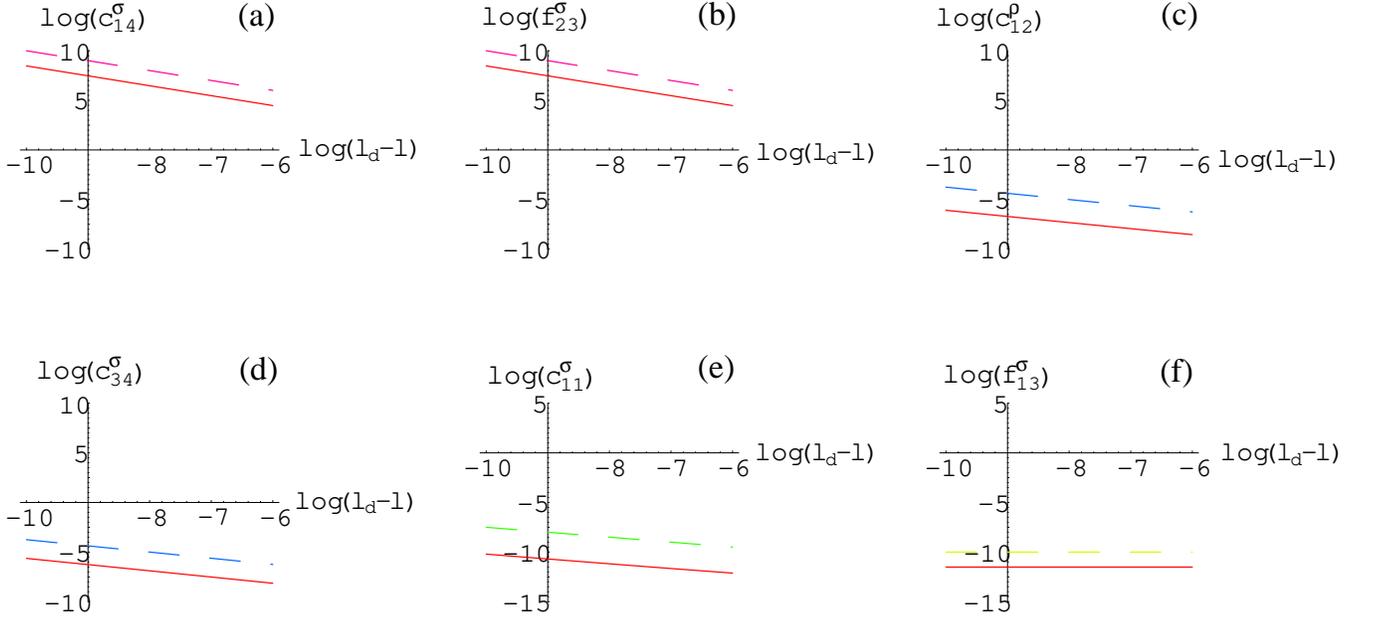}
\end{center}
\caption{This is the log[$\left| g_{i}(l)\right| $] v.s.
log[($l_{d}-l$)] plot for several typical couplings of stripe
fixed ray Eq. (\ref{FRstripe}). The slopes give us the divergent
exponent of each coupling. The solid (red) lines are the numerical
solutions of the RG equations. The dashed lines are
pure straight lines as reference with the predicted $\lambda_{i}^{\max }$: $%
1 $ (pink: $c^\sigma_{14}$ (a), $f^\sigma_{23}$ (b)), $5/8$ (blue: $b^\rho_{12}$ (c), $%
c^\sigma_{34}$ (d)), $1/2$ (green: $c^\sigma_{11}$ (e)), and $0$ (yellow: $%
f^\sigma_{13}$ (f)), respectively. In this case, the numerical
solutions agree very well with the prediction for all the
couplings.} \label{fig:C1S0_stripe}
\end{figure}

Although the subleading terms are also divergent for $0<\lambda
_{i}^{\max }<1$, they are still small compared to the fixed ray
couplings. Therefore, we treat the subleading couplings as the
perturbations to the effective Hamiltonian corresponding to Eq.
(\ref{FRstripe}) in the bosonization method.

We bosonize the relevant couplings to determine the phase diagram.
The full bosonized form of Eq. (\ref{Hintcf}) is
\begin{eqnarray}
H_{int} &=&\frac{1}{4}\sum_{i}\tilde{c}_{ii}^{\sigma }\cos (\sqrt{8\pi }%
\theta _{i\sigma })  \nonumber \\
&&+\frac{1}{4}\sum_{i\neq j}[(\tilde{c}_{ij}^{\rho
}+\tilde{c}_{ij}^{\sigma })\cos \sqrt{4\pi }\phi _{ij}^{\rho
-}\cos \sqrt{4\pi }\theta _{ij}^{\sigma
-}+2\tilde{c}_{ij}^{\sigma }\cos \sqrt{4\pi }\phi _{ij}^{\rho -}\cos \sqrt{%
4\pi }\theta _{ij}^{\sigma +}  \nonumber \\
&&+(\tilde{c}_{ij}^{\rho }-\tilde{c}_{ij}^{\sigma })\cos
\sqrt{4\pi }\phi _{ij}^{\rho -}\cos \sqrt{4\pi }\phi _{ij}^{\sigma
-}-2\tilde{f}_{ij}^{\sigma }\cos \sqrt{4\pi }\phi _{ij}^{\sigma
-}\cos \sqrt{4\pi }\theta _{ij}^{\sigma +}],  \label{BHint}
\end{eqnarray}
where these bosons fields are defined in Eq. (\ref{2-leg BF}). Since $\tilde{%
c}_{ii}^{\rho }$ and $\tilde{f}_{ij}^{\rho }$ only contribute to the
gradient terms, they are not important here. The reason we express
the Hamiltonian by the tilde interactions $\tilde{c},$ $\tilde{f}$
and Eq.(\ref {2-leg BF}) is only for the convenience of notations.
We emphasize that the basis of boson fields should be determined by
the hierarchy of relevant interactions. In other words, the relevant
interactions should not only tell us what boson fields are pinned
but also the basis in terms of which they are pinned.

At some intermediate length scale, the most relevant interactions,
those in Eq. (\ref{FRstripe}), are large but the others, including
the subleading terms, are still small compared to them. In order
to minimize Eq. (\ref {BHint}), the most relevant couplings in Eq.
(\ref{FRstripe}) will pin the values of $\phi _{14}^{\rho -},\phi
_{23}^{\rho -},\theta _{14}^{\sigma
+},\theta _{23}^{\sigma +},\phi _{14}^{\sigma -}$ and $\phi _{23}^{\sigma -}$%
. These pinned bosons, regardless of $\theta $ or $\phi ,$
immediately suggest a basis. In the spin channel, since there are
already four mutually orthogonal combinations of band bosons getting
pinned, it's natural to choose $R_{\sigma }$ corresponding to the
pinned combinations. So the basis for spin fields is fixed. As for
the charge channel, there are two combinations of band bosons which
get pinned and we also know symmetry requires the gapless mode to be
total charge field $(\Phi _{1\rho },\Theta _{1\rho })$ since there
is no interaction involving boson fields not orthogonal to it here.
These three fields and the orthonormal condition will uniquely fix
the only unknown basis field,\begin{equation} \Phi _{2\rho
}=\frac{1}{2}(\phi _{1\rho }-\phi _{2\rho }-\phi _{3\rho }+\phi
_{4\rho }),
\end{equation}
similarly for $\Theta _{2\rho }$ if replace $\phi $ by $\theta $. In
this case, the relevant interactions suggest the basis we should
adopt is:
\\
$R_{\rho }=\left(
\begin{array}{llll}
1/2 & 1/2 & 1/2 & 1/2 \\
1/\sqrt{2} & 0 & 0 & -1/\sqrt{2} \\
0 & 1/\sqrt{2} & -1/\sqrt{2} & 0 \\
1/2 & -1/2 & -1/2 & 1/2
\end{array}
\right) $ , $R_{\sigma }=\left(
\begin{array}{llll}
1/\sqrt{2} & 0 & 0 & 1/\sqrt{2} \\
1/\sqrt{2} & 0 & 0 & -1/\sqrt{2} \\
0 & 1/\sqrt{2} & 1/\sqrt{2} & 0 \\
0 & 1/\sqrt{2} & -1/\sqrt{2} & 0
\end{array}
\right) .$
\\
Now we will rewrite the interactions Eq. (\ref{BHint}) in terms of
the new basis given by $R_{\rho }$ and $R_{\sigma }$. So far we get
six pinned bosons only by considering the fixed ray interactions. At
lower energy scale, the subleading terms also become large, yet
still small compared with the fixed ray. Replacing the pinned bosons
by their pinned values, we get a C2S0 effective Hamiltonian and the
subleading terms will be treated as the perturbations. $\Phi _{1\rho
}$ is absent in Eq. (\ref{BHint}) and it will remain gapless as we
expected. The question is about whether $\Phi _{2\rho }$ will get
pinned due to the perturbation involving it, such as $c_{12}^{\rho
},c_{13}^{\rho },c_{24}^{\rho },c_{34}^{\rho },c_{12}^{\sigma
},c_{13}^{\sigma },c_{24}^{\sigma }$ and $c_{34}^{\sigma }$. At
first glance, one may conclude the gapless bosons $\Phi _{2\rho }$
won't get pinned since all the subleading perturbations also contain
the dual of pinned spin boson $\phi _{14}^{\sigma -}$ and $\phi
_{23}^{\sigma -}$. In other words, these subleading interactions
should be irrelevant and $\Phi _{2\rho }$ can't be pinned. It's true
for this analysis. But the common wisdom tells us that the gapless
mode is usually fragile unless protected by some symmetry or
incommensurability. In fact, there are always some other higher
order interactions which can be generated in the continuum limit as
long as they are allowed by symmetry. We usually don't pay attention
to these higher order terms for they should be much smaller and less
relevant than the interactions in Eq. (\ref{Hintcf}). However, the
scaling dimension of these higher order interactions can be changed
due to the existence of
some other interactions [\onlinecite{Schulz94, Giamarchi}]. For example, consider a $%
4^{th}$ order term in perturbation theory:
\begin{equation}
\delta H\propto \left( c_{12}^{\rho }\right) ^{2}c_{11}^{\sigma
}c_{22}^{\sigma }[\cos \sqrt{4\pi }\phi _{12}^{\rho -}\cos \sqrt{2\pi }%
(\theta _{1\sigma }+\theta _{2\sigma })]^{2}\cos \sqrt{8\pi
}\theta _{1\sigma }\cos \sqrt{8\pi }\theta _{2\sigma }
\end{equation}
Using the operator product expansion, we can replace all factors involving $%
\theta _{1\sigma }$ and $\theta _{2\sigma }$ by a constant. The
remaining operator contains a term:
\begin{equation}
\delta H\propto \cos \sqrt{8\pi }(\phi _{1\rho }-\phi _{2\rho })=\cos \sqrt{%
4\pi }(\sqrt{2}\Phi _{2\rho }+\phi _{14}^{\rho -}-\phi _{23}^{\rho
-})\to \cos \sqrt{8\pi }\Phi _{2\rho },  \label{sublead_stripe}
\end{equation}
where we replaced $\phi _{14}^{\rho -}$ and $\phi _{23}^{\rho -}$ by their
expectation values in the third expression. This operator doesn't depend on
spin fields anymore and we have an effective sine-Gordon
Hamilatonian for $(\Phi _{2\rho },\Theta
_{2\rho })$. The cosine interaction has a scaling dimension of $2/K_{2\rho }$%
. If the renormalized value of the Luttinger parameter for the
$\Phi _{2\rho }$ boson, $K_{2\rho }>1$, then Eq.
(\ref{sublead_stripe}) is relevant. In
general, it's highly nontrivial to determine the renormalized value $%
K_{2\rho }$ after integrating out the gapped modes. However, we
can calculate the renomalization of $K_{2\rho }$ due to the
gradient terms of interactions:
\begin{equation}
K_{2\rho }=\sqrt{\frac{\pi
\overline{v}-\overline{c}+\overline{f}}{\pi
\overline{v}+\overline{c}-\overline{f}}},  \label{K_2rho}
\end{equation}
where
\begin{eqnarray*}
\overline{v} &=&(v_{1}+v_{2}+v_{3}+v_{4})/4, \\
\overline{c} &=&\tilde{c}_{11}^{\rho }+\tilde{c}_{22}^{\rho }+\tilde{c}%
_{33}^{\rho }+\tilde{c}_{44}^{\rho }, \\
\overline{f} &=&\tilde{f}_{12}^{\rho }+\tilde{f}_{13}^{\rho }-\tilde{f}%
_{14}^{\rho }-\tilde{f}_{23}^{\rho }+\tilde{f}_{24}^{\rho }+\tilde{f}%
_{34}^{\rho }.
\end{eqnarray*}
According to the ratios in Eq. (\ref{FRstripe}), we find that $K_{2\rho }>1$%
. Therefore, $\Phi _{2\rho }$ should also be pinned with this
sine-Gordon type interactions. We conclude the final phase should be
C1S0 and the pinned bosons are $\Phi _{2\rho },\phi _{14}^{\rho
-},\phi _{23}^{\rho -},\theta _{14}^{\sigma +},\theta _{23}^{\sigma
+},\phi _{14}^{\sigma -}$ and $\phi _{23}^{\sigma -}$. For this
pinning pattern, the correlation functions of the
$4\overline{k}_{F}$ density operators Eq. (\ref{4KF1}) and
(\ref{4KF2}) with $\{i,j\}=\{1,4\}$ and $\{k,l\}=\{2,3\}$, decay
with a power-law. Also, with $\{s,t\}=\{1,4\}$ or $\{2,3\}$, the
correlation functions of the bipairing operators Eq.
(\ref{bipair_2}) and the term with $\theta _{st}^{\sigma +}$ in Eq.
(\ref{bipair_3}) decay with a power-law. All the pairing operators
Eq. (\ref{P1}) and (\ref{P2}) decay exponentially. This phase has
stripes and bipairing correlations. The result is also consistent
with FSS.

The only question left is how to find the proper initial couplings
so that the RG will flow to the fixed ray. The initial bare
couplings are determined by the interactions in the model. As long
as one includes enough short ranged interactions in the
Hamiltonian, the initial bare couplings can be tuned near the
ratios in Eq. (\ref{FRstripe}). The point is that the fixed ray
should indicate some phase in the strong coupling regions. Thus,
to find the proper initial bare couplings that RG will flow to
this fixed ray may not be the most important issue for our
purpose.

The fact that we need $v_{1}\simeq v_{4}$ and $v_{2}\simeq v_{3}$
in order to see the stripe phase in RG resembles the situation in
the decoupled 2-leg ladders limit we study in section V. It seems
to suggest that the stripe phase should be related to the
renormalization of Fermi velocities from both limits we study.

\subsection{The Weak Coupling Repulsive Hubbard Model}

In the previous section, we found the fixed ray corresponding to the
stripe phase without knowing the exact underlying model. Now we
would like to switch gears and study the fixed rays corresponding to
parameters of the Hubbard model in Eq. (\ref{H0}) and (\ref{Hint}).

In the four-band region, the RG flows to the following fixed ray
[\onlinecite{NlegRG}],

\begin{equation}
\sqrt{2}c_{11}^{\rho }=\sqrt{2}c_{44}^{\rho }=-\frac{c_{14}^{\rho }}{2}=%
\sqrt{\frac{2}{3}}\frac{c_{11}^{\sigma }}{2}=\sqrt{\frac{2}{3}}\frac{%
c_{44}^{\sigma }}{2}=-\frac{c_{14}^{\sigma }}{2\sqrt{3}}=-f_{14}^{\rho }=-%
\frac{1}{16(l_{d}-l)}.  \label{FRU}
\end{equation}
It seems that only the interactions between band 1 and 4 are
relevant and band 2 and 3 are totally decoupled from the system.
Also, the fixed ratios in Eq. (\ref{FRU}) are the same as those of
the C1S0 phase in a doped 2-leg ladder if band 1 and 4 are
regarded as an effective 2-leg system. So the final phase was
referred to as C1S0 + C2S2 = C3S2. Notice that the RG equations,
in the $g_{i}$ basis of Eq. (\ref{RGP}), are invariant under the
permutations of indices but the bare values, $g_{i}(0)$, favor the
phase in which the couplings involving bands 1 and 4 get large.
This can be seen from the factors of $1/v_{i}$ relating the
$g_{i}$'s to the $\tilde{g}_{i}$'s in Eq.
(\ref{gtotildeg1})-(\ref{gtotildeg4}).

However, now we know that the fixed ray solution isn't the whole
story for the RG flow. Bands 2 and 3 are never really decoupled
since there are subleading coupling constants that involve these
bands. Once again, we will plot log[$\left| g_{i}(l)\right| $]
v.s. log[($l_{d}-l$)] from the numerical solution of the RG
equations and we see nice straight lines in the scaling region.
The slopes will be compared with $-\lambda _{i}^{\max }$, which
can be deduced from the eigenvalues $\lambda _{n}$ and
eigenvectors of the matrix $B_{ij}=2M_{jk}^{i}G_{k}$.

With the $G_{i}$ given by Eq. (\ref{FRU}), there are two $-1/2,$ six $-1/16$%
, two $1/2$, two $15/16$, a $1$ and other 0s for $\lambda _{n}$.
Terms corresponding to $-1/2$ and $-1/16$ are irrelevant and not
harmful to
anything. We should carefully look at the terms with $1/2$, $15/16$, and $1$%
. The effects on the phase diagram depend on whether they have
components on the couplings with $G_{i}=0$. This information is
given by their corresponding eigenvectors. One may think in
general $\lambda _{n}=1$ means the fixed ray is unstable. This is
true if the deviations $\delta g_{n}$ have non-zero components on
the couplings besides the fixed ray ones. Fortunately, here the
eigenvector for $\lambda _{n}=1$ only has two components with
negative $c_{11}^{\sigma }$ and positive $c_{44}^{\sigma }$ in
terms of the original coupling basis. It will shift the values of
fixed
ratios regarding $c_{11}^{\sigma }$ and $c_{44}^{\sigma }$ in Eq. (\ref{FRU}%
) a little bit but it won't change the fact that those seven
couplings are
the most relevant ones. This only reflects that the fixed ray Eq. (\ref{FR}%
), is just a special set of the solutions and not the most general
one. Table (\ref{C1S0U}) summarizes $\lambda _{n}$ and their
projections in terms of coupling basis $g_{i}$.

\begin{table}[h]
\begin{tabular}{|c|c|}
\hline & nonzero component \\ \hline $1/2$ & $c_{14}^{\rho
},c_{14}^{\sigma },f_{14}^{\sigma }$ \\ \hline $1/2$ &
$c_{11}^{\rho },c_{44}^{\rho },c_{14}^{\rho },c_{11}^{\sigma
},c_{44}^{\sigma },c_{14}^{\sigma },f_{14}^{\rho }$ \\ \hline
$15/16$ & $c_{13}^{\rho },c_{34}^{\rho },c_{13}^{\sigma
},c_{34}^{\sigma }$
\\ \hline
$15/16$ & $c_{12}^{\rho },c_{24}^{\rho },c_{12}^{\sigma
},c_{24}^{\sigma }$
\\ \hline
$1$ & $c_{11}^{\sigma },c_{44}^{\sigma }$ \\ \hline
\end{tabular}
\caption{This table summarizes the topography in the vicinity of
the fixed
ray Eq.(\ref{FRU}). It shows the eigenvalues $\lambda _{n}=1$ of the matrix $%
B_{ij}$ and their corresponding eigen-direction in terms of the RG
couplings. } \label{C1S0U}
\end{table}

This result can be checked by plotting log[$\left| g_{i}(l)\right|
$] v.s. log[($l_{d}-l$)] for the numerical solutions of RG
equations. As a test, we can artificially tune the ratios of
initial conditions based on Eq. (\ref {FRU}), such that the RG
flow will be really in the vicinity of the fixed ray. In this
case, the slope of each coupling agrees perfectly with the
prediction given above.

Now using the initial conditions as shown in Appendix C,
determined by physical parameters on-site interaction $U$ and the
doping $\delta $, we can do the same analysis. Even in the region
RG flow still controlled by the fixed ray Eq. (\ref{FRU}), now we
find not all the slopes agree with the prediction. Some couplings
with predicted $\lambda _{i}^{\max }=0$, actually have non-zero
slopes in the log-log plot and those slopes may vary according to
the initial conditions. They are new subleading terms besides
those given by the stability check near the fixed ray. However, we
don't find the notable change of the slopes for those couplings
with $\lambda _{i}^{\max }=5/16,1/2$ or $1$, as long as the RG
flow is still dominated by the same fixed ray. That is, the
universal analytic prediction in the vicinity of the fixed ray is
still correct to some extent.

A few selected typical examples are shown in
Fig.~(\ref{fig:C1S0U}). The failure to predict all $\lambda
_{i}^{\max }$ correctly for each couplings doesn't mean we can't
determine the phase diagram. The point is that we should treat all
the divergent terms with the exponent $0<\lambda _{i}^{\max }\leq
1$ as the perturbations to the effective Hamiltonian corresponding
to the fixed ray in the bosonization scheme.

\begin{figure}[th]
\begin{center}
\includegraphics[width=\columnwidth,clip]{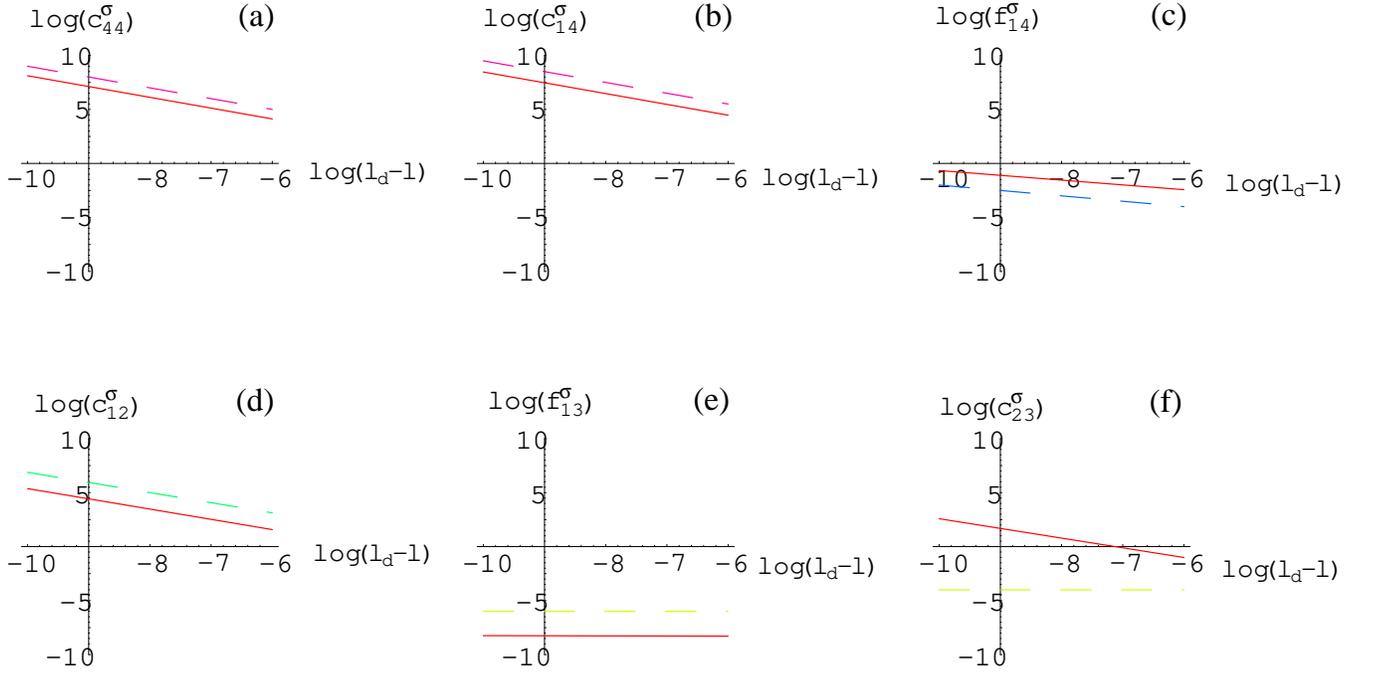}
\end{center}
\caption{log[$\left| g_{i}(l)\right| $] v.s. log[($l_{d}-l$)] plot
for
several typical couplings. The parameters are chosen as $t=t_{i,\perp }=1$, $%
U=0.01$ and the hole doping is 0.135. The slope of it gives us the
divergent exponent of each coupling. The solid (red) lines are the
numerical solutions of the RG equations. The dashed lines are pure
straight lines as reference
with the predicted slopes $1$ (pink: $c_{44}^{\sigma }$ (a), $c_{14}^{\sigma }$ (b)%
), $1/2$ (blue: $f_{14}^{\sigma }$ (c)), $15/16$ (green:
$c_{12}^{\sigma }$ (d)), and $0$ (yellow: $f_{13}^{\sigma }$ (e),
$c_{23}^{\sigma }$ (f)), respectively. As one can see, the numerical
solution of $c_{23}^{\sigma }$ (f) doesn't agree with its predicted
exponent. The number of couplings, whose slopes don't agree with the
stability analysis, depends on the initial conditions. With the
initial conditions used here, there are total 11 couplings,
predicted with zero slope near the fixed ray, but have nonzero
slopes in the numerical solution. However, these new term don't
change the pinned bosons and the final phase is the same as that
when these terms are irrelevant.} \label{fig:C1S0U}
\end{figure}

In order to minimize Eq. (\ref{BHint}), the most relevant couplings in Eq. (%
\ref{FRU}) will pin the values of $\phi _{14}^{\rho -},\theta
_{14}^{\sigma +}$, and $\theta _{14}^{\sigma -}$. The fixed ray
interactions and the symmetry imply we should choose $\Phi
_{2}^{\rho },\phi _{14}^{\rho -},\theta _{14}^{\sigma +}$, and
$\theta _{14}^{\sigma -}$ for the new basis of bosons. Unlike the
previous case of stripe fixed ray, here there are still two
undetermined fields in charge and spin channel each. So the
choices of the basis is not unique anymore. Any two charge (or
spin) fields orthogonal to $\Phi _{2}^{\rho }$ and $\phi
_{14}^{\rho -}$ (or $\theta _{14}^{\sigma +}$, and $\theta
_{14}^{\sigma -}$) can be used. For example, the simplest choice
would be the same as $R_{\rho }$ and $R_{\sigma }$ used in the
previous section.

We then follow the hierarchy of these subleading terms in
repulsive Hubbard model. Pick up the largest one among them and
rewrite it in terms of the new fields according to $R_{\rho }$ and
$R_{\sigma }$. After replace $\phi _{14}^{\rho -},\theta
_{14}^{\sigma +}$, and $\theta _{14}^{\sigma -}$ by constants, The
largest subleading term $(c_{12}^{\rho }+c_{12}^{\sigma })$ in Eq.
(\ref{BHint}) becomes:
\begin{eqnarray}
&&\cos \sqrt{4\pi }\phi _{12}^{\rho -}\cos \sqrt{4\pi }\theta
_{12}^{\sigma
-}  \nonumber \\
&=&\cos \sqrt{\pi }(\sqrt{2}\Phi _{2\rho }+\phi _{14}^{\rho
-}-\phi _{23}^{\rho -})\cos \sqrt{\pi }(\theta _{14}^{\sigma
+}+\theta _{14}^{\sigma
-}-\theta _{23}^{\sigma +}-\theta _{23}^{\sigma -})  \nonumber \\
&\to &\cos (\sqrt{2\pi }\Phi _{2}^{\rho }-\sqrt{\pi }\phi
_{23}^{\rho -})\cos \sqrt{\pi }(\theta _{23}^{\sigma +}+\theta
_{23}^{\sigma -}). \label{subleadU_1}
\end{eqnarray}
Similarly, next largest term $(c_{13}^{\rho }+c_{13}^{\sigma })$
becomes:
\begin{eqnarray}
&&\cos \sqrt{4\pi }\phi _{13}^{\rho -}\cos \sqrt{4\pi }\theta
_{13}^{\sigma
-}  \nonumber \\
&\to &\cos (\sqrt{2\pi }\Phi _{2}^{\rho }+\sqrt{\pi }\phi
_{23}^{\rho -})\cos \sqrt{\pi }(\theta _{23}^{\sigma +}-\theta
_{23}^{\sigma -}). \label{subleadU_2}
\end{eqnarray}

With the perturbations like Eq. (\ref{subleadU_1}) and
(\ref{subleadU_2}), four more boson fields $\Phi _{2}^{\rho },\phi
_{23}^{\rho -},\theta _{23}^{\sigma +}$ and $\theta _{23}^{\sigma
-}$ will get pinned. Thus, as long as the RG flow is dominated by
Eq. (\ref{FRU}), the final phase should be C1S0 and the pinned
bosons are $\Phi _{2}^{\rho },\phi _{14}^{\rho -},\phi _{23}^{\rho
-},\theta _{14}^{\sigma +},\theta _{14}^{\sigma -},\theta
_{23}^{\sigma +}$ and $\theta _{23}^{\sigma -}$. This pinning
pattern will make the pair operator Eq. (\ref{P1}) decay with a
power law and $4k_{F}$ density operators Eq. (\ref{4KF1}) and
(\ref{4KF2}) exponentially decay. So it's a pairing phase with no
stripes. Since in this phase the pinned charge or spin bosons are all $\phi $ or $%
\theta $ fields, respectively, we can use other choices for $R_{\rho }$ and $%
R_{\sigma }$ and the result will be the same.

Strictly speaking, the analysis in this section is only valid in the weak coupling region.
Other phases might occur at strong coupling or in the $t-J$ model. In the next section,
we will study a different limit that may reveal some strong coupling physics.

\section{LIMIT OF 2 DECOUPLED 2-LEG LADDERS}

One interesting limit in which it is relatively easy to understand
the stripe phase is the limit of two 2-leg ladders weakly coupled
by electron hopping and density-density interaction. Essentially
this limit was discussed in Ref. [\onlinecite{NlegRG}], sub-section
(VII-B-1) in the context of a 2-dimensional array of 2-leg
ladders. As we discuss below, the low energy effective Hamiltonian
in this limit is the same one describing the 2-leg
\textit{bosonic} ladder which was discussed in Ref. [\onlinecite{Orignac}]
based on the ``bosonization'' for 1D bosons [\onlinecite{Haldane}]. See
Fig.~(\ref{fig:decoupled}) for illustration.

This limit corresponds to $t_{2,\perp }$ and $V_{2,\perp }$ very
small, is the following Hamiltonian:
\begin{equation}
H_{0}=-\sum_{\alpha =\pm }\left[
\sum_{x=1}^{L-1}\sum_{a=1}^{N}tc_{a,\alpha }^{\dagger
}(x)c_{a,\alpha }(x+1)+\sum_{x=1}^{L}\sum_{a=1}^{N-1}t_{a,\perp
}c_{a,\alpha }^{\dagger }(x)c_{a+1,\alpha }(x)\right] +h.c.
\end{equation}
and
\begin{equation}
H_{int}=U\sum_{x=1}^{N}\sum_{a=1}^{L}n_{a,\uparrow
}(x)n_{a,\downarrow }(x)+\sum_{\alpha =\pm
}\sum_{x=1}^{L}\frac{V_{2,\perp }}{2}n_{2,\alpha }(x)n_{3,\alpha
}(x).
\end{equation}

We set $t_{1,\perp }=t_{3,\perp }=t$ for simplicity, but this is
not essential. Thus we may begin by considering the behavior of 2
decoupled 2-leg Hubbard ladders. Over a wide range of parameters,
the 2-leg Hubbbard ladder is expected to be in a C1S0 phase
[\onlinecite{Ian2leg, 2leg}]. Introducing band bosons,
$\phi _{1\nu }^{U}$, $\phi _{2\nu }^{U}$ for the upper 2-leg
ladder (on legs $1$ and $2$ and $\nu =\rho $ or $\sigma $) and
then changing variables to
\begin{equation}
\phi _{\pm \nu }^{U}\equiv (\phi _{1\nu }^{U}\pm \phi _{2\nu
}^{U})/\sqrt{2},
\end{equation}
this phase is expected to have $\phi _{-\rho }^{U}$ and $\theta
_{\pm \sigma }^{U}$ pinned. The lowest Friedel oscillation
wave-vector is $4\bar{k}_{F}$, corresponding to the 2-leg version
of stripes, namely equally spaced pairs of holes (1 on each leg)
forming a ``quasi charge density wave'' near the boundary. The
$\phi _{+\rho }^{U}$ boson is, of course, gapless. Let $\Delta $
be the minimum gap for the other three bosons. In the limit
$U<<t$, we expect $\Delta \propto t\exp [-\hbox{const}\times
t/U]$. For $U\geq t$ we expect $\Delta $ to be $O(t)$ or larger.
Of course the lower 2 legs have a gapless boson $\phi _{+\rho
}^{L}$.

We now turn on small $t_{2\perp }$ and $V_{2\perp }$, coupling
together the
two 2-leg ladders. Both these interactions involve duals of pinned bosons, $%
\theta _{-\rho }^{U/L}$ and $\phi _{\pm \sigma }^{U/L}$ and hence
are ultimately irrelevant. On the other hand, a pair-hopping term,
with amplitude $t^{\prime }\propto t_{2,\perp }^{2}$ and an
interaction between
the $4\bar{k}_{F}$ density operators in the two 2-leg ladders, of strength $%
V^{\prime }\propto V_{2,\perp }$ are generated perturbatively.
[This linear dependence of $V^{\prime }$ on $V_{2,\perp }$ follows
since, (by analogy with the calculation in Appendix B), the
$4\bar{k}_{F}$ term in the density operators for each 2-leg ladder
is $O(U)$. Here we disagree slightly with Ref. [\onlinecite{NlegRG}] which
finds this interaction to be $\propto (V_{2,\perp })^{2}$.]
Neither of these interactions involves the dual of any pinned
bosons. For sufficiently weak $t_{2,\perp }$ and $V_{2,\perp }$ we
may analyze the low energy theory by simply replacing all pinned
bosons from the 2 2-leg ladders by their expectation values and
writing an effective Hamiltonian for $\phi _{+\rho }^{U}$ and
$\phi _{+\rho }^{L}$. The effective Hamiltonian describes the
physics at energy scales $\ll \Delta $ and for it
to be valid the energy scales characterizing the pair hopping and $4\bar{k}%
_{F}-4\bar{k}_{F}$ density interactions must also be $\ll \Delta
$. It is now convenient to change boson variables to:
\begin{equation}
\phi _{\pm }\equiv [\phi _{+\rho }^{U}\pm \phi _{+\rho
}^{L}]/\sqrt{2},
\end{equation}
since the pair hopping and $4\bar{k}_{F}$-$4\bar{k}_{F}$ density
interactions only involve $\phi _{-}$ and its dual, $\theta _{-}$.
The effective Hamiltonian at energy scales $\ll \Delta $ can be
written:
\begin{eqnarray}
H_{eff} &=&\int dx\{\frac{v_{+}}{2}[K_{+}(\partial _{x}\phi _{+})^{2}+\frac{1%
}{K_{+}}(\partial _{x}\theta _{+})^{2}]  \nonumber \\
&&+\frac{v_{-}}{2}[K_{-}(\partial _{x}\phi _{-})^{2}+\frac{1}{K_{-}}%
(\partial _{x}\Theta _{-}^{\rho })^{2}]  \nonumber \\
&&+t^{\prime }\cos (\sqrt{2\pi }\phi _{-})+V^{\prime }\cos (\sqrt{8\pi }%
\theta _{-})\}.  \label{Heff}
\end{eqnarray}
Here, to lowest order in $t_{2,\perp }$ and $V_{2,\perp }$,
$K_{+}=K_{-}$ is simply the Luttinger parameter of the $\phi
_{+\rho }^{U/L}$ bosons on the upper and lower 2-leg ladders and
likewise $v_{+}=v_{-}$ is the corresponding velocity. The
Luttinger parameter of the 2-leg ladder is expected to approach
$1$ at half-filling and to decrease as the density moves away from
half-filling.

\begin{figure}[t]
\begin{center}
\includegraphics[width=\columnwidth,clip]{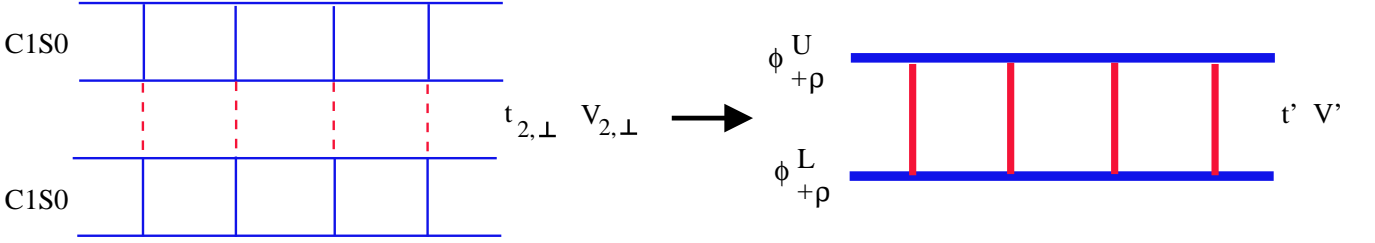}
\end{center}
\caption{In the limit that $t_{2,\perp }$ and $V_{2,\perp }$ are
much smaller than the minimum gap of the bosons, $\Delta $, each
2-leg ladder is well-described by the C1S0 phase, which has pairing
and $4\bar{k}_{F}$ density oscillation. The direct electron hopping
$t_{2,\perp }$ becomes an irrelevant process yet pair hopping,
$t^{\prime },$ generated by the higher order process will appear and
$4\bar{k}_{F}-4\bar{k}_{F}$ component, $V^{\prime },$ is the lowest
order relevant term in the interaction $V_{2,\perp }$. The phase is
determined by the competition between $t^{\prime }$ and $V^{\prime
}$. If $t^{\prime }$ dominates, the system has pairing (boson
superfluid) and $8\bar{k}_{F}$ (4$\pi \rho _{0}$) density
oscillation in fermion (boson) language, where $\rho _{0}$ is the
average boson density. If $V^{\prime }$ dominates, the system has
bipairing (boson pair superfluid) and $4\bar{k}_{F}$ (2$\pi \rho
_{0}$) density. } \label{fig:decoupled}
\end{figure}

We observe that the connection between $\phi _{\pm }$ and the
fields $\Phi _{A}^{\rho }$, in Eq. (\ref{4leg-BF}) is not so
straightforward, even in the
limit $t_{2,\perp }\to 0$. Ignoring for the moment all interactions, when $%
t_{2,\perp }$ is strictly zero the bands come in two identical
pairs, one member of each pair from the upper 2 legs and one from
the lower 2 legs. However, as soon as $t_{2,\perp }\neq 0$, these
bands are mixed. This band-mixing is not taken into account in the
present approach in which bosonization fields are introduced
separately for the bands on the upper 2 and lower 2 legs. The
present approach should be the correct one in the limit considered
of very small $t_{2,\perp }$, but it is non-trivial to connect the
results with those obtained from the standard weak coupling
approach. We note that a very analogous situation occurs even in
the much simpler and well-studied 2-leg spinless fermion model. If
the inter-chain hopping is sufficiently weak one normally
bosonizes the fermions on each leg, whereas in the weak coupling
limit, bosons are introduced for each band. Because of the
exponentials entering the bosonization formulas the relationship
between the ``leg boson'' and ``band bosons'' is very non-linear.
Nonetheless, it appears that the same phase diagram can be
obtained using either approach. This can be seen by comparing the
various features of the phases obtained using either method
[\onlinecite{2leg}].

It is not so straightforward to estimate the conditions on
$t_{2,\perp }$ and $V_{2,\perp }$ for this effective Hamiltonian
to be valid. Fortunately,
this is not important for our purposes. Normalizing the operators in Eq. (%
\ref{Heff}) so that:
\begin{eqnarray}
<e^{i\sqrt{2\pi }\phi _{-}(x)}e^{-i\sqrt{2\pi }\phi _{-}(y)}> &=&{\frac{1}{%
|x-y|^{1/K_{-}}}}  \nonumber \\
<e^{i\sqrt{8\pi }\theta _{-}(x)}e^{-i\sqrt{8\pi }\theta _{-}(y)}> &=&{\frac{1%
}{|x-y|^{4K_{-}}}},
\end{eqnarray}
we see that $t^{\prime }$ has a scaling dimension of (energy)$%
^{2-1/(2K_{-})} $ and $V^{\prime }$ has a scaling dimension of (energy)$%
^{2-2K_{-}}$ (after setting $v_{-}=1$). These energies scales must
be much less than the cut-off scale, $\Delta $, i.e.
\begin{eqnarray}
t^{\prime } &\ll &\Delta ^{2-1/(2K_{-})}  \nonumber \\
V^{\prime } &\ll &\Delta ^{2-2K_{-}}.  \label{dimt'V'}
\end{eqnarray}
Here we assume $1/4<K_{-}<1$, which is certainly true near
half-filling. As mentioned above, $t^{\prime }\propto t_{2,\perp
}^{2}$ and $V^{\prime }\propto V_{2,\perp }$. A more complete
estimate of these parameters is more difficult to make and could
be quite different depending on whether the 2-leg ladders are in
the weak or strong coupling domain.

The phase diagram of the model in Eq. (\ref{Heff}) has been
discussed in Ref. [\onlinecite{NlegRG}] in the context of a 2D array of
2-leg ladders. Precisely the same model also arises from a
treatment of a 2-leg ladder of spinless \textit{bosons} in Ref.
[\onlinecite{Orignac}]. The boson annihilation operators on the upper and
lower legs are represented as:
\begin{equation}
\Psi ^{U/L}\propto e^{-i\sqrt{\pi }\phi _{+\rho }^{U/L}}+\ldots ,
\end{equation}
and the boson density operators as:
\begin{equation}
\Psi ^{U/L\dagger }(x)\Psi ^{U/L}(x)\propto n_{b}+{\frac{1}{\sqrt{\pi }}}%
\partial _{x}\theta _{+\rho }^{U/L}+\hbox{constant}\times \{\exp [i2\pi
n_{b}x+i\sqrt{4\pi }\theta _{+\rho }^{U/L}]+h.c.\}+\ldots .
\end{equation}
Here $n_{b}$ is the density of bosons on each leg. Of course, it
is hardly surprising that this low energy Hamiltonian describes a
2-leg bosonic ladder; in our low energy approximation, the
fermionic degrees of freedom on each 2-leg ladder have been
discarded keeping only the spinless pairs, corresponding to
bosons. $t^{\prime }$ represents (single) boson hopping between
the chains and $V^{\prime }$ represents inter-chain boson
back-scattering. It follows from Eq. (\ref{dimt'V'}) that both
$t^{\prime }$ and $V^{\prime }$ are relevant for $1/4<K_{-}<1$
(and in general at least
one of them is relevant for all $K_{-}$). Thus one of $\phi _{-}$ and $%
\theta _{-}$ boson is always gapped, yielding a C1S0 phase. There
are two possible phases in which either $\phi _{-}$ or $\theta
_{-}$ is pinned [\onlinecite{Orignac}]. These two phases have evident
physical interpretations in the various underlying models from
which $H_{eff}$ arises. In the 2-leg boson model, the phase in
which $t^{\prime }$ is relevant and $\phi _{-}$ is pinned
corresponds to a standard 1D superfluid phase in which the boson
creation operator has a power-law decaying correlation function
but the term in the boson density operator oscillating at
wave-vector $2\pi n_{b}$ decays exponentially. On the other hand,
the phase in which $\theta _{-}$ is pinned corresponds to a boson
pairing phase. Now the boson creation operator has an
exponentially decaying correlation function. There is a
corresponding gap to create a single boson. The 2-boson creation
operator $\Psi (x)\Psi (x)$ has a power law decaying correlation
function as does the term in the boson density operator
oscillating at wave-vector $2\pi n_{b}$. In the 2D array of 2-leg
fermionic ladders, discussed in Ref. [\onlinecite{NlegRG}], the phase in
which $\phi _{-}$ is pinned is a conventional 2D superconducting
phase and the one phase in which $\theta _{-}$ is pinned is an
incommensurate charge density wave phase. (The power-law decay in
the single or double 2-leg ladder system is expected to become
true long range order in the 2D system.) The physical
interpretation of these phases in our model of 2 weakly coupled
2-leg (fermionic) ladders is now also clear. The phase in which
$\phi _{-}$ is pinned is a conventional pairing phase. Note that
the density of bosons (average number of bosons per site in the
2-leg bosonic ladder) should be identified with the density of
electrons (average number of electrons per site in the 4-leg
ladder). This follows since there are half as many sites per unit
length in the bosonic 2-leg ladder as in the fermionic 4-leg
ladder. Equivalently, we may identify the boson density with the
fermion density measured from half-filling
\begin{equation}
n_{b}=\delta =1-n.
\end{equation}
Thus we see that there are no density oscillations at $2\pi \delta
$ or equivalently $4\bar{k}_{F}$ in the pairing phase. The phase
in which $\theta _{-}$ is pinned is a bipairing phase with
stripes.

Which phase occurs depends on $K_{-}$ and also the relative size of $%
t^{\prime }$ and $V^{\prime }$. When $K_{-}$ is in the range
$1/4<K_{-}<1$, where both $t^{\prime }$ and $V^{\prime }$ are
relevant, we may estimate the phase boundary by the condition that
the corresponding energy scales, determined from Eq.
(\ref{dimt'V'}) are equal. Thus we expect the stripe phase to
occur, for this range of $K_{-}$ where:
\begin{equation}
V^{\prime }>(t^{\prime })^{[2-2K_{-}]/[2-1/(2K_{-})]}.
\end{equation}
Decreasing $K_{-}$ favors the stripe phase. Noting that we expect
$K_{-}$ to decrease from 1 as we dope away from half-filling, it
is natural to expect that, for large enough $V^{\prime }/t^{\prime
}$, the pairing phase will occur close to half-filling and the
stripe phase at larger doping, which is consistent with the DMRG
result [\onlinecite{White4leg}].

At this point it is appropriate to point out that much of the same
physics was discussed in two other earlier papers [\onlinecite{Troyer, Ledermann}]. In Ref. [\onlinecite{Troyer}] an effective 2-leg bosonic model was also discussed
as an approximation to the 4-leg fermionic ladder. In that case
the derivation was more heuristic than what appears here. A
somewhat longer range interaction was chosen in the 2-leg bosonic
model (up to separations of 3 lattice sites along the chain
direction) in order to partially match the numerical results on
the 4-leg fermionic ladder with those on the 2-leg bosonic ladder.
In particular, the Friedel oscillations were compared in the 2
models and shown to exhibit stripes in both models at higher
doping. The advantage of considering 2 nearly decoupled 2-leg
fermionic ladders ($t_{2,\perp }$, $V_{2,\perp }$ small) is that
we can make this mapping more rigorous. In Ref. [\onlinecite{Ledermann}]
the limit of very small $\delta $ is studied, for fixed $U$. It
was argued that, starting with half-filling, at small $\delta ,$ 2
of the bands remain in a gapped state with an average filling of
1/2 while the other band pair is doped. At higher doping, both
band pairs are doped. It was assumed that each of these doped band
pairs yields fermion pairs. At somewhat higher doping they argue
that these pairs form 4-hole clusters. Friedel oscillations were
not discussed. Although both of these papers discuss 4-hole
clusters, as does the earlier DMRG work of Ref. [\onlinecite{White4leg}],
none of them discuss the implications of such 4-hole clusters that
follow from 1D field theory considerations: exponentially decaying
pair correlations and a gap to add a single pair of holes to the
system. In particular, it seems likely that the effective 2-leg
bosonic ladder model studied in Ref. [\onlinecite{Troyer}] was in the
boson pairing phase, with a gap to add a single boson and
exponential decay of the boson creation operator correlation
function, but this point was not commented on. In this regard,
Figures (16) and (17) of Ref. [\onlinecite{Troyer}] are very interesting.
Fig. (16)
appears to be a plot of $E(N_{b})-E(N_{b}-1)$ versus $(N_{b}-1/2)/2L$ where $%
N_{b}$ is the number of bosons in the 2-leg bosonic ladder. In a
boson pairing phase, we expect:
\begin{equation}
E(N_{b})\to \mu N_{b}+{\frac{\Delta _{b}}{2}}(-1)^{N_{b}}+O(1/L),
\end{equation}
where $2\mu $ is the chemical potential for boson pairs and
$\Delta _{b}$ is the single boson gap. Thus:
\begin{equation}
E(N_{b})-E(N_{b}-1)\to \mu +(-1)^{N_{b}}\Delta _{b}+O(1/L).
\end{equation}
Both $\mu $ and $\Delta _{b}$ will evolve smoothly with density
but this zig-zag structure of $E(N_{b})-E(N_{b}-1)$ is the signal
of a boson gap, i.e. of boson pairing. Such a zig-zag is seen for
the last three points in Fig. (16), implying that
$E_{8}+E_{10}<2E_{9}$ (for $L=23$) and a corresponding boson gap
at $\delta \approx .2$ of $\Delta _{b}\approx .05t$. A zig-zag is
not seen in Fig. (16) at smaller $N_{b}$, despite the fact that
the change in Friedel oscillations to stripes appears to occur at
$\delta _{c}\approx .125$. Possibly this is because the boson gap
is too small
relative to the finite size gap to be observable for $\delta $ closer to $%
\delta _{c}$. Fig. (17) shows the analogous quantity for the 4-leg
fermionic ladder, $E(N_{h})-E(N_{h}-2)$ plotted versus
$(N_{h}-1)/(4L)$ for even $N_{h} $. In this case no clear zig-zag
is seen, which would indicate a gap to add a single fermionic
pair, up to $\delta =.2$. Possibly the problem is again that the
gap is too small relative to the finite size gap. This may
indicate that the heuristic mapping is not working in great detail
since the bosonic gap appears to be significantly larger than the
fermionic pair gap. Clearly more numerical work on both 2-leg
bosonic and 4-leg fermionic models would be interesting, either or
larger $L$ or for a different choice of interaction parameters, to
clarify whether or not a bosonic gap (and corresponding fermionic
pair gap) exists in the stripe phase, $\delta
>\delta _{c}$.

\section{Conclusion}

We have taken a number of different approaches to the 4-leg generalized
Hubbard ladder based on bosonization and RG. We gave general arguments about
possible phases based on possible ways of pinning bosons and the finite size
spectrum. We studied particular phases from solving the weak coupling RG
equations. We determined the phases which occur in the limit of two weakly
coupled 2-leg ladders, using the connection with a 2-leg bosonic ladder. All
of these approaches led to the same conclusion. It is not possible to find
any C1S0 phases that have both stripes and pairing. On the other hand, it
\textit{is} entirely possible to find phases in which stripes coexist with
bipairing. Whether or not 4 leg ladders, for physically reasonable and
numerically accessible ranges of parameters have such a phase remains an
open question. DMRG results have suggested a phase with stripes, but have,
so far, found no evidence for bipairing. We can see three resolutions of
this paradox.

\begin{itemize}
\item  {Our methods are based on certain approximations: either weak
coupling, or weakly coupled pairs of 2-leg ladders. It is entirely possible
that other phases may exist for these systems which are inaccessible to
these methods. Possibly these phases include ones with coexisting stripes
and pairing. We remark, however, that these field theory methods have been
remarkably successful in the past at describing many types of 1D strongly
correlated systems, including, for example, 2-leg ladders [\onlinecite{Ian2leg, 2leg}]. It would be an important discovery that they break down for
the 4-leg ladder.}

\item  {Possibly these systems \textit{do not} really have a stripe phase in
the sense that we are using. We have given a precise meaning to ``stripes''
in the limit of a very long 4-leg ladder. We mean Friedel oscillations at a
dominant wave-vector of $4\bar{k}_{F}=2\pi n$ where $n$ can be taken to be
the hole density. Existing DMRG results certainly suggest this but it is
possible that careful extrapolation to larger systems might not confirm this
result.}

\item  {Possibly the stripe phase apparently observed with DMRG \textit{is}
a bipairing phase. We remind the reader that we define bipairing precisely
to be a phase in which correlation functions of all pair operators decay
exponentially but correlation functions of some charge 4 operators exhibit
power law decay. Furthermore, such a phase has a gap to add one or two
particles, but no gap for four particles. The limited published DMRG results
have suggested that the decay of the pair correlation function may be power
law and have not seen a gap to add two particles. Possibly the correlation
length for the exponential decay is too large, and the corresponding gap to
add two particles too small, to be observed so far. Further DMRG
calculations could clarify this point. One could either study larger systems
or else change the parameters of the model in an attempt to make the
correlation length and inverse gap smaller. In this regard, numerical work
on 2-leg bosonic ladders would also be useful to confirm that, as expected
from field theory arguments, a boson pairing phase occurs in a wide range of
parameters. As has been emphasized before, it may be crucial to include long
range Coulomb interactions to understand stripe phases in real materials. }
\end{itemize}

We encourage further DMRG and analytical work to decide which of these
possibilities is correct. Confirming any of them would be an important
advance.

Assuming, for the moment, that 4 leg ladders \textit{do} exhibit stripes and
bipairing, we speculate on the implications for the 2-dimensional Hubbard
model. One might think that if 2-hole clusters form on 2-leg ladders and
4-hole clusters form on 4-leg ladders then perhaps, extrapolating to an
infinite number of legs would simply give an incommensurate charge density
wave. Such a state is perhaps not conducive to superconductivity. Stripes
have also been observed in 6 leg ladders {[\onlinecite{White6leg,
Jeckelmann6leg}]}. In this case, it appears that the number of holes per
rung is 4, rather than 6, as might have been expected. This is suggestive of
more exotic behavior than a simple CDW, closer to ideas about fluctuating 1D
conducting wires, that have been proposed for stripe phases in 2D.
Developing a field theory description of this stripe phase in 6-leg ladders
is an important open problem.

\begin{acknowledgements}
We are grateful to D.J. Scalapino and S.R. White for helpful discussions. We are indebted
to H.-H. Lin for informing us of the improved analysis of RG equations prior to his
publication. We also thank R. Pereira for the interesting discussion on $4k_{F}$ density operators and
S. Capponi for useful correspondences. This research is supported by NSERC and CIfAR.
\end{acknowledgements}

\appendix

\section{More Details of the Proof}

In this Appendix, we will demonstrate how to use the conditions Eq. (\ref
{IP1})-(\ref{IP5}) and Eq. (\ref{IPP1})-(\ref{IPP2}) to prove that the
correlation functions of pair operators Eq. (\ref{P1})-(\ref{P2}) and $4\bar{%
k}_{F}$ density operators Eq. (\ref{4KF1})-(\ref{4KF2}) can not both decay
with a power-law.

Before we start, please note that we adopt the usual convention for
set theory in mathematics. Two sets A and B are said to be equal, if they
have the same elements.

\subsection{Total Charge Mode Is Gapless}

In the following discussion in this subsection, we will
only consider the renormalization of $k_{Fi}$ such that the pinned $\theta
_{i\rho }^{\prime }$ bosons are orthogonal to $\Theta _{1\rho }$. In other
words, the conditions such as $2(k_{Fi}+k_{Fj})=2\pi $, won't occur but
other possibilities such as $k_{Fi}-k_{Fj}=0$ or $%
k_{F1}-k_{F2}+k_{F3}-k_{F4}=0$ are allowed. Then $\theta _{ij}^{\rho -}$ and
$(\theta _{1\rho }-\theta _{2\rho }+\theta _{3\rho }-\theta _{4\rho })/2$
may get pinned but they are still orthogonal to $\Theta _{1\rho }$. In this
case, the gapless charge mode will be $(\Theta _{1\rho },\Phi _{1\rho })$
since all pinned $\theta _{i\rho }^{\prime }$ or $\phi _{i\rho }^{\prime }$
fields are orthogonal to $\Theta _{1\rho }$ or $\Phi _{1\rho }$,
respectively. We can reorder the transformed basis, $\phi _{i\rho }^{\prime }
$, so that $\phi _{1\rho }^{\prime }=\Phi _{1\rho }$ and $\theta _{1\rho
}^{\prime }=\Theta _{1\rho }.$

Let's first consider whether the pair operator in Eq. (\ref{P1}) and $4k_{F}$
density operators in Eq. (\ref{4KF1}) or (\ref{4KF2}) could both have power
law decay. In order to satisfy Eq. (\ref{IP1}), $\vec{v}_{\rho \Delta }\cdot
\vec{u}_{\rho n}=-\pi $, the index $a$ in Eq. (\ref{P1}) must be chosen the
same as precisely one of the indices from the set $\{i,j,k,l\}$ in Eq. (\ref
{4KF1}) or (\ref{4KF2}). For example if $\{i,j,k,l\}=\{1,1,2,3\},$ then $a$
must be either 2 or 3. If such choice of $a$ exists (one counter example is $%
\{i,j,k,l\}=\{1,1,2,2\},$ $\vec{v}_{\rho \Delta }\cdot \vec{u}_{\rho n}\neq
-\pi $ for any $a$), however, it implies $\vec{u}_{\sigma \Delta }\cdot \vec{%
v}_{\sigma n}\neq 0$ between the term $\theta _{a\sigma }$ in Eq. (\ref{P1})
and $(\phi _{ij}^{\sigma -}\pm \phi _{kl}^{\sigma -})$ in Eq. (\ref{4KF1})
or (\ref{4KF2}). Thus, the condition Eq. (\ref{IP3}) is not satisfied and
the pair operator Eq. (\ref{P1}) can't coexist with any $4k_{F}$ density
operator of arbitrary $\{i,j,k,l\}$, including stripes.

Things are less trivial for the other type of pair operators Eq. (\ref{P2}).
The discussion depends on the situations of the indices set $\{i,j,k,l\}$.
Eq. (\ref{4KF1}) and (\ref{4KF2}) are reduced into different operator forms
depending on the different choices of the indices and the inner products
need to be discussed separately. For example, if $i=j$, $k=l$ and $i\neq k$,
then Eq. (\ref{4KF1}) and (\ref{4KF2}) contain neither charge nor spin $\phi
$ fields. One can easily find that Eq. (\ref{P2}) can not coexist with stripes. We will skip the details since they are similar to what we just showed above [\onlinecite{thesis}].

\subsection{Gapless Field Is Not the Total Charge Mode}

Now we should address the issue about the possibility of some pinned $\theta
_{i\rho }^{\prime }$ fields which are \textit{not} orthogonal to $\Theta
_{1\rho }$. In this case, $\Theta _{1\rho }$ can not be a basis field and
thus the gapless charge mode won't be $\Theta _{1\rho }$ anymore. As we
discussed at the beginning of this appendix, whether a $\theta _{i\rho
}^{\prime }$ field can be pinned or not depends on the renormalization of
Fermi momenta. Fermi momenta are important regarding both translational
symmetry and the interactions in the Hamiltonian. A certain $\theta _{i\rho
}^{\prime }$ field can be pinned if it's allowed by symmetry and there is a
relevant interaction involving it in the Hamiltonian. The oscillating
factors of four fermion interactions only involve the Fermi momentum
combinations like $\pm (k_{Fi}\pm k_{Fj}\pm k_{Fk}\pm k_{Fl})$ whereas any
arbitrary combination can be considered from symmetry's point of view.
Therefore, it will be much easier to discuss the possible pinned $\theta
_{i\rho }^{\prime }$ fields, not orthogonal to $\Theta _{1\rho }$, through
the possible new interactions containing $\theta _{i\rho }^{\prime }$ in the
Hamiltonian.

In order to check through all the possible pinning patterns efficiently, our
strategy is to start with one Fermi momentum renormalization condition so
that the new interaction involving a $\theta _{i\rho }^{\prime }$ field not
orthogonal to $\Theta _{1\rho }$ is present. Next, we will assume that
interaction is relevant and $\theta _{i\rho }^{\prime }$, as one of the
basis fields, is indeed pinned. We further assume the phase has pairing,
that is, either Eq. (\ref{P1}) or (\ref{P2}) has power-law decaying
correlations. So there will be two types of pairing to be discussed. Each
type of pairing, will require that some boson fields get pinned. If another $%
\theta _{i\rho }^{\prime }$ field needs to be pinned in order to have
pairing, it's allowed but one more Fermi momentum renormalization condition
must occur. For a phase with pairing, we will have a pinning pattern for
some boson fields but the rest of it will still be undetermined. Then the
question is if it's possible to have stripes by arbitrarily choosing the
pinning pattern for the rest of the bosons. To answer it, we have to check
the correlation function of all $4k_{F}$ density operators carefully since
some of them may correspond to stripes if additional Fermi momentum
renormalization conditions are satisfied. After this is done, we repeat this
procedure but start with a new $\theta _{i\rho }^{\prime }$ field not
orthogonal to $\Theta _{1\rho }$, that is, another Fermi momentum
renormalization condition. When all the possible Fermi momentum conditions
associated with the interactions are discussed, we will have completed the
proof.

Recall that a interaction can appear in the Hamiltonian if its oscillating
factor becomes a constant, that is $\pm (k_{Fi}\pm k_{Fj}\pm k_{Fk}\pm
k_{Fl})=0$ or $2\pi $, here $i,j,k$ and $l$ are arbitrary band indices (see
the oscillating factor in Eq. (C.2) in Appendix C). These interactions
contain the charge field like $\pm (\theta _{i\rho }\pm \theta _{j\rho }\pm
\theta _{k\rho }\pm \theta _{l\rho })$ after bosonized. We have to consider
the possible combination not orthogonal to the total charge mode $\Theta
_{1\rho }$. If $i,j,k$ and $l$ are all different, we only need to consider $%
k_{Fi}+k_{Fj}+k_{Fk}-k_{Fl}=0$ or $2\pi $ since $%
k_{Fi}+k_{Fj}+k_{Fk}+k_{Fl}= $ $2\pi n$ and $k_{Fi}+k_{Fj}-k_{Fk}-k_{Fl}$
doesn't result in the field not orthogonal to the total charge mode. If two
indices are the same, say $i=l$, we need to consider $2k_{Fi}+k_{Fj}+k_{Fk}=$
$2\pi $ and $2k_{Fi}+k_{Fj}-k_{Fk}=$ $0$ or $2\pi $ where $i,j$ and $k$ are
different. If three indices are the same, say $i=k=l$, we need to consider $%
3k_{Fi}+k_{Fj}= $ $2\pi $ and $3k_{Fi}-k_{Fj}=$ $0$ or $2\pi $ where $i\neq
j $. If two pairs of indices are the same, then we need to consider $%
2k_{Fi}+2k_{Fj}=$ $2\pi $ where $i\neq j$. When each of above
renormalization condition is satisfied, there will be a new interaction
containing a $\theta _{\rho }$ field not orthogonal to the total charge
mode. In each case, we also allow other Fermi momentum renormalization
conditions to occur so as to change the wave vector of density operators or
pin some other $\theta _{\rho }$ field orthogonal to the total charge mode.
We have to discuss the cases for $2k_{Fi}+2k_{Fj}=$ $2\pi $ and $%
2k_{Fi}+k_{Fj}+k_{Fk}=$ $2\pi $ in details. As for other conditions, there
will be a general argument.

In the following, we only show the details for the case when $%
2k_{Fi}+k_{Fj}+k_{Fk}=$ $2\pi $ and skip the case when $2k_{Fi}+2k_{Fj}=$ $%
2\pi $. For further details please see Ref. [\onlinecite{thesis}].
We consider all possible four fermions interactions to determine the
phases (or pinning patterns). We will show that pairing and stripes still
can't coexist even if the gapless mode is not the total charge field.

\subsubsection{New Interactions Due to $2k_{Fp}+k_{Fq}+k_{Fr}=2\pi$}

Now we consider different renormalization conditions on Fermi momenta, $%
2k_{Fp}+k_{Fq}+k_{Fr}=2\pi $, which can also result in a pinned charge
field not orthogonal to the total charge mode. Without loss of generality,
we choose $2k_{F1}+k_{F2}+k_{F3}=2\pi $. In this case, for example, the
interaction like $\psi _{R1\alpha }^{\dagger }\psi _{R2\overline{\alpha }%
}^{\dagger }\psi _{L1\alpha }\psi _{L3\overline{\alpha }}$ can be present in
the Hamiltonian because its oscillating factor becomes a constant. If this
new interaction is relevant, the charge field $(2\theta _{1\rho }+\theta
_{2\rho }+\theta _{3\rho })/\sqrt{6}$, not orthogonal to the total charge
mode, will get pinned. Assume it's really pinned and therefore $\Theta
_{1\rho }$ can't be an element of the pinning basis anymore. What's the
gapless mode in this case? We know that the field $\Theta _{1\rho }$ is
never pinned even though the total charge field is not an element of the
pinning basis. This implies that the combination like $\theta _{4\rho
}-\theta _{1\rho }$ is also unpinned. If it's pinned, combining with the
pinned field $(2\theta _{1\rho }+\theta _{2\rho }+\theta _{3\rho })/\sqrt{6}$%
, will imply that $\Theta _{1\rho }$ is pinned. One may think that the field
$(\theta _{4\rho }-\theta _{1\rho })/\sqrt{2}$ should be the gapless mode,
yet it's not orthogonal to $(2\theta _{1\rho }+\theta _{2\rho }+\theta
_{3\rho })/\sqrt{6}$, and can't be chosen as an element of the pinning
basis. So what gapless field can let $\Theta _{1\rho }$ and $(\theta _{4\rho
}-\theta _{1\rho })/\sqrt{2}$ both be unpinned? It could be either ($\theta
_{1\rho },\phi _{1\rho }$) or ($\theta _{4\rho },\phi _{4\rho }$). However,
any pinning basis field should be orthogonal to the pinning basis field $%
(2\theta _{1\rho }+\theta _{2\rho }+\theta _{3\rho })/\sqrt{6}$. Thus we
conclude that ($\theta _{4\rho },\phi _{4\rho }$) is the gapless mode in
this case. Besides $(2\theta _{1\rho }+\theta _{2\rho }+\theta _{3\rho })/%
\sqrt{6}$, the interaction $\psi _{R1\alpha }^{\dagger }\psi _{R2\overline{%
\alpha }}^{\dagger }\psi _{L1\alpha }\psi _{L3\overline{\alpha }}$ also
contains another charge boson field $\phi _{23}^{\rho -}$. Then we should
assume that $\phi _{23}^{\rho -}$ is also pinned by this relevant
interaction. We find that orthogonal matrix representing the basis in the
charge channel should be \newline
\begin{equation}
R_{\rho }=\left(
\begin{array}{llll}
0 & 0 & 0 & 1 \\
2/\sqrt{6} & 1/\sqrt{6} & 1/\sqrt{6} & 0 \\
0 & 1/\sqrt{2} & -1/\sqrt{2} & 0 \\
1/\sqrt{3} & -1/\sqrt{3} & -1/\sqrt{3} & 0
\end{array}
\right) .\newline
\end{equation}

With the condition that ($\theta _{4\rho },\phi _{4\rho }$) is gapless and $%
(2\theta _{1\rho }+\theta _{2\rho }+\theta _{3\rho })/\sqrt{6}$ and $\phi
_{23}^{\rho -}$ are pinned, the correlation functions of the pair operators
Eq. (\ref{P2}) decay exponentially. The only possible pairing operator that
might have power-law correlations correspond to $a=4$ in Eq. (\ref{P1}). In
addition we must assume that $\theta _{4\sigma }$ is also pinned. Now we
should check the $4k_{F}$ density operators. We have to consider four
situations for the indices in Eq. (\ref{4KF1}) and (\ref{4KF2}), where the
density operators reduce to different forms. These four situations are: (1)
all the indices are different; (2) two indices are the same but different
from the other two; (3) two pairs of indices are the same; (4) three indices
are the same but different from the last one.

For case (1): When $i,j,k$ and $l$ are all different in Eq. (\ref{4KF1}) and
(\ref{4KF2}), Eq. (\ref{P1}) can not coexist with both Eq. (\ref{4KF1}) and (%
\ref{4KF2}) since $\vec{u}_{\sigma \Delta }\cdot \vec{v}_{\sigma n}\neq 0$
between the term $\theta _{a\sigma }$ in Eq. (\ref{P1}) and $(\phi
_{ij}^{\sigma -}\pm \phi _{kl}^{\sigma -})$ in Eq. (\ref{4KF1}) or (\ref
{4KF2}).

For case (2): We assume only two indices are the same in Eq. (\ref{4KF1})
and (\ref{4KF2}). Then Eq. (\ref{4KF1}) and (\ref{4KF2}) will reduce to the
operators with three band indices. We just have to work out all the operator
forms when two out of four indices are the same in Eq. (\ref{4KF1}) and (\ref
{4KF2}). The resultant operators have the general forms:
\begin{eqnarray}
&&e^{-i\sqrt{\pi }[(\sqrt{2}\theta _{i\rho }+\theta _{jk}^{\rho +})+(\phi
_{jk}^{\rho -})+(\sqrt{2}\theta _{i\sigma }\pm \theta _{jk}^{\sigma
+})+(\phi _{jk}^{\sigma -})]},  \label{three4KF1} \\
&&e^{-i\sqrt{\pi }[(\sqrt{2}\theta _{i\rho }+\theta _{jk}^{\rho +})+(\phi
_{jk}^{\rho -})+(\theta _{jk}^{\sigma -})+(\sqrt{2}\phi _{i\sigma }-\phi
_{jk}^{\sigma +})]},  \label{three4KF2} \\
&&e^{-i\sqrt{\pi }[(\sqrt{2}\theta _{i\rho }+\theta _{jk}^{\rho +})+(\sqrt{2}%
\phi _{i\rho }-\phi _{jk}^{\rho +})+(\sqrt{2}\theta _{i\sigma }+\theta
_{jk}^{\sigma +})+(\sqrt{2}\phi _{i\sigma }-\phi _{jk}^{\sigma +})]},
\label{three4KF3} \\
&&e^{-i\sqrt{\pi }[(\sqrt{2}\theta _{i\rho }+\theta _{jk}^{\rho +})+(\sqrt{2}%
\phi _{i\rho }-\phi _{jk}^{\rho +})+(\theta _{jk}^{\sigma -})+(\phi
_{jk}^{\sigma -})]},  \label{three4KF4}
\end{eqnarray}
where $i,j$ and $k$ are all different. In fact, Eq. (\ref{three4KF1})-(\ref
{three4KF4}) don't cover all the possible operators forms. However, these
other cases correspond to simply changing the signs in front of last three
parentheses in Eq. (\ref{three4KF1})-(\ref{three4KF4}). We only care about
whether the inner products between coefficient vectors are zero or not.
Therefore, we can ignore the signs.

The correlation functions of Eq. (\ref{three4KF3}) and (\ref{three4KF4})
will decay exponentially because $(2\theta _{1\rho }+\theta _{2\rho }+\theta
_{3\rho })/\sqrt{6}$ is pinned. Due to the $\phi _{jk}^{\rho -}$ term in Eq.
(\ref{three4KF1}) and (\ref{three4KF2}), we can only choose $\{j,k\}=\{2,3\}$%
. The index $i$ could be either 1 or 4. For $i=4$, the correlation functions
of Eq. (\ref{three4KF1}) and (\ref{three4KF2}) could decay with a power-law
if and only if $(\theta _{1\rho }-\theta _{2\rho }-\theta _{3\rho })/\sqrt{3}
$ is also pinned, which requires an extra condition on the Fermi momenta but
it's possible. The point here is that the corresponding wave-vector is $%
k_{F2}+k_{F3}+2k_{F4}$ and it only corresponds to $2\pi n$ if $k_{F1}=k_{F4}$%
. However, we started with the condition $2k_{F1}+k_{F2}+k_{F3}=2\pi $,
which implies $k_{F1}-k_{F4}=2\pi -2\pi n\neq 0$. Therefore, these are no
stripes in that case. For $i=1$, Eq. (\ref{three4KF1}) and (\ref{three4KF2})
correspond to the wave-vector $2k_{F1}+k_{F2}+k_{F3}=2\pi $, which doesn't
correspond to that of stripes, either.

For case (3):Next we discuss $4k_{F}$ density operators where two pairs of
indices are the same in Eq. (\ref{4KF1}) and (\ref{4KF2}). Recall that we
are considering the pinning pattern in which ($\theta _{34}^{\rho +},\phi
_{34}^{\rho +}$) is gapless and $\theta _{12}^{\rho +}$,$\phi _{34}^{\rho -}$
and $\theta _{3\sigma }$ (or $\theta _{4\sigma }$) are pinned. The $4k_{F}$
density operators reduce to the forms:
\begin{eqnarray}
&&e^{-i\sqrt{4\pi }(\theta _{ij}^{\rho +}\pm \theta _{ij}^{\sigma \pm })},
\label{two4KF1} \\
&&e^{-i\sqrt{4\pi }(\theta _{ij}^{\rho +}\pm \phi _{ij}^{\sigma -})},
\label{two4KF2} \\
&&e^{-i\sqrt{4\pi }(\theta _{ij}^{\rho +}\pm \phi _{ij}^{\rho -}\pm \theta
_{ij}^{\sigma +}\pm \phi _{ij}^{\sigma -})},  \label{two4KF3} \\
&&e^{-i\sqrt{4\pi }(\theta _{ij}^{\rho +}\pm \phi _{ij}^{\rho -})},
\label{two4KF4}
\end{eqnarray}
where $i\neq j$. In order to have power-law decaying correlation functions,
these operators should contain the gapless mode ($\theta _{4\rho },\phi
_{4\rho }$), that is, one of $\{i,j\}$ should be $4$. Then Eq. (\ref{two4KF2}%
), (\ref{two4KF3}) and (\ref{two4KF4}) will have exponentially decaying
correlations since they contain the dual of the pinned boson, $(2\theta
_{1\rho }+\theta _{2\rho }+\theta _{3\rho })/\sqrt{6}$ or $\theta _{4\sigma }
$. The correlation function of Eq. (\ref{two4KF1}) with $\{i,j\}=\{1,4\}$
can decay with a power-law if $(\theta _{1\rho }-\theta _{2\rho }-\theta
_{3\rho })/\sqrt{3}$ is pinned. So what's the interaction we need to pin the
field $(\theta _{1\rho }-\theta _{2\rho }-\theta _{3\rho })/\sqrt{3}$? We
started with the interaction $\psi _{R1\alpha }^{\dagger }\psi _{R2\overline{%
\alpha }}^{\dagger }\psi _{L1\alpha }\psi _{L3\overline{\alpha }}$ and $%
(2\theta _{1\rho }+\theta _{2\rho }+\theta _{3\rho })/\sqrt{6}$ and $\phi
_{23}^{\rho -}$ are pinned. Is it possible that $(2\theta _{1\rho }+\theta
_{2\rho }+\theta _{3\rho })/\sqrt{6}$ ,$\phi _{23}^{\rho -}$ and $(\theta
_{1\rho }-\theta _{2\rho }-\theta _{3\rho })/\sqrt{3}$ are all pinned? It's
possible if there is another relevant interaction involving the the fields $%
(\theta _{1\rho }-\theta _{2\rho }-\theta _{3\rho })/\sqrt{3}$ but not
involving $\phi _{23}^{\rho -}$. This means that the $\theta $ field in that
interaction must be orthogonal to $\phi _{23}^{\rho -}$ but not orthogonal
to $(\theta _{1\rho }-\theta _{2\rho }-\theta _{3\rho })/\sqrt{3}$. The $%
\theta $ field in that interaction also has to be orthogonal to the gapless
mode ($\theta _{4\rho },\phi _{4\rho }$). Then the interaction containing $%
\theta _{23}^{\rho +}$, such as $\psi _{R2\alpha }^{\dagger }\psi _{R2%
\overline{\alpha }}^{\dagger }\psi _{L3\overline{\alpha }}\psi _{L3\alpha }$%
, is what we need. This interaction can appear in the Hamiltonian if the
condition $k_{F2}+k_{F3}=\pi $ is satisfied. However, the necessary
condition that the $4k_{F}$ density operator with $\{i,j\}=\{1,4\}$ can
correspond to stripes is $\{k_{F1},k_{F4}\}=\{k_{F2},k_{F3}\}$. This will
lead to the contradiction that $k_{F1}+k_{F2}+k_{F3}+k_{F4}=2\pi $ $\neq $ $%
2\pi n$. For the choices $\{i,j\}\neq \{1,4\}$ in Eq. (\ref{two4KF1}), the
correlation functions of Eq. (\ref{two4KF1}) will decay exponentially since $%
\phi _{23}^{\rho -}$ is pinned.

For case (4): When three indices are the same in Eq. (\ref{4KF1}) and (\ref
{4KF2}), their correlation functions will decay exponentially since the
condition $\vec{u}_{\sigma n}\cdot \vec{v}_{\sigma n}=0$ is never satisfied
for those operators. This is in general true and independent of the pinning
patterns and we don't have to consider this situation for $4k_{F}$ density
operators.

In this subsection, we have shown that pairing and stripes can't coexist
when the gapless mode is ($\theta _{4\rho },\phi _{4\rho }$) and at least
one Fermi momentum condition $2k_{F1}+k_{F2}+k_{F3}=2\pi $ is satisfied. We
will skip the details for the case when there are new interactions due to $%
k_{Fp}+k_{Fq}=\pi $ since they are similar to the case we just showed.

\subsubsection{New Interactions Due to Other Conditions}

There are four more types of Fermi momentum renormalization conditions that
can lead to the presence of new interactions that may pin the boson
orthogonal to $\Theta _{1\rho }$. They are:
\begin{eqnarray}
3k_{Fi}+k_{Fj} &=&2\pi ,  \label{cond1} \\
2k_{Fi}-k_{Fj}+k_{Fk} &=&0\mathrm{\ or\ }2\pi ,  \label{cond2} \\
3k_{Fi}-k_{Fj} &=&0\mathrm{\ or\ }2\pi ,  \label{cond3} \\
k_{Fi}+k_{Fj}+k_{Fk}-k_{Fl} &=&0\mathrm{\ or\ }2\pi ,  \label{cond4}
\end{eqnarray}
where the indices $i,j,k$ and $l$ are all different. Although these
conditions will make some interactions non-oscillating, they can't be
relevant since they will inevitably contain some charge boson and its dual
at the same time, in other words, $\vec{u}_{\rho n}\cdot \vec{v}_{\rho
n}\neq 0$ for these interactions.

This can be seen easily. According to our convention to define the boson
fields, we know that the relation between chiral fermions and charge bosons
are:
\begin{eqnarray}
\varphi _{Ri\alpha } &=&\frac{1}{2\sqrt{2}}(\theta _{i\rho }+\phi _{i\rho
}+\cdots ),  \label{chiR} \\
\varphi _{Li\alpha } &=&\frac{1}{2\sqrt{2}}(\theta _{i\rho }-\phi _{i\rho
}+\cdots ).  \label{chiL}
\end{eqnarray}
When Eq. (\ref{cond1}) is satisfied, the interaction like $\psi _{Ri\alpha
}^{\dagger }\psi _{Li\alpha }\psi _{Ri\overline{\alpha }}^{\dagger }\psi _{Lj%
\overline{\alpha }}$ (or $i$ and $j$ exchanged) can appear in the
Hamiltonian. However, this interaction contains $(-3\theta _{i\rho }+\theta
_{j\rho })$ and $(-\phi _{i\rho }+\phi _{j\rho })$ which are not orthogonal
to each other. Thus, it can not be relevant since a field and its dual can't
be both pinned.

As for Eq. (\ref{cond2}), (\ref{cond3}) and (\ref{cond4}), the corresponding
interactions will have unequal numbers of right and left chiral fermions.
For right moving fermions $\psi _{Ri\alpha }$, is associated with the Fermi
momentum $+k_{Fi}$ while $-k_{Fi}$ is associated with the left moving
fermions $\psi _{Li\alpha }$. Eq. (\ref{cond2}), (\ref{cond3}) and (\ref
{cond4}) are composed of three positive and one negative Fermi momentum.
Therefore, the corresponding four fermion interactions with zero charge must
contain three right and one left (or three left and one right) moving
fermion fields. These interactions will inevitably result in $\vec{u}_{\rho
n}\cdot \vec{v}_{\rho n}\neq 0$ and as a consequence they can't be relevant.

\subsubsection{More Than One New Interaction}

In the above discussion, although more than one Fermi momentum
renomalization condition is allowed to occur, we always implicitly assume
that the new interaction associated with the first Fermi momentum
renomalization condition is the most relevant one. This assumption is
related to how we determine the basis field. We always choose the basis
fields guided by the interactions. For example, if the interaction $\psi
_{R1\alpha }^{\dagger }\psi _{R1\overline{\alpha }}^{\dagger }\psi _{L2%
\overline{\alpha }}\psi _{L2\alpha }$ is relevant, then $\theta _{12}^{\rho
+}$ will be pinned. Whether $\theta _{12}^{\rho +}$ is an element of the
pinning basis is the issue here. If $\theta _{12}^{\rho +}$ is not a basis
field, $\theta _{12}^{\rho +}$ should be expressed in terms of the
combination of the pinning basis fields. If the interaction $\psi _{R1\alpha
}^{\dagger }\psi _{R1\overline{\alpha }}^{\dagger }\psi _{L2\overline{\alpha
}}\psi _{L2\alpha }$ is the most relevant interaction, there is no reason to
write $\theta _{12}^{\rho +}$ in terms of other fields and we will choose $%
\theta _{12}^{\rho +}$ as an element of the pinning basis fields. However,
what if the interactions such as $\psi _{R1\alpha }^{\dagger }\psi _{R1%
\overline{\alpha }}^{\dagger }\psi _{L2\overline{\alpha }}\psi _{L2\alpha }$
and $\psi _{R1\alpha }^{\dagger }\psi _{R1\overline{\alpha }}^{\dagger }\psi
_{L3\overline{\alpha }}\psi _{L3\alpha }$ both appear in the Hamiltonian and
are equally relevant? We know that $\theta _{12}^{\rho +}$ and $\theta
_{13}^{\rho +}$ should be pinned but how do we choose the pinning basis?
First of all, since $\Theta _{1\rho }$ is unpinned, then $\theta _{34}^{\rho
+}$ and $\theta _{24}^{\rho +}$ are also unpinned. Then the gapless mode is (%
$\theta _{4\rho },\phi _{4\rho }$) in this case. Now we can obtain the other
three pinning basis fields by applying an orthogonal transformation to the
band boson fields $\theta _{1\rho },$ $\theta _{2\rho }$ and $\theta _{3\rho
}$. Rewrite $\theta _{12}^{\rho +}$ and $\theta _{13}^{\rho +}$ in terms of
the pinning basis fields and see which pinning basis fields should be
pinned. However, there are three basis fields but only two constraints ($%
\theta _{12}^{\rho +}$ and $\theta _{13}^{\rho +}$ are pinned). To get a
C1S0 phase, we need another interaction. Is it possible to pin a $\phi $
field in this case? If such charge $\phi $ field exists, it has to be
orthogonal to $\theta _{12}^{\rho +}$ and $\theta _{13}^{\rho +}$. Due to
the charge conservation, any pinned charge $\phi $ field also has to be
orthogonal to $\Phi _{1\rho }$. This is impossible for the $\phi $ field
obtained from the linear combination of $\phi _{1\rho },\phi _{2\rho }$ and $%
\phi _{3\rho }$. So the third pinned charge field is also a $\theta $ field.

Now comes the question: Is it possible that pairing and stripes can coexist
under the condition when ($\theta _{4\rho },\phi _{4\rho }$) is gapless and
all three pinned charge boson are $\theta $ fields? The charge fields in the
$4k_{F}$ density operators Eq. (\ref{two4KF1})-(\ref{two4KF2}) are $\theta $
fields. The pair operator Eq. (\ref{P1}) with the choice $a=4$, has no other
$\theta $ dependent charge fields since the gapless mode $\phi _{4\rho }$ is
the only charge field in Eq. (\ref{P1}). Then the correlation functions of
Eq. (\ref{two4KF1})-(\ref{two4KF2}) and Eq. (\ref{P1}) may decay with a
power-law at the same time. Indeed, this is what happens. However, in order
to pin three $\theta $ fields, three conditions of Fermi momenta regarding
the band indices $\{1,2,3\}$, such as $k_{F1}+k_{F2}=\pi $, $%
k_{F1}+k_{F3}=\pi $ and $k_{F2}=k_{F3}$, or $k_{F1}+k_{F2}=\pi $, $%
k_{F1}=k_{F3}$ and $k_{F2}=k_{F3}$ etc, need to be satisfied. Recall that $4%
\overline{k}_{F}=2\pi n$ is always satisfied. It also needs one additional
Fermi momentum renormalization condition so that the wave-vectors of Eq. (%
\ref{two4KF1}) and (\ref{two4KF2}) correspond to $4\overline{k}_{F}$. There
are five equations to solve four unknown Fermi momenta. We find there is no
solution for all the possible cases. Thus pairing and stripes can't coexist
under these situations.

\section{$4k_{F}$ density operators}

The higher order components of density operator is known in 1D [%
\onlinecite{Haldane}] but it's not clear what's the generalization in ladder
systems. Here we will derive the higher Fourier modes of density
operators through the process of integrating out the large momentum modes in
the perturbative fashion. The density operator on the $a^{\hbox{th}}$ leg
can be written:
\begin{equation}
n_{a}(x)=\sum_{\alpha }c_{a,\alpha }^{\dagger }c_{a,\alpha }=\sum_{\alpha
,i,j}S_{ai}S_{aj}\psi _{i,\alpha }^{\dagger }\psi _{j,\alpha }.
\end{equation}
Using Eq. (\ref{chiral}), we decompose $n_{a}(x)$ into components that
oscillate with various phase factors $k_{Fi}\pm k_{Fj}$. We refer
generically to all components that oscillate with phases $\pm (k_{Fi}+k_{{Fj}%
})$ as ``$2k_{F}$'' terms. Naively, these appear to be all components of the
density operators. However, there are actually additional $4k_{F}$ (and
higher) components. These arise from considering more carefully the RG
transformation which leads to the low energy effective Hamiltonian. This
transformation corresponds to integrating out, within the Feynman path
integral, the ``fast modes'' of the fermion fields; i.e. all Fourier modes
except for narrow bands, of width $\Lambda $, near each Fermi point, $\pm
k_{Fi}$. We consider in detail how this produces $4k_{F}$ terms in $n_{a}(x)$%
, in lowest order in the Hubbard interaction, $U$. Consider calculating some
Green's function involving $n_{a}(x)$, or $<n_{a}(x)>$ with open boundary
conditions. Expanding the exponential of the action, to first order in $U$,
inside the path integral effectively adds an extra term to $n_{a}(x)$:
\begin{equation}
n_{a}(x,\tau )\to n_{a}(x,\tau )[1+U\int d\tau ^{\prime }\sum_{x^{\prime
}=1}^{L}\sum_{b=1}^{4}n_{b,\uparrow }(x^{\prime },\tau ^{\prime
})n_{b,\downarrow }(x^{\prime },\tau ^{\prime })].
\end{equation}
We now expand the second term, of $O(U)$ in band fermions using:
\begin{equation}
n_{a}(x,\tau )=\sum_{i,j,\alpha ,p,q}S_{ai}S_{aj}e^{iqx}\psi _{i\alpha
}^{\dagger }(p)\psi _{j\alpha }(q+p),
\end{equation}
\begin{equation}
\sum_{x^{\prime }=1}^{L}\sum_{b=1}^{4}n_{b,\uparrow }(x^{\prime
})n_{b,\downarrow }(x^{\prime
})=%
\sum_{b,i_{1},i_{2},i_{3},i_{4},p_{1},p_{2},p_{3}}C_{i_{1},i_{2},i_{3},i_{4}}\psi _{i_{1}\uparrow }^{\dagger }(p_{1}-p_{2}+p_{3})\psi _{i_{2}\uparrow }(p_{1})\psi _{i_{3},\downarrow }^{\dagger }(p_{2})\psi _{i_{4},\downarrow }(p_{3}).
\end{equation}
Here:
\begin{equation}
C_{i_{1},i_{2},i_{3},i_{4}}\equiv S_{ai_{1}}S_{ai_{2}}S_{ai_{3}}S_{ai_{4}},
\end{equation}
and we have suppressed the imaginary time labels $\tau $, $\tau ^{\prime }$
which are not too important. Each fermion field, $\psi _{i\alpha }(p)$, may
either be a slow mode with $|p-k_{Fi}|<\Lambda $ or $|p+k_{Fi}|<\Lambda $ or
it may be a fast mode with $|p\pm k_{Fi}|>\Lambda $. Doing the functional
integral over the fast modes eliminates some of the fermion fields from the
correction, $\delta n_{a}(x)$, to the density operator, $n_{a}(x)$,
replacing them by their expectation value. To generate $4k_{F}$ terms in $%
n_{a}(x)$ we take the case where four of the six fields in $n_{a}H_{int}$
are slow modes and two of them are fast modes. For instance consider the
case where:
\begin{eqnarray}
p &=&-k_{F1}+\tilde{p},\ \ i=1  \nonumber \\
q &=&k_{F1}+k_{F2}+k_{F3}+k_{F4}+\tilde{q}  \nonumber \\
p_{1} &=&k_{F4}+\tilde{p}_{1},\ \ i_{1}=4  \nonumber \\
p_{2} &=&-k_{F2}+\tilde{p}_{2},\ \ i_{2}=2  \nonumber \\
p_{3} &=&k_{F3}+\tilde{p}_{3},\ \ i_{3}=3  \label{kshifts}
\end{eqnarray}
where all the $\tilde{p}$ and $\tilde{p}_{i}$ obey $|\tilde{p}|<\Lambda $
and $\tilde{q}$ is also small, of $O(\Lambda )$. Then four of the fields are
slow modes but $\psi _{j\alpha }(p+q)$ and $\psi _{i_{1}\uparrow }^{\dagger
}(p_{1}-p_{2}+p_{3})$ are fast modes. Note that we have chosen the band
index to correspond to the momentum range for all slow modes. This would be
necessary if we assume that the momentum range $\Lambda $ around each Fermi
momentum is smaller than the difference of Fermi momenta between different
bands. The product of fast mode fields gets replaced by its expectation
value during the RG transformation:
\begin{equation}
<\psi _{j\alpha }(p+q)\psi _{i_{1}\uparrow }^{\dagger
}(p_{1}-p_{2}+p_{3})>\propto \delta _{ji_{1}}\delta _{\alpha \uparrow
}\delta (p+q-p_{1}+p_{2}-p_{3}).  \label{deltas}
\end{equation}
From Eq. (\ref{kshifts}) we see that the last $\delta $-function in Eq. (\ref
{deltas}) can be written:
\begin{equation}
\delta (p+q-p_{1}+p_{2}-p_{3})=\delta (\tilde{p}+\tilde{q}-\tilde{p}_{1}+%
\tilde{p}_{2}-\tilde{p}_{3}).
\end{equation}
Thus the extra term in the density operator can be written schematically as:
\begin{eqnarray}
\delta n_{a}(x) &\propto &U\exp [i(k_{F1}+k_{F2}+k_{F3}+k_{F4})x]\sum_{j,%
\tilde{p},\tilde{p}_{1},\tilde{p}_{2},\tilde{p}_{3}}S_{a1}S_{aj}C_{j423}\exp
[i(-\tilde{p}+\tilde{p}_{1}-\tilde{p}_{2}+\tilde{p}_{3})x]  \nonumber \\
&&\psi _{1\uparrow }^{\dagger }(-k_{F1}+\tilde{p})\psi _{4\uparrow }(k_{F4}+%
\tilde{p}_{1})\psi _{2\downarrow }^{\dagger }(-k_{F2}+\tilde{p}_{2})\psi
_{3\downarrow }(k_{F3}+\tilde{p}_{3})  \nonumber \\
&=&U\exp [i(k_{F1}+k_{F2}+k_{F3}+k_{F4})x]S_{a1}S_{a2}S_{a3}S_{a4}\psi
_{L1\uparrow }^{\dagger }(x)\psi _{R4\uparrow }(x)\psi _{L2\downarrow
}^{\dagger }(x)\psi _{R3\downarrow }(x).
\end{eqnarray}
Naturally, a large number of other such $4k_{F}$ terms are generated by
choosing other momentum ranges for the slow and fast modes. It turns out all
the terms allowed by the symmetry will be generated in the low energy
continuum limit, which is what we expected.

\section{Initial Values For RG equations}

Here we explicitly give the bare coupling values in Eq. (\ref{Hintcf}) in
terms of the interactions in Eq.(\ref{H0}) and (\ref{Hint}) for general
doped $N$-leg ladders. By using $\overrightarrow{\sigma }_{\alpha \beta
}\cdot \overrightarrow{\sigma }_{\gamma \delta }=2\delta _{\alpha \delta
}\delta _{\beta \gamma }-\delta _{\alpha \beta }\delta _{\gamma \delta }$,
Eq. (\ref{Hintcf}) can be written as
\begin{eqnarray}
&&H_{int}=\sum_{\alpha \beta }\sum_{ij}\sum_{x}[\frac{1}{4}(\tilde{c}%
_{ij}^{\rho }+\tilde{c}_{ij}^{\sigma })\psi _{Ri\alpha }^{\dagger }\psi
_{Rj\alpha }\psi _{Li\beta }^{\dagger }\psi _{Lj\beta }-\frac{1}{2}\tilde{c}%
_{ij}^{\sigma }\psi _{Ri\alpha }^{\dagger }\psi _{Rj\beta }\psi _{Li\beta
}^{\dagger }\psi _{Lj\alpha }  \nonumber \\
&&+\frac{1}{4}(\tilde{f}_{ij}^{\rho }+\tilde{f}_{ij}^{\sigma })\psi
_{Ri\alpha }^{\dagger }\psi _{Ri\alpha }\psi _{Lj\beta }^{\dagger }\psi
_{Lj\beta }-\frac{1}{2}\tilde{f}_{ij}^{\sigma }\psi _{Ri\alpha }^{\dagger
}\psi _{Ri\beta }\psi _{Lj\beta }^{\dagger }\psi _{Lj\alpha }],
\label{int1CF}
\end{eqnarray}
where $i$ and $j$ are running from 1 to $N$. Note that we have a factor of
1/2 difference from the definition of operator $J_{ij}$ in Ref. [%
\onlinecite{NlegRG}]. Expand Eq. (\ref{H0}) and (\ref{Hint}) in terms of
the chiral fermions Eq. (\ref{chiral}) and we have
\begin{eqnarray}
H_{U,V} &=&\sum_{\alpha \beta }\sum_{ijkl}\sum_{x}\sum_{P_{i}}[\frac{V}{2}%
A_{ijkl}e^{(-iP_{3}k_{F_{k}}+iP_{4}k_{F_{l}})}+\frac{V_{\perp }}{2}%
(B_{ijkl}^{1}+B_{ijkl}^{2}+B_{ijkl}^{3})]\times  \nonumber \\
&&e^{i(-P_{1}k_{F_{i}}+P_{2}k_{F_{j}}-P_{3}k_{F_{k}}+P_{4}k_{F_{l}})x}\psi
_{P_{1}i\alpha }^{\dagger }\psi _{P_{2}j\alpha }\psi _{P_{3}k\beta
}^{\dagger }\psi _{P_{4}l\beta }  \nonumber \\
&&+\sum_{ijkl}\sum_{x}%
\sum_{P_{i}}A_{ijkl}Ue^{i(-P_{1}k_{F_{i}}+P_{2}k_{F_{j}}-P_{3}k_{F_{k}}+P_{4}k_{F_{l}})x}\psi _{P_{1}i\uparrow }^{\dagger }\psi _{P_{2}j\uparrow }\psi _{P_{3}k\downarrow }^{\dagger }\psi _{P_{4}l\downarrow },
\label{UVCF}
\end{eqnarray}
where $P_{i}=\pm $ for $R/L$ fermions and with the $S_{jm}$ in Eq. (\ref
{bandFermion})
\begin{eqnarray*}
A_{ijkl} &=&\sum_{m=1}^{N}S_{im}^{*}S_{jm}S_{km}^{*}S_{lm}, \\
B_{ijkl}^{m} &=&S_{mi}^{*}S_{mj}S_{m+1,k}^{*}S_{m+1,l}.
\end{eqnarray*}
The $1/2$ factor for $V$ and $V_{\perp }$ in Eq. (\ref{Hint}) will make them
in the equal footing as $U$. Now we just have to compare the coefficients in
Eq. (\ref{Hintcf}) and (\ref{UVCF}) for the same interaction then we can
obtain the bare initial values of the RG interactions. Recall that $\tilde{f}%
_{ij}=\tilde{f}_{ji}$, $\tilde{c}_{ij}=\tilde{c}_{ji}$ and $\tilde{f}_{ii}=0$%
. Following the convention in Ref. [\onlinecite{NlegRG}], the RG equations
are written down for $\tilde{c}_{ii},$ $\tilde{c}_{ij}$ and $\tilde{f}_{ij}$
where $i<j$. It will be convenient to define the following quantity for OBC:
\[
S_{ijkl}=\sum_{m=1}^{N-1}B_{ijkl}^{m}.
\]

In this basis, we write down the general form for the initial values in the
RG equations:
\begin{eqnarray}
\tilde{c}_{ii}^{\rho } &=&2[(2-\cos 2k_{F_{i}})VA_{iiii}+V_{\perp
}S_{iiii}+UA_{iiii}],  \label{Arho} \\
\tilde{c}_{ii}^{\sigma } &=&2(VA_{ijij}\cos 2k_{F_{i}}+V_{\perp
}S_{iiii}+UA_{iiii}), \\
\tilde{c}_{ij}^{\rho } &=&4\{VA_{ijij}[2\cos (k_{F_{i}}-k_{F_{j}})-\cos
(k_{F_{i}}+k_{F_{j}})]+V_{\perp }S_{ijij}+UA_{ijij}\}, \\
\tilde{c}_{ij}^{\sigma } &=&4[VA_{ijij}\cos (k_{F_{i}}+k_{F_{j}})+V_{\perp
}S_{ijij}+UA_{ijij}], \\
\tilde{f}_{ij}^{\rho } &=&4\{VA_{ijij}[2-\cos
(k_{F_{i}}+k_{F_{j}})]+V_{\perp }(2S_{iijj}-S_{ijij})+UA_{ijij}\}, \\
\tilde{f}_{ij}^{\sigma } &=&4[VA_{ijij}\cos (k_{F_{i}}+k_{F_{j}})+V_{\perp
}S_{iijj}+UA_{ijij}].  \label{Asigma}
\end{eqnarray}
where $i<j$ here. Eq. (\ref{Arho})-(\ref{Asigma}) are not the basis so that
the RG potential exists. In practice, we always deal with the RG equations
in the potential basis. Therefore, we rescale Eq. (\ref{Arho})-(\ref{Asigma}%
) into the RG potential basis (without tilde) by
\begin{eqnarray}
\tilde{c}_{ii}^{\rho } &=&4\sqrt{2}(2\pi v_{i})c_{ii}^{\rho },
\label{gtotildeg1} \\
\tilde{c}_{ii}^{\sigma } &=&4\sqrt{\frac{2}{3}}(2\pi v_{i})c_{ii}^{\sigma },
\label{gtotildeg2} \\
\tilde{a}_{ij}^{\rho } &=&4\sqrt{v_{i}v_{j}}(2\pi )a_{ij}^{\rho },
\label{gtotildeg3} \\
\tilde{a}_{ij}^{\sigma } &=&\frac{4}{\sqrt{3}}\sqrt{v_{i}v_{j}}(2\pi
)a_{ij}^{\sigma },  \label{gtotildeg4}
\end{eqnarray}
here $a$ is $c$ or $f$ and again $i<j$. With all above results, we have the
bare initial values for the RG equations derived from simply taking the
derivative of the RG potential in Ref. [\onlinecite{MSChang}].

\end{document}